\documentclass[aps,11pt,preprintnumbers,axodraw,nofootinbib]{revtex4}
\usepackage{multirow}
\usepackage{amsmath}
\usepackage{bm}
\usepackage{times}
\usepackage{graphicx}
\usepackage{color}
\usepackage{dsfont}
\usepackage{wrapfig}
\def\nn{\nonumber}
\def\bea{\begin{eqnarray}}
\def\eea{\end{eqnarray}}
\def\ba{\begin{eqnarray}}
\def\ea{\end{eqnarray}}
\def\be{\begin{equation}}
\def\ee{\end{equation}}
\def\beq{\begin{equation}}
\def\eeq{\end{equation}}
\def\gsim{{~\raise.15em\hbox{$>$}\kern-.85em
          \lower.35em\hbox{$\sim$}~}}
\def\lsim{{~\raise.15em\hbox{$<$}\kern-.85em
          \lower.35em\hbox{$\sim$}~}}

\newcommand{\slashed}{\slash \hspace{-0.23cm}}

\def\bsbsR{\bar b_R\gamma^\mu s_R\;\bar b_R\gamma_\mu s_R}
\def\bsbs{\bar b_L\gamma^\mu s_L\;\bar b_L\gamma_\mu s_L}

\def\Tr{{\rm Tr}}
\def\dzero{D\O}
\def\yudyu{Y_U^\dagger \,Y_U^{\phantom{\dagger}}}
\def\yddyd{Y_D^\dagger \,Y_D^{\phantom{\dagger}}}
\def\yuyud{Y_U^{\phantom{\dagger}} \, Y_U^\dagger}
\def\ydydd{Y_D^{\phantom{\dagger}} \, Y_D^\dagger}

\begin{document}
\title{\Large Flavor Symmetric Sectors and Collider Physics}

\author{Benjam\'\i{}n Grinstein}
\email[Electronic address:]{bgrinstein@ucsd.edu} 
\affiliation{Department of Physics, University of California at San Diego, La Jolla, CA 92093}

\def\Cincy{Department of Physics, University of Cincinnati, Cincinnati, Ohio 45221,USA}

\author{Alexander L. Kagan}
\email[Electronic address:]{kaganalexander@gmail.com} 
\affiliation{\Cincy}

\author{Michael Trott}
\email[Electronic address:]{mtrott@perimeterinstitute.ca} 
\affiliation{Perimeter Institute for Theoretical Physics, Waterloo, ON N2J-2W9, Canada}

\author{Jure Zupan\footnote{On leave of absence from University of Ljubljana, Depart. of Mathematics and Physics, Jadranska 19,
 1000 Ljubljana, Slovenia and Josef Stefan Institute, Jamova 39, 1000 Ljubljana, Slovenia.\\[-1mm]}}
\email[Electronic address:]{jure.zupan@cern.ch} 
\affiliation{\Cincy}

\date{\today}
\begin{abstract}
We discuss the phenomenology of effective field theories with new scalar or vector representations 
of 
the Standard Model quark flavor symmetry group, allowing for large flavor breaking involving the third generation.
Such field content can have a relatively low mass scale $\lesssim {\rm TeV}$ and $O(1)$ couplings to quarks, while being 
naturally consistent with both flavor violating and flavor diagonal constraints.   These theories therefore have the potential for early discovery at LHC, and provide a flavor safe ``tool box" for addressing anomalies at colliders and low energy experiments. 
We catalogue the possible flavor symmetric representations, and consider applications 
to the
anomalous Tevatron $t \, \bar{t}$ forward backward asymmetry and $B_s$ mixing measurements, individually or concurrently.
Collider signatures and constraints on flavor symmetric models are also studied more generally.
In our examination of the $t \, \bar{t}$ forward backward asymmetry
we determine model independent acceptance corrections appropriate for comparing against CDF data 
that can be applied to any model seeking to explain the $t \, \bar{t}$ forward backward asymmetry.
\end{abstract}
\maketitle
\newpage

\tableofcontents

\section{Introduction}
In recent years, the study of New Physics (NP) that lies close to the electroweak (EW) energy scale has been motivated primarily by the 
hierarchy problem.  However, it is possible that the correct
solution to this problem 
or 
the detailed nature of EW symmetry breaking remain to be proposed.  
Experimental input, as expected from the LHC, is crucial.
Furthermore, hints for new physics (NP) may have already emerged from the Tevatron.
In this paper we are motivated by recent experimental anomalies at the Tevatron
and the strong discovery potential at LHC to
explore collider signatures of new physics (NP) sectors that are flavor symmetric.   
They will  
be taken to be invariant under the global flavor symmetry group $\rm G_F= U(3)_{U_R}\times U(3)_{D_R}\times U(3)_{Q_L}$, or its subgroup $\rm H_F= U(2)_{U_R}\times U(2)_{D_R}\times U(2)_{Q_L}\times U(1)_3$ (where the quarks of the first two families are in doublets of the corresponding ${\rm SU(2)}$ factors).
The group $\rm G_F$
is the global symmetry of the Standard Model (SM) in the limit where one can neglect the Yukawa interactions
\begin{equation}
{\cal L}_Y=Y_{U} \, {\bar u}_{R}  \, H^T \, i \sigma_2 \,  Q_L - Y_{D} \, {\bar d}_{R} \,H^\dagger \, Q_L+{\rm h.c.}\,,
\end{equation} 
where $Y_U$ and $Y_D$ are the up and down quark Yukawa matrices, respectively.

The NP sectors will contain scalar or vector fields that have masses $\lesssim {\rm TeV}$ and $\mathcal{O}(1)$ couplings to quarks.  At the same time they will be consistent with flavor changing neutral current (FCNC) constraints precisely because of their flavor structure, as long as the breaking of the flavor symmetries is sufficiently small.
In the SM the top and bottom Yukawa couplings break the flavor group $\rm G_F$ to its $\rm H_F$ subgroup.
We take this breaking into account in our analysis of NP sectors that are initially $\rm G_F$ symmetric. The existence of the $\rm H_F$ symmetry at low energies protects these theories against dangerously large FCNC's, e.g., in neutral meson mixing. This protection satisfies the naturalness criteria of Glashow and Weinberg \cite{Glashow:1976nt} and is not the result of simply tuning parameters.
Note that NP models that have an approximate $\rm H_F$ symmetry are sometimes referred to as models of next to 
minimal flavor violation \cite{Agashe:2005hk}. 
The breaking of $\rm H_F$ in the SM is due to the other quark masses and the CKM mixing angles, and is thus small. 
The precise mechanism by which $\rm H_F$ is broken 
in the NP sector will not be important when we explore flavor diagonal collider signatures.
However, the nature of $\rm H_F$ breaking will be relevant to our discussion of low energy FCNC's, see below.

Scalar and vector fields with dimension four $\rm G_F$ invariant direct  couplings to quarks are limited in their allowed charge assignments by flavor and the SM gauge symmetry. There are only 14 different nontrivial flavor representations allowed in each case.
In the case of 
$\rm H_F$ symmetric models the possible NP fields are conveniently classified in terms of these representations, with the understanding that they need not come in complete $\rm G_F$ multiplets. 
A systematic exploration of new flavor symmetric sectors is therefore feasible, either in general, or with the aim of explaining a particular anomaly.\footnote{Color symmetric fermion content that mixes with the SM fermion fields is not as constrained in its allowed representations. Initial studies of vector-like fermions have also been undertaken in Refs.  \cite{Grossman:2007bd,Arnold:2010vs}. Of the models studied only two were natural in the Glashow-Weinberg sense \cite{Arnold:2010vs}.}

New flavor symmetric sectors that are perturbatively coupled to quarks are particularly interesting to consider as candidate explanations for 
Tevatron anomalies. 
In the first part of this paper, 
we focus our attention on two $>3 \sigma$ anomalies:
(i) the CDF measurement of the $t \bar t$ forward backward asymmetry, $A_{FB}^{t \bar t}$, for $m_{\bar{t} \, t} \geq 450 \,{\rm GeV}$ \cite{Aaltonen:2011kc}
is $3.4\sigma$ away from the NLO SM prediction. 
 (A recent D\O\, analysis~\cite{Abazov:2011rq}  does not observe a significant $m_{t\bar t}$ dependence in the ``folded" detector level asymmetries, but it appears to be consistent with the CDF detector level measurements within errors.)
The inclusive $t \bar t$ forward backward asymmetry, averaged over the CDF
semileptonic \cite{Aaltonen:2011kc} and hadronic \cite{CDF-dilepton} $t\bar t$ decay samples and the recent 
\dzero~measurement \cite{Abazov:2011rq}, is $\approx 3 \sigma$ from the NLO SM prediction;
(ii) the like sign dimuon asymmetry measured by  \dzero\  is $3.9\sigma$ away from the SM expectation \cite{Abazov:2010hv,Williams:2011nc}. 
Each of these anomalies, if confirmed, points to a relatively low scale of NP with a significant coupling to quarks.
We identify flavor symmetric models that have the potential to explain them either individually or simultaneously, and study related constraints. 
In the case of $\rm G_F$ symmetric models, under the assumption that the NP only couples to quarks, some hierarchies among these couplings would be required in order to consistently explain the $A_{FB}^{t \bar t}$ anomaly, e.g., due to the absence of dijet or $t\bar t $ resonances at the Tevatron and LHC.  Thus, 
breaking of the $\rm G_F$ symmetry to $\rm H_F$ would be necessary.  
Alternatively, one could consider $\rm H_F$
symmetric models where the more constrained quark couplings would simply be absent.

It is possible that the above anomalies could be due to statistical fluctuations or underestimates of
theoretical or experimental errors.  Even if this turns out to be the case, the models we explore in this paper are interesting
in their own right, 
as they have strong discovery prospects at LHC.  Again, this is because
their flavor symmetric structures allow 
for sub-TeV NP mass scales.
In the second part of this paper, we address the phenomenology of flavor symmetric sectors more generally. 
The global flavor symmetries we consider could be accidental, or they could be a remnant of the underlying mechanism generating the 
SM flavor structure (such as non-Abelian  horizontal symmetries). 
We do not concern ourselves with the UV origin of these symmetries, but instead focus on the collider and low energy phenomenology of the new sectors.
This approach is inspired by effective field theories (EFT), where one generally constructs all possible interactions 
consistent with the symmetries of interest.  The analysis of flavor diagonal collider constraints and signatures can then be kept quite general, i.e., independent of the way ${\rm H_F}$ is broken, as already mentioned.   
For simplicity, in attempts to explain a measured deviation from the SM
we will only consider the phenomenology of single $\rm G_F$ multiplets 
(or the corresponding $\rm H_F$ multiplets), effectively assuming that 
there is a significant mass gap with other possible representations.  Moreover, we will only consider their couplings to quarks. Note that more generally, these fields could couple
to additional states transforming under $\rm G_F$ or $\rm H_F$, possibly providing them with additional decay channels.

The determination of low energy flavor physics constraints on flavor symmetric models generally requires the breaking of $\rm H_F$ to be specified. 
When determining these constraints we assume the Minimal Flavor Violation (MFV) hypothesis \cite{Chivukula:1987py,D'Ambrosio:2002ex}, i.e., that {\it all} breaking of $\rm H_F$ is due to the SM Yukawas.  
This enforces maximal consistency with FCNC constraints through 
a symmetry principle, and allows us to explore how low the NP mass scale can be.  In all the models we consider, the new states can have EW scale masses.
In MFV models that lead to class-2 operators (those that involve right handed fields) in the language of \cite{Kagan:2009bn}, the breaking of $\rm H_F$ can actually 
be orders of magnitude larger then assumed in MFV, while still obeying the FCNC bounds. 

The paper is organized as follows. In Section \ref{MFVreps} we list
all vector and scalar representations of the form we have motivated, and write down
in detail the vector field Lagrangians for two examples.
In Section \ref{Tevatron} we systematically discuss the potential of models of this form 
to explain the $A_{FB}^{t \bar{t}}$ anomaly, the \dzero\  dimuon anomaly, and related phenomenology.
In Section \ref{lepEWPD} we explore existing bounds on these models from LEP,
electroweak precision data (EWPD), FCNCs and dijet studies at the LHC and Tevatron.  In Section \ref{LHC} we
discuss additional LHC phenomenology. Finally, in Section
\ref{conc} we give our conclusions.  Many details have been relegated to the Appendices.
In Appendix \ref{AppendixA} we list the details of flavor symmetric vector Lagrangians, in Appendix \ref{Apptevatron} we gives the details of $2\to 2$ scattering calculations and phenomenology, and in Appendix \ref{AppFCNCs} we give a detailed discussion of constraints from meson mixing amplitudes.


\section{$\rm G_F$ Symmetric Representations} \label{MFVreps}
\label{section:MFVrepresentations}
\begin{table}[t]
\begin{center}
\begin{tabular}[t]{c|ccccc}
  \hline
  \hline
   Case &$ {\rm SU(3)_c}$ & ${\rm SU(2)_{L}}$ &${\rm U(1)_Y}$  & ${\rm U(3)_{U_R} \times U(3)_{D_R} \times U(3)_{Q_L}}$ & Couples to \\
  \hline
$\rm I_{s,o}$ &  1,8 & 1& 0  & (1,1,1) & ${\bar{d}_R} \, \gamma^\mu \, d_R$ \\
$\rm II_{s,o}$ &  1,8 & 1& 0  & (1,1,1) & ${\bar{u}_R} \, \gamma^\mu \, u_R$ \\
$\rm III_{s,o}$ &  1,8 & 1& 0  & (1,1,1) & ${\bar{Q}_L} \, \gamma^\mu \, Q_L$ \\
$\rm IV_{s,o}$ &  1,8 & 3& 0  & (1,1,1) & ${\bar{Q}_L} \, \gamma^\mu \, Q_L$ \\
  \hline
  \hline
$\rm V_{s,o}$ &  1,8 & 1& 0  & (1,8,1) & ${\bar{d}_R} \, \gamma^\mu \, d_R$ \\
$\rm VI_{s,o}$ &  1,8 & 1& 0  & (8,1,1) & ${\bar{u}_R} \, \gamma^\mu \, u_R$ \\
$\rm VII_{s,o}$ &   1,8 & 1& -1  & ($\bar{3}$,3,1) & ${\bar{d}_R} \, \gamma^\mu \, u_R$ \\
$\rm VIII_{s,o}$ &  1,8 & 1& 0  & (1,1,8) & ${\bar{Q}_L} \, \gamma^\mu \, Q_L$ \\
$\rm IX_{s,o}$ &  1,8 & 3& 0  & (1,1,8) & ${\bar{Q}_L} \, \gamma^\mu \, Q_L$ \\
$\rm X_{\bar{3},6}$ &  $\bar{3}$,6 & 2& -1/6  & (1,3,3) & ${\bar{d}_R} \, \gamma^\mu \, Q_L^c$ \\
$\rm XI_{\bar{3},6}$ & $\bar{3}$,6 & 2& 5/6  & (3,1,3) & ${\bar{u}_R} \, \gamma^\mu \, Q_L^c$ \\
  \hline
  \hline
\end{tabular}
\end{center}
\caption{The flavor and gauge representations for vector fields that can couple directly to quarks through ${\rm G_F}$ symmetric dimension four interactions
without the insertion of a Yukawa matrix. 
$Q_L^c$ denotes the right handed conjugate representation of the left handed SM doublet.}
\label{int-vector}
\end{table}

We are interested in scalars and vectors that couple directly to
quarks through dimension four interactions. The scalar fields of this form are renormalizable models,
while the vector fields are nonrenormalizable.
In this section we list all the possible representations 
of $\rm G_F$ and the SM gauge group that such fields can have when $\rm G_F$ is unbroken. 
The vector field representations are listed in Table \ref{int-vector} and
the scalar field representations are listed in Table \ref{int-scalar} (the latter have been studied
and classified in \cite{Arnold:2009ay,Manohar:2006ga}). This completes the program initiated in \cite{Arnold:2009ay}.
The complete set of $\rm H_F$ symmetric representations which can couple directly to
quarks through dimension four interactions appear in these tables as submultiplets of the $\rm G_F$ representations.
For example, in model $\rm VI_{s(o)}$, the corresponding $\rm H_F$ symmetric vector representations would consist of  
a triplet, a complex doublet and a singlet of $\rm SU(2)_{U_R}$, which are color singlets (octets), or a subset of these.
  
\begin{table}
\begin{center}
\begin{tabular}[t]{c|ccccc}
  \hline
  \hline
   Case &${\rm SU(3)_c}$ & ${\rm SU(2)_{L}}$ &${\rm U(1)_Y}$  & ${\rm U(3)_{U_R} \times U(3)_{D_R} \times U(3)_{Q_L}}$ & Couples to \\
  \hline
$\rm S_I$ &      1& 2& 1/2  & (3,1,${\bar 3}$) & ${\bar{u}_R}$ \, $Q_L$ \\
$\rm S_{II}$ &      8& 2& 1/2  & (3,1,${\bar 3}$) & ${\bar{u}_R}$ \, $Q_L$ \\
$\rm S_{III}$ &         1& 2& -1/2  &(1,3,${\bar 3}$)& ${\bar{d}_R}$ \, $Q_L$ \\   
$\rm S_{IV}$ &         8& 2& -1/2  &(1,3,${\bar 3}$)& ${\bar{d}_R}$ \, $Q_L$ \\   
$\rm S_{V}$&    3& 1& -4/3 & (3,1,1) & $u_R$ \, $u_R$ \\
$\rm S_{VI}$&  $\bar{6}$ &1& -4/3 & (${\bar 6}$,1,1) & $u_R$ \, $u_R$  \\
$\rm S_{VII}$&    3& 1& 2/3 & (1,3,1)  & $d_R$ \, $d_R$ \\
$\rm S_{VIII}$&    $\bar{6}$ &1& 2/3 & (1,${\bar 6}$,1)  & $d_R$ \, $d_R$  \\
$\rm S_{IX}$&  3 & 1& -1/3 & (${\bar 3}$,${\bar 3}$,1) & $d_R$ \, $u_R$ \\
$\rm S_{X}$&   $\bar{6}$ & 1& -1/3 & (${\bar 3}$,${\bar 3}$,1) & $d_R$ \, $u_R$ \\
 $\rm S_{XI}$& 3  & 1 & -1/3 & (1,1,${\bar 6}$) & $Q_L$ \, $Q_L$ \\ 
 $\rm S_{XII}$&  $\bar{6}$  & 1 & -1/3 & (1,1,3) & $Q_L$ \, $Q_L$ \\  
 $\rm S_{XIII}$ & 3&3&-1/3&(1,1,3)&$Q_L$ \, $Q_L$ \\  
 $\rm S_{XIV}$&  $\bar{6}$  & 3 & -1/3 & (1,1,${\bar 6}$) & $Q_L$ \, $Q_L$ \\  
 \hline
 \hline
  $\rm S_{H,8}$&  $1,8$  & 2 & 1/2 & (1,1,1) & $\bar Q_Lu_R$,  \,\,$\bar Q_Ld_R$\\   
  \hline
  \hline
\end{tabular}
\end{center}
\caption{Different scalar representations that are not singlets under the flavor group that are $\rm G_F$ symmetric \cite{Arnold:2009ay} (the upper rows). The two flavor singlet representations are in the last row
and were discussed in \cite{Manohar:2006ga}.}
\label{int-scalar}
\end{table}

Several remarks are in order before we construct the Lagrangians. 
\begin{itemize}
\item  $\rm I_{s,o}$, $\rm II_{s,o}$
  and $\rm III_{s,o}$ carry the same quantum numbers and are thus 
  a sub-classification of interactions of a single field. For instance, in the case
  of a color singlet vector with the same couplings to $Q_L, u_R, d_R$ this is just the baryonic $Z'$. 
  We found it useful to split the interactions into three subgroups.  At
  colliders there is no interference among these interactions up to
  effects suppressed by light quark masses (but if there are relations between their couplings this can 
  have important consequences for the predicted cross section; for example, for a purely axial gluon the NP interference with the SM amplitude does not contribute to the top pair 
  production cross section \cite{Cao:2010zb,Blum:2011up,Tavares:2011zg}). In the treatment of
  FCNCs the interference effects are trivial to include in the analysis.
\item  Many of the scalar and vector fields do not lead to proton decay at any order in perturbation theory
due to the SM gauge
  symmetry and $\rm G_F$. 
  The vectors X--XI and scalars $\rm S_V$--$\rm S_{XIV}$ 
  carry baryon number $\pm 2/3$ and may lead to proton decay if they also couple to
  lepto-quark bilinears, {\it e.g.,} $\bar L^c_L\gamma^\mu u_R$ and $\bar
  L^c_L\gamma^\mu d_R$ for fields X and XI, respectively. This type of
  coupling is not possible for scalars or vectors in the color $\bf 6$
  representation and can be forbidden for the $\bf 3$ color representation fields by extending the flavor group to the lepton sector of the SM \cite{Arnold:2009ay} .
\item
We assume that the new quanta have weak scale masses and that the cut-off of the theory is well above the weak scale so that we only need to focus on dimension four interactions for most of our discussion. 
Other dimension four couplings such as $B_{\mu\nu} \Tr(V^\mu V^\nu)$, with $B_{\mu \nu}$ the hypercharge field strength and $V_\mu$ the vector fields, are not directly relevant to the 
phenomenology of interest in this paper. We leave the exploration of these interactions to a future publication.
\item
The kinetic terms (with flavor breaking insertions) can always be made universal through field redefinitions. Below, we only write down the interaction terms. 
\item 
Tree-level exchanges  of  fields in a single representation of $\rm G_F$ cannot explain {\it both} of the Tevatron anomalies simultaneously for any of the models considered. Models $\rm VII_{s,o}$ and $\rm S_I$ do, however, lead to enhanced $A_{FB}^{t\bar t}$, while not modifying the $t\bar t$ differential spectrum. At the same time they give new CP violating contributions to $B_s$ and $B_d$ mixing of the right order of magnitude to yield the observed like-sign dimuon asymmetry.
\end{itemize}

The interaction Lagrangians for the color triplet and sextet scalar fields are given in \cite{Arnold:2009ay}. For vector fields the ${\rm G_F}$ symmetric interactions are given by $\bar q\gamma^\mu T^a q'
V_\mu^a $, where $T^a$ represents a product of generators of color, flavor, and
weak SU(2), or some subset thereof, while $q^{(\prime)}$ are the $u_R$, $d_R$ or $Q_L$ 
family triplets, as appropriate  
\footnote{In $\rm H_F$ symmetric models the Lagrangians are trivially obtained from the corresponding $\rm G_F$ symmetric Lagrangians, allowing for the possibility that only particular submultiplets of the $\rm G_F$ symmetric representations are present.}.  To write down the $\rm G_F$ breaking interactions it is useful to (initially) adopt some of the formalism of MFV and promote $Y_{U,D}$ to spurions 
that formally transform as bi-fundamentals of $\rm G_F$
\begin{equation}
Y_U \rightarrow V_U \, Y_U \, V_Q^{\dagger},~~~~~~~~~~Y_D \rightarrow V_D \, Y_D \, V_Q^{\dagger}.
\end{equation}
Here $V_{U,D,Q}$ are elements of $\rm SU(3)_{U,D,Q}$, respectively. Assuming full MFV breaking of $\rm G_F$, all
interactions are then formally invariant under $\rm G_F$ even for nonzero
Yukawa couplings.  We will mostly work to the first nontrivial order in top Yukawa insertions (the resummation to all orders can 
be done using a nonlinear representation of ${\rm G_F}$, see \cite{Feldmann:2008ja,Kagan:2009bn,Feldmann:2009dc,Barbieri:2011ci}). 
The explicit forms of the interactions for all the vector models are given in 
Appendix \ref{AppendixA}. Here we show two examples, models $\rm VI_{s,o}$ and  $\rm X_{\bar{3},6}$.

\underline{\bf Fields $\rm VI_{s(o)}$} are ${\rm SU(3)_{U_R}}$ flavor octets, color singlets (octets). The individual field components are $V_\mu^B$
$(V_\mu^{A,B})$, where the color label $A$ and flavor label $B$ both run over $1..8$. To compress the expressions we introduce
\begin{equation}\label{Vmus}
V^s_\mu= T^B V_\mu^B , \qquad V^o_\mu= {\cal T}^A T^B V_\mu^{A,B}.
\end{equation}
with flavor (color) Gell-Mann matrices $T^B$($\mathcal{T}^A$) normalized to ${\rm Tr}[T^A T^B] = \delta^{AB}/2$. The renormalizable
interactions between quarks and fields $\rm VI_{s,o}$ are
\beq\label{symmetricV}
\mathcal{L}_{\rm VI_{s,o}} =  \eta_1^{s,o} \, \bar{u}_R \,  \slashed V^{s,o} u_R.
\eeq
There are also terms that break ${\rm G_F} \rightarrow {\rm H_F}$,
\beq \label{breakVI}
\Delta
\mathcal{L}_{\rm VI_{s,o}} = [\eta_2^{s,o} \, \bar{u}_R (\slashed V^{s,o}\Delta_U) u_R
+h.c.]
+\tilde \eta_3^{s,o}\, \bar{u}_R (\Delta_U \slashed V^{s,o}\Delta_U)  u_R+\dots.  
\eeq 
We kept only the breaking due to $\Delta_U\equiv \yuyud $ insertions. These are diagonal in the up-quark basis and change the coupling of third generation quarks to the vector fields (explicitly written out in Appendix \ref{AppendixA}, Eqs. \eqref{VIoEq}, \eqref{VIsEq}). Note that $\eta_2^{s,o}$ can be complex and possibly a source of beyond the SM
CP violation, of interest when considering $B_s$ mixing, while $\tilde \eta_3^o$ is real. Insertions of
$Y_U^{\phantom{\dagger}} Y_D^\dagger Y_D^{\phantom{\dagger}}
Y_U^\dagger$ are also possible to break the symmetry further and are almost diagonal in the up
quark basis, while the off-diagonal elements lead to FCNC's. We postpone the
discussion of these until Section
\ref{FCNC}.

The ${\rm G_F} \rightarrow {\rm H_F}$ breaking also splits the vector mass spectrum. The flavor invariant mass terms are 
\bea\label{mass1s}
\mathcal{L}_{VI_{a}}^{mass}  = (1+\delta_{a,o})\left\{m_{V_a}^2 \, {\rm Tr} \left[ \tilde{V}_\mu^{a} \,  \tilde{V}^{\mu a} \right] + \lambda \, (H^\dagger \, H) \, {\rm Tr} \left[ \tilde{V}_\mu^a \,  \tilde{V}^{\mu a} \right]\right\}, \qquad a=o,s,
\eea
where the color and $\rm SU(3)$ indices are suppressed, and there is no summation over $a=o,s$ (the Kronecker delta $\delta_{a,o}$ insures the proper normalization for the color octet fields). 
Note that $\tilde{V}$ is defined when rotating to the mass eigenstate basis; see Appendix A.
Adding the ${\rm G_F} \rightarrow {\rm H_F}$ breaking terms $(1+\delta_{a,o})  \zeta_1 m_V^2 {\rm Tr} \big[ \tilde{V}_\mu \, \Delta_U  \tilde{V}^{\mu} \big] $
and $(1+\delta_{a,o}) \zeta_2 m_V^2 {\rm Tr} \big[\Delta_U  \,  \tilde{V}_\mu \, \Delta_U  \tilde{V}^{\mu} \big]$,   
the mass spectrum of the vector states is (suppressing the $o,s$ labels)
\beq
\begin{split}
m_{1,2,3}^2&= m_V^2+\frac{\lambda}{2} v^2, \qquad m_{4,5,6,7}^2= m_{1}^2+m_V^2\frac{{\zeta}_1}{2} \, y_t^2,\qquad
 m_8^2= m_{4}^2 + m_V^2 \frac{{2 \zeta}_2}{3} \, y_t^4. \label{Vimass}
\end{split}
\eeq
Note that $\lambda, \zeta_{1,2}$ are all real.  The vectors $V_{1,2,3}$ and $V_{4,5,6,7}$
are degenerate since $\rm SU(2)_U$ is only broken by light quark Yukawas, not by $y_t$.

\underline{\bf Fields $\rm X_{\bar{3},6}$} are weak doublets in the bi-fundamental representation $(1,3,3)$ of the flavor group. The color
anti-triplets have field components $(V_\mu)^{r
  \gamma}_{i,j}$,  and color sextets the
field components $(V_\mu)^{r}_{i,j,\alpha,\beta}=(V_\mu)^{r}_{i,j,\beta,\alpha}$, with $r$ the weak $\rm SU(2)_L$ index, $\alpha, \beta,\gamma$ the color
indices, while $i$ and $j$ are the indices of the $(1,3,1)$ and $(1,1,3)$ representations respectively. 
The tree level quark coupling Lagrangian terms are (suppressing all the indices apart from color, see also Eqs. \eqref{X3:Eq},  \eqref{X6:Eq})
\bea
\mathcal{L}_{X_{\bar{3}}} =  \eta_1 \, \epsilon_{\alpha \beta \gamma}\, \bar{d}_R^{\alpha} \,  \slashed V^{\gamma}  \, Q_L^{c\beta} + h.c., \qquad 
\mathcal{L}_{X_{6}} =  \eta_1 \, \bar{d}_R^{\alpha}   ( \slashed V)_{\alpha, \beta}  \, Q_L^{c\beta}+ h.c.
\eea
Note that the $V_\mu$ fields transform as $V_\mu \to V_D V_\mu V_Q^T$, where $d_R\to V_D d_R$ and $Q_L^c\to V_Q^\ast Q_L^c$. The mass terms are
\bea\label{massX}
\mathcal{L}_{X_{o,s}}^{\rm mass}  = m_V^2 {\rm Tr}(V_\mu V^{\mu\dagger}) + \lambda \,( H^\dagger \, H ){\rm Tr}(V_\mu V^{\mu\dagger})+ \lambda' {\rm Tr}(H_r V_\mu^r H^{s \dagger} V_s^{\mu\dagger}) +\lambda' {\rm Tr}(H^C_r V_\mu^r H^{C s \dagger} V_s^{\mu\dagger}),
\eea
where we have suppressed all the traced over flavor, color and weak indices (except in the last two terms where we show explicitly the weak contractions). We use $[(V_\mu)^{\phantom{\dagger}r \gamma}_{i,j}]^\ast=(V_\mu^\dagger)_{r \gamma}^{j,i}$, and similarly for the sextet. Note that the last two terms break the mass degeneracy between charge $1/3$  and charge $-2/3$ components of $V_\mu$ weak doublets. The flavor is broken through Yukawa insertions 
\bea\label{massflavorX}
\Delta \mathcal{L}_{X_{o,s}}^{\rm mass}/m_V^2=  \zeta_1 {\rm Tr}(V_\mu Y_D^\dagger Y_D^{\phantom{\dagger}}  V^{\mu\dagger}) + \zeta_2 {\rm Tr}(Y_D^{\phantom{\dagger}}  Y_D^\dagger V_\mu  V^{\mu\dagger}) +\zeta_3 {\rm Tr}(V_\mu Y_U^{\phantom{\dagger}}  Y_U^\dagger V^{\mu\dagger}) +\cdots,
\eea
where we do not write down the terms with more than two Yukawa insertions or the similar terms with Higgs fields. The resulting ${\rm H_F}$ symmetric spectrum for charge $2/3$  vectors is
\beq
\begin{split}
m^2_{11,12,21,22} &=  m_V^2 + \frac{1}{2} (\lambda+\lambda') \, v^2 , \qquad m^2_{13,23,32,31,33} = m^2_{11}+ {\zeta_3} m_V^2 y_t^2.
 \end{split}
\eeq
The interactions of mass eigenstates $\tilde{V}_{k l}$  with mass eigenstate quarks (denoted with primes) are given by (showing explicitly only color contractions, see also Eqs. \eqref{X3:EqBreak},  \eqref{X3:EqBreak})
\beq
\begin{split} \label{XosEq}
\mathcal{L}_{X_{\bar{3}}} &=  \eta_1 \, \epsilon_{\alpha \beta \gamma}\, (\bar{d_R'})^{\alpha} \,  \slashed \tilde V_1^{\gamma}  \, (u_L')^c{}^{\beta} + \eta_1 \, \epsilon_{\alpha \beta \gamma}\, (\bar{d_R'})^{\alpha} \,  \slashed \tilde V_2^{\gamma} V_{\rm CKM} \, (d_L^{'c}){}^{\beta} +h.c., \\
\mathcal{L}_{X_{6}} &=  \eta_1 \, (\bar{d_R'})^{\alpha}   ( \slashed V)_{\alpha, \beta}  \, (u_L')^{c\beta}+ \eta_1 \, (\bar{d_R'})^{\alpha}   ( \slashed V)_{\alpha, \beta}  V_{\rm CKM}  \, (u_L')^{c\beta}+ h.c., 
\end{split}
\eeq
where $\tilde V_{1}$ and $\tilde V_2$ are the mass eigenstate vector fields of the $\rm SU(2)$ doublet. 
The residual $\rm H_F$ flavor universality of these interactions can be broken by insertions of the spurions $\yudyu $ and  $\yddyd $. In MFV
this is the only form of further flavor breaking. The rest of the Lagrangian constructions are collected in the Appendix \ref{AppendixA}.

\section{Phenomenology of Tevatron anomalies}\label{Tevatron}
We now discuss two recent experimental anomalies observed at the Tevatron: the large forward-backward asymmetry $A_{FB}^{t\bar t}$, and the like-sign dimuon anomaly
in $B_s$ decays. 
In this section, we systematically address the following questions: 
\begin{itemize}
\item
Is it possible to explain either of the two anomalies assuming $\rm H_F$ symmetric models? By which charge and flavor assignments?
\item
Are closely related experimental constraints simultaneously obeyed?
\item
Is it possible to explain both anomalies using just a single $\rm H_F$ symmetric field? 
\end{itemize}

A common feature of models put forward to explain the $A_{FB}^{t \bar{t}}$ anomaly  \cite{Gresham:2011fx,Nelson:2011us,Jung:2011id,Ferrario:2009bz,Arhrib:2009hu,Arnold:2009ay,AguilarSaavedra:2011vw,Kamenik:2011wt,Cheung:1995nt,Antipin:2008zx,Gupta:2009wu,Hioki:2009hm,Choudhury:2009wd,Hioki:2010zu,HIOKI:2011xx,Blum:2011up,Jung:2009pi,Zhang:2010dr,Delaunay:2011gv,Jung:2011zv,Jung:2009jz,Cheung:2009ch,Shu:2009xf,Dorsner:2009mq,Cao:2009uz,Barger:2010mw,Cao:2010zb,Xiao:2010hm,Cheung:2011qa,Cao:2011ew,Shelton:2011hq,Berger:2011ua,Barger:2011ih,Bhattacherjee:2011nr,Patel:2011eh,Barreto:2011au,Craig:2011an,Buckley:2011vc,Shu:2011au,Jung:2011ua,Fox:2011qd,Cui:2011xy,Duraisamy:2011pt,AguilarSaavedra:2011ug,Dorsner:2010cu,Blum:2011fa,Gresham:2011dg,Grinstein:2011yv,Delaunay:2011vv,Babu:2011yw,Ligeti:2011vt,Tavares:2011zg,Isidori:2011dp,Frampton:2009rk,Chivukula:2010fk,Bai:2011ed,Xiao:2010ph,Ferrario:2009ee,Martynov:2010ed,Bauer:2010iq,Chen:2010hm,Burdman:2010gr,Degrande:2010kt,Choudhury:2010cd,Cao:2010nw,Foot:2011xu,Haisch:2011up} is that they have $O(1)$ NP couplings to the up quark.  They fall roughly into two classes, those with $s$-channel exchange above a TeV\footnote{For a recent exception with a sub-TeV axigluon, see \cite{Tavares:2011zg}.}  \cite{Frampton:2009rk,Ferrario:2009bz,Ferrario:2009ee}, in which case 
the axial vector NP couplings to the top quarks and up quarks must be of opposite sign \cite{Ferrario:2009bz,Frampton:2009rk}, and those with sub-TeV $t$-channel exchange \cite{Jung:2009jz,Cheung:2009ch,Frampton:2009rk,Shu:2009xf,Arhrib:2009hu,Dorsner:2009mq,Cao:2009uz}, in which case large inter-generational couplings are required (for additional possibilities, see \cite{Kamenik:2011wt}).
The couplings in either class could arise from large flavor violation in the underlying theory, which may lead to 
violations of FCNC constraints in 
$K^0-{\bar K}^0, B_d^0-{\bar B_d}^0, D^0-{\bar D}^0$ mixing, $B\to K\pi$, or $b \rightarrow s\gamma$, 
unless the couplings are carefully aligned (see, e.g., \cite{Shelton:2011hq,Jung:2011zv,Shu:2011au}).
Moreover, the $t$-channel models can lead to excessive (flavor violating) single top or same sign top pair production at the Tevatron and LHC \cite{Jung:2009jz}.

However, flavor violation is not necessary for large $A_{FB}^{t\bar t}$ \cite{Grinstein:2011yv}. Note that in $t\bar t$ production no net top quark flavor charge is generated. 
Furthermore, models with
an unbroken $\rm H_F$ subgroup do not generate FCNCs in processes with light quarks. 
The exact size of FCNCs 
then depends on the size of $\rm H_F$ breaking. If this breaking is MFV-like the FCNCs are generically suppressed below present bounds. 
The flavor symmetries also eliminate single top and same sign top production.

Many new models have also been put forward to explain the $\rm DO \! \! \! \!/$  dimuon anomaly \cite{Dobrescu:2010rh,Buras:2010mh,Jung:2010ik,Chen:2010aq, Blum:2010mj,Buras:2010pz,Buras:2010zm,Trott:2010iz}. 
Together with possible indications for deviations from the SM in  $B_s \rightarrow J/\psi \, \phi$ decays and $B^- \rightarrow \tau^- \, \nu$ decays, 
it may point to a NP phase in $B_{s,d}$ mixing.  Intriguingly, MFV suffices to explain the dimuon anomaly \cite{Ligeti:2010ia}.  After discussing flavor symmetric fields and the $A_{FB}^{t \bar{t}}$ anomaly, we will examine whether these fields can also give large enough contributions to $B_s$ mixing, under the assumption of MFV breaking of ${\rm H_F}$. 

\subsection{General analysis of the $t \, \bar{t}$ forward backward asymmetry}

We entertain the possibility that $A_{FB}^{t \bar{t}}$ is enhanced above SM levels via
tree level exchanges of flavor symmetric scalars or vectors.
The experimental evidence for such enhancement is as follows. Using $5.3 {\rm fb}^{-1}$ of data CDF measured an inclusive asymmetry $A_{FB}^{t \bar{t}} = 0.158 \pm0.072 \pm 0.017$ in the $t \, \bar{t}$ rest frame  (fixing $m_t = 172.5 \, {\rm GeV}$)~\cite{Aaltonen:2011kc}. In a channel with both $t$ and $\bar t$ decaying semileptonically an even larger asymmetry was found, $A_{FB}^{t \bar{t}}= 0.42 \pm0.15 \pm 0.05$ \cite{CDF-dilepton}. Similarly, a recent D\O\, analysis finds $A_{FB}^{t \bar{t}} = 0.196 \pm0.060^{+0.018}_{-0.026}$ using $5.4 {\rm fb}^{-1}$ of data ~\cite{Abazov:2011rq}.   
Combining in quadrature the statistical and systematic errors of the three measurements gives $A_{FB}^{t \bar{t}}= 0.200\pm 0.047$. 
This is to be compared to the SM prediction $A_{FB}^{t \bar{t}, SM} = 0.072^{+0.011}_{-0.007}$ from an approximate NNLO QCD calculation~\cite{Ahrens:2011uf}  with $m_t = 173.1 \, {\rm GeV}$ and using the MSTW2008 set of PDFs~\cite{Martin:2009iq}. Inclusion of electroweak corrections leads to an enhancement of the asymmetry, with 
 $A_{FB}^{t \bar{t}, SM} = 0.09 \pm 0.01$ recently obtained in Ref. \cite{Hollik:2011ps}.
In the $p\bar p$ frame, a recent approximate NNLO calculation~\cite{Kidonakis:2011zn} gives
 $A_{FB}^{t \bar{t} \, SM} = 0.052^{+0.000}_{-0.006}$ with $m_t = 173 \, {\rm GeV}$, to be compared with the CDF value of $A_{FB}^{t \bar{t}} = 0.150 \pm0.058 \pm 0.024$~\cite{Aaltonen:2011kc}. 
  The approximate NNLO SM predictions use the known NLO results~\cite{Antunano:2007da,Bowen:2005ap,Kuhn:1998kw} 
  and build on recent progress in NNLO calculations~\cite{Moch:2008ai,Czakon:2009zw,Beneke:2009ye,Kidonakis:2008mu,Cacciari:2008zb,Kidonakis:2010dk}.  
D\O \, also reports a leptonic asymmetry $A_{FB}^l = 0.152\pm0.038^{+0.010}_{-0.013}$ to be compared to the MC@NLO prediction of $A_{FB}^{l,\rm SM} = 0.021\pm0.001$~\cite{Abazov:2011rq}.

CDF reported evidence that the anomalously large asymmetry rises with the invariant mass of the $t \, \bar{t}$ system, with  $A_{FB}^{t \bar{t}}(M_{t\bar t}> 450 {\rm ~ GeV})=0.475\pm0.114$,  while $A_{FB}^{t \bar{t}}(M_{t\bar t}< 450 {\rm ~ GeV})=-0.116\pm0.153$~\cite{Aaltonen:2011kc}. 
A similar rise of the asymmetry with respect to the absolute top vs. anti-top rapidity difference $|\Delta y| = |y_t-y_{\bar t}|$ was also reported by CDF with $A_{FB}^{t \bar{t}}(| \Delta y | < 1.0) = 0.026 \pm 0.104 \pm 0.056$ and  $A_{FB}^{t \bar{t}}(| \Delta y | > 1.0) = 0.611 \pm 0.210 \pm 0.147$~\cite{Aaltonen:2011kc}. The recent D\O\, analysis~\cite{Abazov:2011rq}  does not observe a significant rise of the ``folded" detector level asymmetry with respect to $M_{t\bar t}$ and $|\Delta y|$.  However, until these results are unfolded they can not be directly compared to the CDF measurements, although at the detector level they appear to be consistent within errors.
We collect the above results in Table \ref{table:partondata}.

\begin{table*}
\center
\begin{tabular}{ c | c  c } 
\hline \hline 
Observable & Measurement  &  SM predict. \\ 
\hline  
$A_{FB}^{t\bar t}$ & $
  \left.
\begin{matrix}
0.158 \pm 0.072 \pm 0.017\, \text{\cite{Aaltonen:2011kc}}\\
 0.42 \pm0.15 \pm 0.05 \, \text{\cite{CDF-dilepton}}\\
0.196 \pm0.060^{+0.018}_{-0.026} \, \text{\cite{Abazov:2011rq}}\\
 \end{matrix}
 \right\} \simeq 0.200\pm0.047
 $ & 
 $(7.24^{+1.04}_{-0.67}{}^{+0.20}_{-0.27})\cdot 10^{-2}  \text{\cite{Ahrens:2011uf}}$\\
$A_{FB}^{t\bar t}(M_{t\bar t}>450$GeV)  &  $0.475 \pm 0.101 \pm 0.049$ \cite{Aaltonen:2011kc}  & 
$
(11.1^{+1.7}_{-0.9})\cdot 10^{-2}  \text{\cite{Ahrens:2011uf}}$\\
$A_{FB}^{t\bar t}(M_{t\bar t}<450$GeV) &  $-0.116 \pm 0.146 \pm 0.047$ \cite{Aaltonen:2011kc} & $(5.2^{+0.9}_{-0.6})\cdot 10^{-2} $  \cite{Ahrens:2011uf}\\
$A_{FB}^{t\bar t}(| \Delta y | < 1.0$) &  $0.026 \pm 0.104 \pm 0.056$  \cite{Aaltonen:2011kc}  & $(4.77^{+0.39}_{-0.35})\cdot 10^{-2} $  \cite{Ahrens:2011uf} \\
$A_{FB}^{t\bar t}(| \Delta y | > 1.0$) &  $0.611 \pm 0.210 \pm 0.147$  \cite{Aaltonen:2011kc} & $(14.59^{+2.16}_{-1.30})\cdot 10^{-2} $  \cite{Ahrens:2011uf} \\
$\sigma_{t\bar t}$&  $(6.9 \pm 1.0)$pb \cite{Aaltonen:2009iz}  & 
$\left\{
\begin{matrix}
(6.63^{+0.00}_{-0.27})\text{pb  \cite{Ahrens:2011mw}}  \\ 
(7.08^{+0.00}_{-0.24}{}^{+0.36}_{-0.27})\text{pb  \cite{Kidonakis:2011jg}}
\end{matrix}
\right.
$
\\
\hline\hline
\end{tabular}
\caption{Measurements and predictions for observables in $t\bar t$ production at the Tevatron. We quote the approximate NNLO QCD prediction of $A_{FB}^{t \, \bar{t}}$ from~\cite{Ahrens:2011uf} using MSTW2008 PDFs~\cite{Martin:2009iq}.  The two other choices for PDFs give results in agreement with these~\cite{Ahrens:2011uf}. Among the cross section predictions obtained in~\cite{Ahrens:2011mw} we quote the ${\rm 1PI}_{\rm SCET}$ one.}
\label{table:partondata}
\end{table*}

Any NP enhancement of $A_{FB}^{t \bar{t}}$ must not spoil the 
agreement between the measured
production cross section,  $\hat{\sigma}_{t \, \bar{t}}$, and the SM predictions. At NLO with NNLL summation of threshold logarithms, the SM prediction is  $\sigma_{t\bar t}=(6.63^{+0.00}_{-0.27})$pb  \cite{Ahrens:2011mw} (using MSTW2008 pdf sets and ${\rm 1PI_{SCET}}$ choice of kinematic variables and resummations -- the other choices give consistent results but with larger error bars). This is  
somewhat smaller than the approximate NNLO result (for $m_t = 173 \,\, {\rm GeV}$),
$\sigma_{t\bar t}=7.08^{+0.00}_{-0.24}{}^{+0.36}_{-0.27}$ pb \cite{Kidonakis:2011jg} (see also \cite{Cacciari:2008zb,Kidonakis:2008mu,Moch:2008ai}). Both of these results
agree well, within errors, with the measured CDF result based on $4.6 fb^{-1}$ of data \cite{Aaltonen:2009iz}
$\sigma_{t \, \bar{t}}(m_t = 173.1 \, {\rm GeV}) = 6.9 \pm 1.0 {\rm pb}$.
Thus the NP contribution to the $t\bar
t$ cross section, $\sigma_{t \, \bar{t}}^{\rm NP}$, is tightly
constrained. 

Good agreement between experiment and SM predictions is
also seen in the differential cross section $d\sigma/d M_{t\bar
  t}$. This has important implications for the viability of different NP models. For instance, comparing the measured and predicted cross sections together with the measured and predicted $A_{FB}^{t \bar{t}}$, for $M_{t\bar t}>450$ GeV,  one finds that  
the NP contributions need to reduce the backward $t\bar t$-scattering cross section (a statement valid at $2\sigma$). This can only happen if NP interferes with the SM \cite{Grinstein:2011yv}. NP in the $s$-channel
which interferes with the single gluon exchange amplitude must therefore be due to color octet fields. 
In general $s$-channel resonances lead to significant effects in $d \sigma / d M_{t \, \bar{t}}$.  However, this may be avoided for a purely axial gluon that is broad \cite{Tavares:2011zg}, in particular regions of parameter space.
There are no such clear requirements on the charge assignments of possible $t$-channel NP
contributions.  However, a characteristic high mass tail in the spectrum could lead to tension with the Tevatron and future LHC cross section measurements at large $M_{t\bar t}$.

We collect expressions to be used in our analysis below. The total cross section  $\sigma_{t \, \bar{t}}$, forward-backward asymmetry $A_{FB}^{t \bar{t}}$, and the cross sections for forward and backward scattering $\sigma_{F,B}$ are
defined as
\beq
 \sigma_{t \, \bar{t}} =\sigma_F+\sigma_B, \qquad A_{FB}^{t \bar{t}} = \frac{\sigma_F-\sigma_B}{\sigma_F+\sigma_B},\qquad \sigma_{F(B)}=\int_{0(-1)}^{1(0)} d \cos\theta \,\frac{d\sigma}{d\cos\theta}, 
\eeq
where $\theta$ is the angle between incoming proton and outgoing top quark. 
We use NLO SM predictions for $\sigma_{t \, \bar{t}}$ and $d\sigma_{t \, \bar{t}}/dM_{t\bar t}$, and LO predictions for the NP corrections (including interference with the SM). 
To obtain $A_{FB}^{t \bar{t}}$ we define a partonic level asymmetry,
\bea
A_{FB}^{NP+ SM} = \frac{\sigma^{NP}_F - \sigma^{NP}_B}{(\sigma^{\rm NP+ SM}_F)_{LO}  + (\sigma^{\rm NP+ SM}_B)_{LO}} + A_{FB}^{SM} \left(\frac{\sigma^{SM}}{\sigma^{SM} + \sigma^{NP}} \right),
\eea
which is to be compared against the binned unfolded partonic level results of \cite{Aaltonen:2011kc}. 
We use the NLO + NNLL SM predictions for the forward, backward and total cross sections \cite{Ahrens:2010zv}, the $t\bar t$ spectrum \cite{Ahrens:2010zv}
and $A_{FB}^{t \bar{t}, SM}$ \cite{Ahrens:2011uf}. For concreteness, for
$M_{t \, \bar{t}} < 450$ GeV we take the central values $\sigma^{SM} = 4.23$ pb and $A_{FB}^{t \bar{t}, SM} = 0.052$, while for $M_{t \, \bar{t}}> 450$ GeV we take $\sigma^{SM} = 2.40$ pb and $A_{FB}^{t \bar{t}, SM} = 0.111$ (MSTW08 pdfs); for the inclusive asymmetry (in the $t\bar t$ rest frame) we take $A_{FB}^{t \bar{t}}=0.0724$.

\subsubsection{Acceptance effects}
\label{Acceptance}
As pointed out in \cite{Gresham:2011pa,Jung:2011zv}, care is needed when comparing NP predictions to the experimental parton level $A_{FB}^{t\bar t}$ and $M_{t\bar t}$ differential spectrum deduced by CDF \cite{Aaltonen:2009iz,Aaltonen:2011kc}, since the deconvolution was done assuming the SM. The acceptance corrections are especially important if NP enhances top production in the very forward region 
This is because the SM $t\bar t$ event distribution is more central.  We take into account the CDF experimental cuts using correction factors $\epsilon_i$. For the $i-$th bin in $M_{t \bar t}$ one needs to multiply the calculated partonic $t\bar t$ cross section $d\sigma^{\rm NP}/dM_{t\bar t}$ by $\epsilon_i$ in order to compare with the CDF measured partonic cross section $(d\sigma^{\rm NP}/dM_{t\bar t})^{\rm CDF}$
\beq\label{eq:conversion}
\left(\frac{d\sigma^{\rm NP}}{dM_{t\bar t}}\right)^{\rm CDF}_i=\epsilon_i \times \left(\frac{d\sigma^{\rm NP}}{dM_{t\bar t}}\right)_i.
\eeq
There is no summation over $i$ in this equation.
Since CDF is using SM acceptances and no bins in $\Delta y$ in the deconvolution of the $d\sigma/dM_{t\bar t}$ measurement in \cite{Aaltonen:2011kc}, the $\epsilon_i$ are given by the ratio of acceptances for the NP model and the SM
\beq
\epsilon_i=\frac{\epsilon_i^{\rm NP}}{\epsilon_i^{\rm SM}},
\eeq
where $\epsilon_i^{\rm NP (SM)}$ are calculated by splitting each $i$-th $M_{t\bar t}$ bin into $j$ bins in $\Delta y=y_t-y_{\bar t}$
\beq\label{eq:epsilon_i}
\epsilon_i^{\rm NP (SM)}=\frac{\sum_{j} \epsilon_{ij} \sigma_{ij}^{\rm NP (SM)}}{\sum_{j} \sigma_{ij}^{\rm NP (SM)}},
\eeq
and the sum is over the bins in $\Delta y$. Here $\epsilon_{ij}$ is the acceptance for each $(M_{t\bar t},\Delta y)$ bin, and $\sigma_{ij}^{\rm NP (SM)}$ is the corresponding cross section integrated over the bin.  The above expressions are approximate in so far as the bins have finite sizes, and the spill-over of events between different bins is not taken into account.  The acceptances are calculated by simulating the partonic $t -\bar t$ cross section using {\tt MadGraph4.4.30} \cite{Alwall:2007st}, decaying the top quarks in {\tt Pythia6.4} \cite{Sjostrand:2006za}, which also simulates the LO showering and hadronization, and using {\tt PGS} for detector simulation. The events were read into {\tt Mathematica}, where the cuts from \cite{Aaltonen:2008hc,Aaltonen:2011kc} were implemented. The resulting values for the acceptances $\epsilon_{ij}$ are collected in Table \ref{tab:acceptances}.

\begin{table}[tb]
\begin{center}
\begin{tabular}[t]{c|ccccc}
  \hline
  \hline
  $ M_{t\bar t}$ (GeV)$: /\Delta y:$&$0-0.6$ & $0.6-1.2$ &$1.2-1.8$  & $1.8-2.4$ & $2.4-3.0$ \\
  \hline
$350-400$ & $2.42$ & $2.23$ & $-$& $-$  & $-$ \\
$400-450$ & $3.44$ & $2.74$ & $1.95$& $-$  & $-$ \\
$450-500$ & $4.29$ & $3.32$ & $1.75$& $-$  & $-$ \\
$500-550$ & $5.25$ & $3.50$ & $2.34$& $1.18$  & $-$ \\
$550-600$ & $5.61$ & $4.39$ & $2.67$& $1.15$  & $-$ \\
$600-700$ & $6.59$ & $4.81$ & $3.01$& $1.09$  & $0.42$ \\
$700-800$ & $7.38$ & $6.06$ & $3.49$& $1.34$  & $0.45$ \\
$800-1400$ & $6.68$ & $6.26$ & $4.30$& $1.83$  & $0.42$ \\
  \hline
  \hline
\end{tabular}
\end{center}
\caption{Acceptances $\epsilon_{ij}$ (in \%) to be used for emulating the CDF deconvolution to parton level measurements (cf. eqs. \eqref{eq:conversion} - \eqref{eq:epsilon_i}). Note that the $\epsilon_{ij}$ do not depend on the NP model and are symmetric in $\Delta y\leftrightarrow -\Delta y$.}
\label{tab:acceptances}
\end{table}

\begin{figure}
\includegraphics[width=0.5\textwidth]{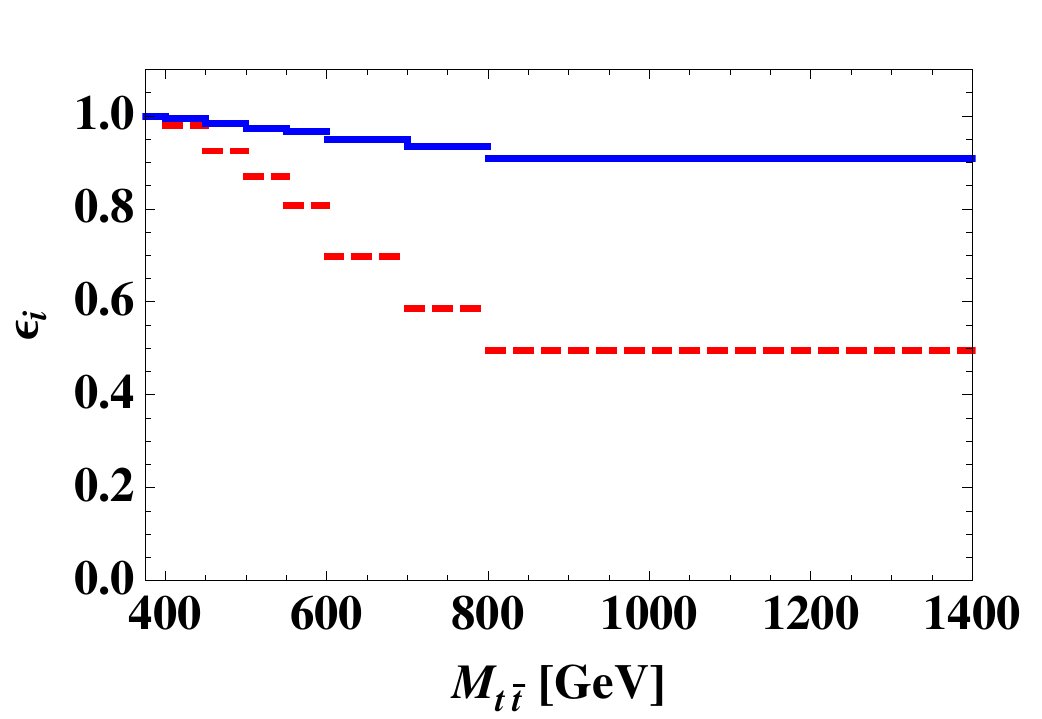}%
\includegraphics[width=0.5\textwidth]{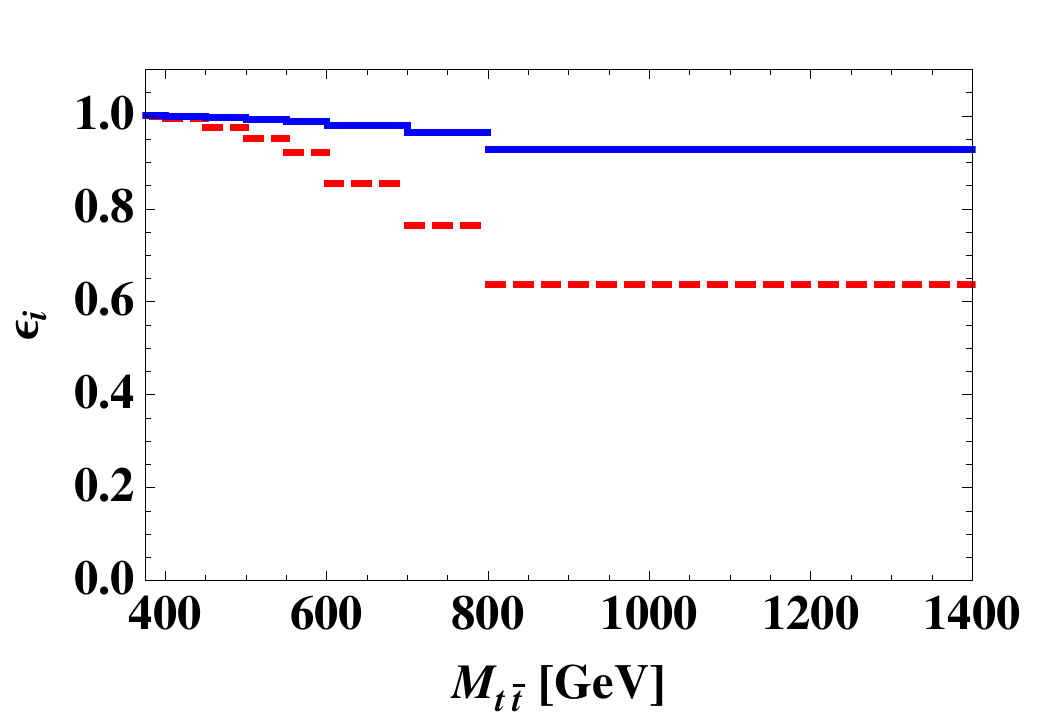}
\caption{The correction factors $\epsilon_i$ for $t$-channel dominated vector models ${\rm VI_{s}}$ (left) and ${\rm VI_o}$ (right), where ($m_{V},  \sqrt{\eta_{ij}\eta_{33}}, \eta_{i3}, \Gamma_V/m_V$)=($300 \, {\rm GeV}, 1,1.33, 0.08$)[dashed red];
($1200 \, {\rm GeV}, 2.2, 4.88, 0.5$)[solid blue].}\label{Fig-eff2}
\end{figure}

\begin{figure}[t]
\centerline{
\includegraphics[width=0.50\textwidth]{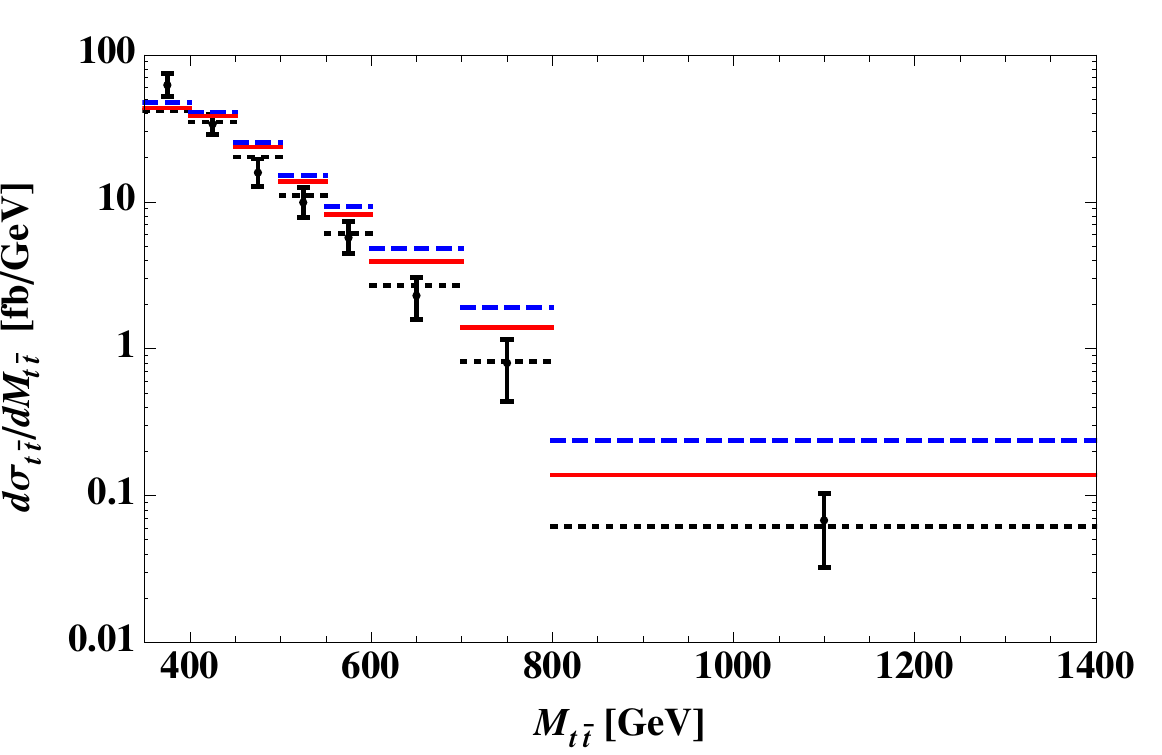}%
\includegraphics[width=0.50\textwidth]{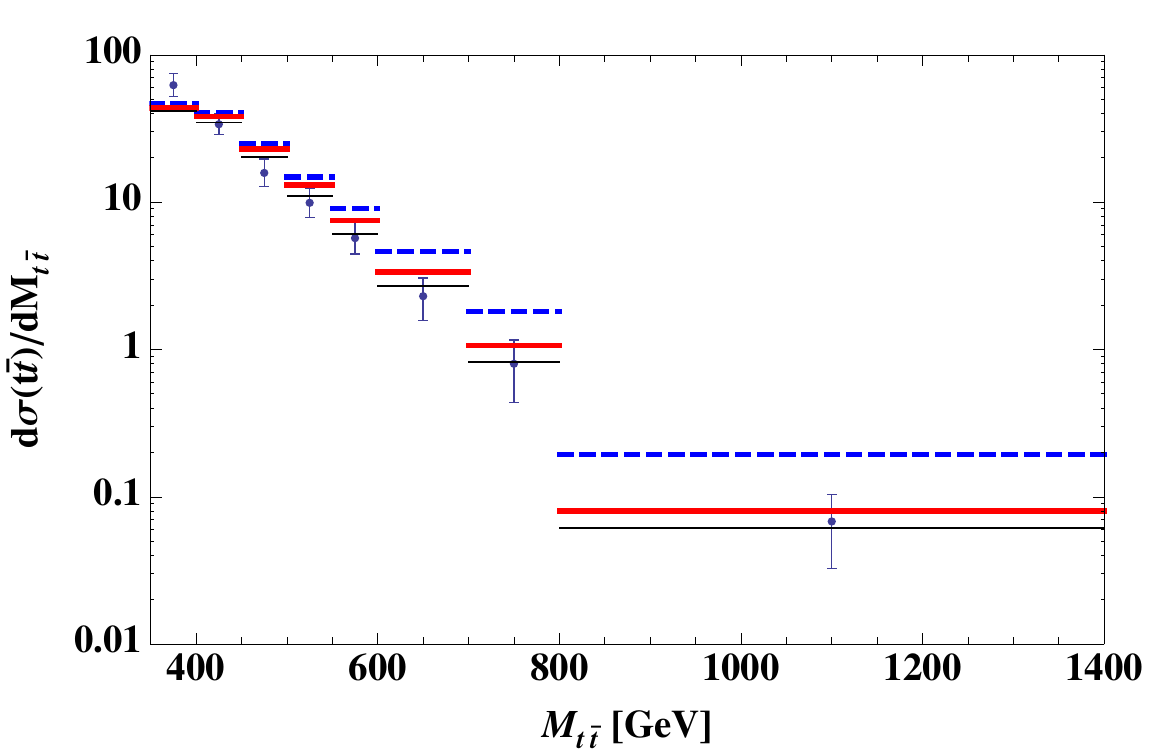}}
\caption{The effect of acceptance corrections on the vector model discussed in \cite{Grinstein:2011yv}. Before acceptance corrections (left), after corrections applied (right).  For the parameters  ($m_{V},  \sqrt{\eta_{ij}\eta_{33}}, \eta_{i3}, \Gamma_V/m_V$): solid red ($300 \, {\rm GeV}, 1,1.33, 0.08$);
dashed blue ($1200 \, {\rm GeV}, 2.2, 4.88, 0.5$).}\label{fig:vecXacc}
\end{figure}

In Fig.~\ref{Fig-eff2} the correction factors $\epsilon_i$ are shown as a function of the $M_{t\bar t}$ bin for two benchmarks points (corresponding to illustrative couplings and mass valuers) in models ${\rm VI_{s,o}}$ which exhibit substantial departures from the SM acceptances. 
Results for $d\sigma /d M_{t\bar t} $ and $A_{FB}^{t\bar t}$  
for the two model $\rm VI_o$ benchmark points in Fig.~\ref{Fig-eff2} were shown previously without acceptance corrections~\cite{Grinstein:2011yv}.
The differential distributions with acceptance corrections are compared to those in ~\cite{Grinstein:2011yv} in Fig.~\ref{fig:vecXacc}.  We see that the corrections bring the predicted spectrum for the light vector example into good agreement with experiment.
For the remaining $\rm G_F$ representations discussed below, the correction factors are not as important. For models $\rm S_V$, ${\rm VIII_s}$ and ${\rm IX_s}$ the $\epsilon_i $ are below $15\%$, and for the others
they are below $5\%$, for all $M_{t\bar t}$ bins and all benchmarks points considered.

The pattern of acceptance corrections $\epsilon_i$ can be understood from 
the angular dependence of the NP contribution to the differential cross section in each model.   
For instance, the $t$-channel exchange of a vector with mass $m_V$
leads to a Rutherford scattering peak in the forward direction for $m^2_{t\bar t} >> m^2_V$.  Specifically, the expressions for the NP cross sections contain 
characteristic $t$-channel $(1-\cos\theta)^{2,4} $ factors in the denominators, whose angular dependence is reinforced by $(1+ \cos\theta)^2 $ factors in the numerators, where $\theta$ is the top quark scattering angle in the $t\bar t $ center of mass frame.
Thus, models ${\rm VI_{s,o}}$ with light vector masses favor forward top-quark production at large $M_{t\bar t}$,
 yielding $\epsilon_i $ that are substantially less than 1 in the high $M_{t \bar t }$ bins.
In models $\rm S_I$ ($t$-channel) and $\rm S_V$, $\rm S_{VI}$ ($u$-channel),
the angular dependence introduced by the characteristic $(1\mp \cos\theta )^{2,4}$ factors in the denominators is offset by $(1 \mp \cos\theta)^2$ 
factors in the numerators, which leads to central NP top-quark production, as in the SM.
The result is that the $\epsilon_i$ in these models are actually slightly larger than 1.

To apply the acceptance corrections to $A_{FB}^{t\bar t}$ we follow CDF, where $A_{FB}^{t\bar t}$ was obtained using four bins in $M_{t\bar t}$ and $\Delta y$: $M_{t\bar t}$ above or below $450$~GeV and $\Delta y$ positive or negative. The correction factor for each of the four bins is given as in Eqs. \eqref{eq:conversion}-\eqref{eq:epsilon_i}, except that the sum in \eqref{eq:epsilon_i} now runs over all $j$ with either $\Delta y>0$ or $\Delta y<0$, and over the appropriate values of $i$ with either $M_{t\bar t}>450$~GeV or $M_{t\bar t}<450$~GeV. 
We find that the corrections are small for all of the models we consider.  
For instance, for the light vector color octet example in Fig. \ref{Fig-eff2} the shift is from an uncorrected $A_{FB}^{t\bar t}=(0.10949,0.357)$ to a corrected $A_{FB}^{t\bar t}=(0.10953,0.339)$, where the first and second numbers are the low and high mass bins in $A_{FB}^{t\bar t}$.  
The small shifts in $A_{FB}^{t\bar t}$ are due to the coarse binning in $M_{t\bar t}$ \cite{Gresham:2011pa}.
In particular, the high mass bin is dominated by events with $M_{t \bar t }$ near 450 GeV, which are more central, as in the SM.

\subsubsection{The $t \, \bar{t}$ phenomenology of  $\rm H_F$ symmetric  scalar fields}
The flavor symmetric models introduced in Section \ref{section:MFVrepresentations} and collected in Tables \ref{int-vector}, \ref{int-scalar} can couple to light and heavy quark bilinears with unsuppressed couplings. They are thus interesting candidates to explain the  $A_{FB}^{t \bar{t}}$ anomaly, as noted in  \cite{Arnold:2009ay,Grinstein:2011yv,Ligeti:2011vt}.
In the case of scalars, $\rm SU(2)_L$ singlet color triplets or color sextets and 
$\rm SU(2)_L$ doublet color singlets have previously been identified as being promising for  
explaining the $A_{FB}^{t \bar{t}}$ anomaly \cite{Shu:2009xf,Grinstein:2011yv,Ligeti:2011vt,Nelson:2011us,Babu:2011yw,AguilarSaavedra:2011hz,Blum:2011fa}.  
Here, we will  focus on some of the flavored versions listed in Table  \ref{int-scalar}.
Our results overlap with past studies,
but we also include 
color triplet and sextet scalars that couple to initial state down quarks, that have not
been studied as extensively. We find that these models may also be viable, although they generically require a coupling that is a factor of $\sim 2$ larger
than if the up quark is in the initial state. This rule of thumb also holds for the vector models that we will study in next subsection.

\begin{figure}[t]
\includegraphics[width=0.48\textwidth]{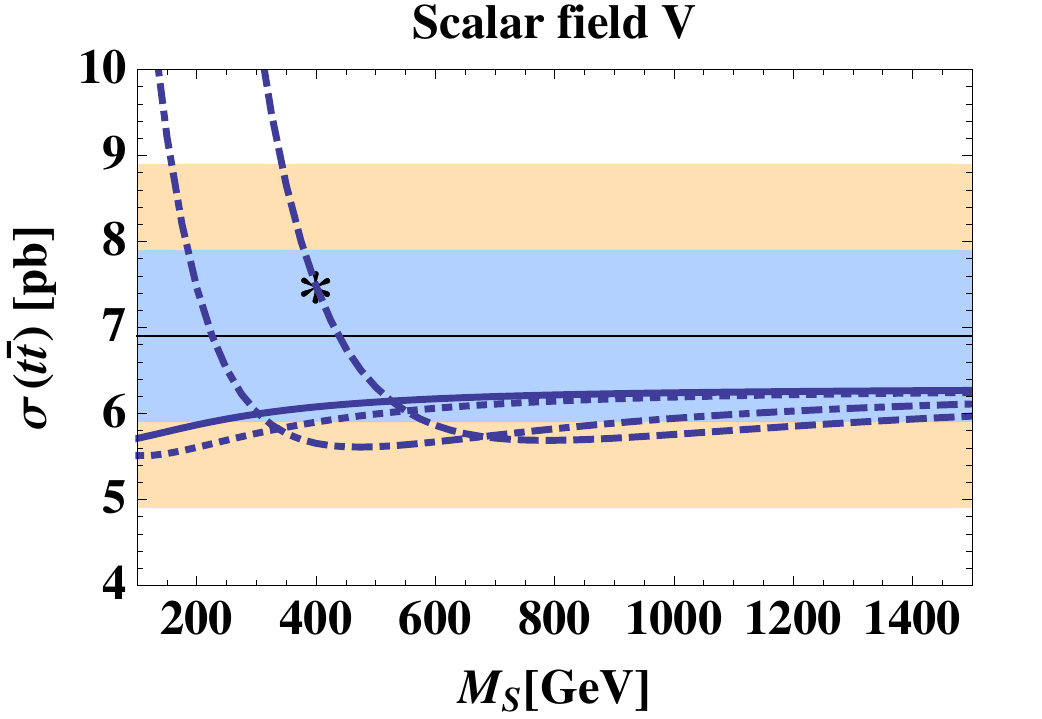}
\includegraphics[width=0.48\textwidth]{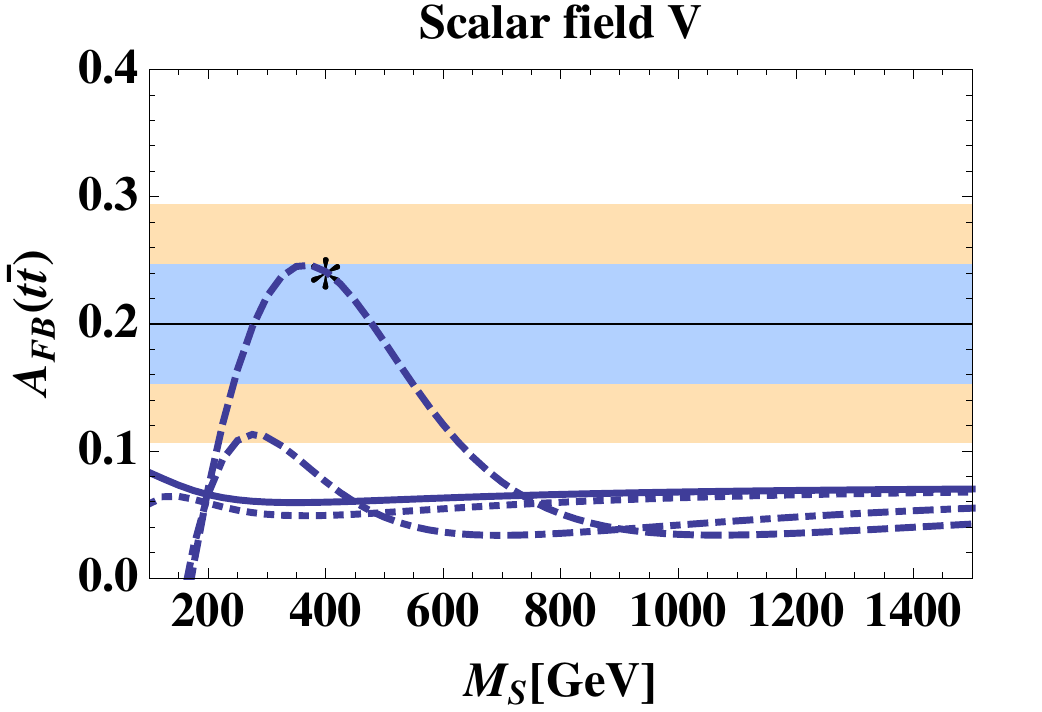}\\[2mm]
\includegraphics[width=0.48\textwidth]{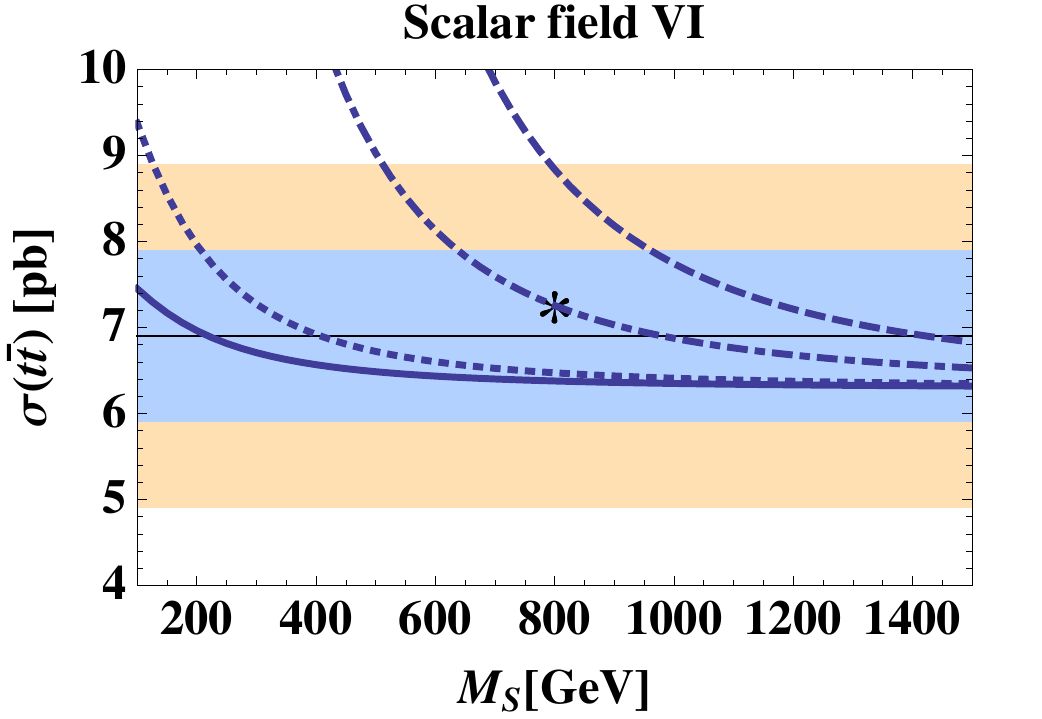}
\includegraphics[width=0.48\textwidth]{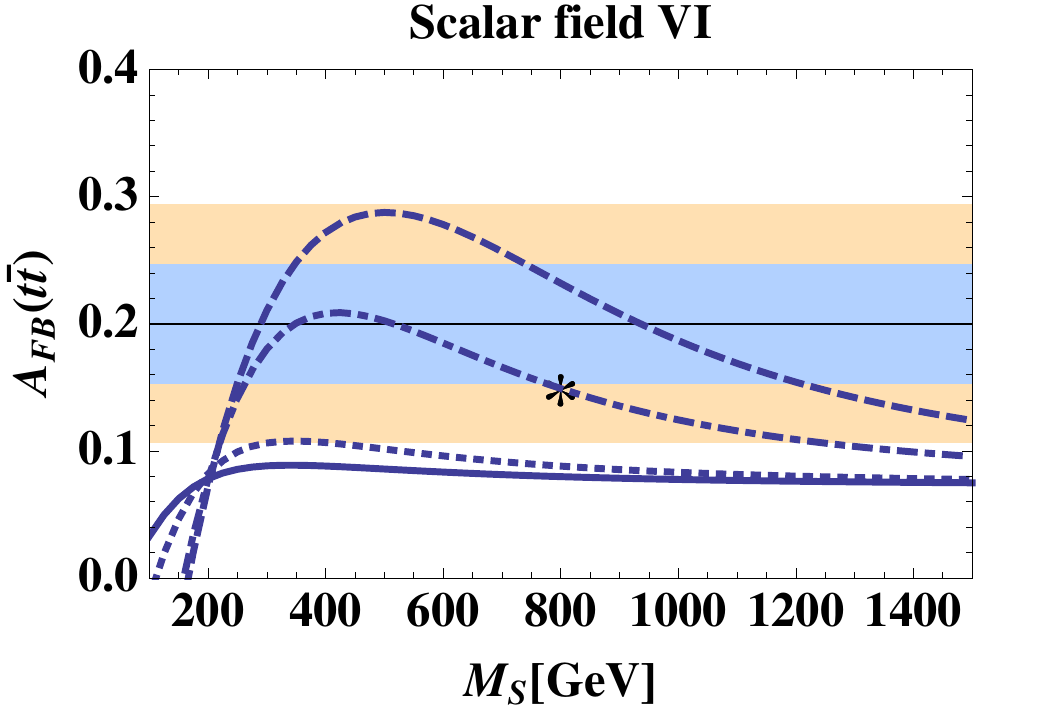}\\[2mm]
\includegraphics[width=0.48\textwidth]{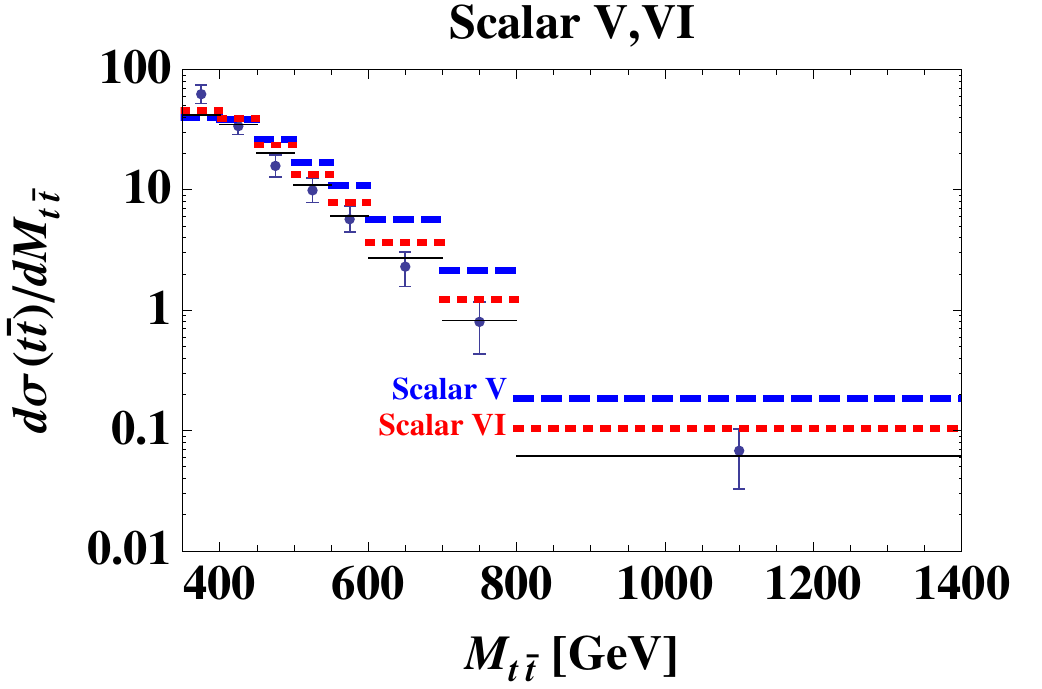}
\includegraphics[width=0.48\textwidth]{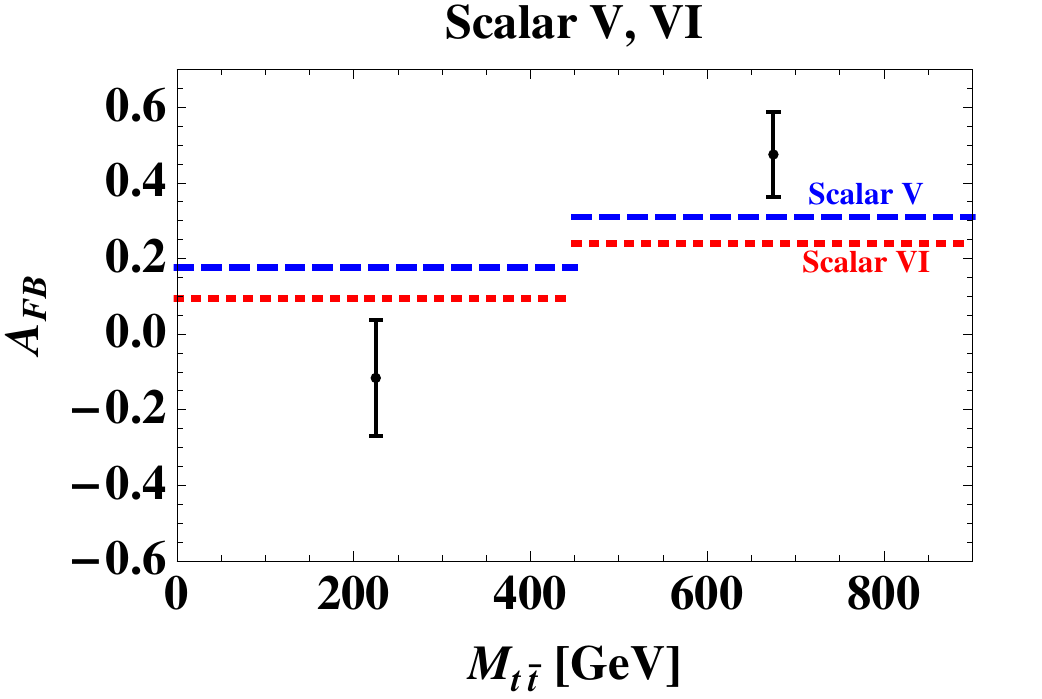}
\caption{Upper two rows: predictions for the inclusive cross sections $\sigma(t\bar t)$ and inclusive forward-backward asymmetry $A_{FB}(t\bar t)$ as a function of scalar mass $m_S$ for models $\rm S_V, S_{VI}$ and couplings $\eta =1/4 $ (solid line), $1/2 \sqrt{2}$ (dotted), $1/\sqrt2$ (dot-dashed), $1$ (dashed) compared to $1\sigma$ and $2\sigma$ experimental (shaded) bands. Predictions for $A_{FB}^{t \bar{t}}$ with $M_{t\bar t}$ below and above 450 GeV and for $d \sigma_{t \, \bar{t}}/dM_{t \, \bar{t}}$ are compared to the experimental data from \cite{Aaltonen:2009iz,Aaltonen:2011kc} in the last row for benchmark points labeled with a $\star$ in the inclusive predictions ($m_S = 400$ GeV for $\rm S_V$ and 800 GeV for $\rm S_{VI}$).
\label{Fig:ScalarVVI}   }
\end{figure}

We show results below for models  $\rm S_{V,VI}$  that couple to initial state up quarks, for models $\rm S_{IX,X}$ that couple to initial state down quarks, and model $\rm S_{I}$ that couples to initial state up and down quarks.  (Flavored color sextets and color triplets were considered previously in \cite{Grinstein:2011yv,Ligeti:2011vt}). 
We expect models $\rm S_{XI}-S_{XIV}$ to yield results similar to those for $\rm S_{IX},S_{X}$.
The scalar models ${\rm S_{III}}$
and ${\rm S_{IV}}$ 
are in the $\bf 1$ and $\bf 8$ color representations respectively. These models are known to generically suppress $A_{FB}^{t \bar{t}}$ when interfering with the SM \cite{Shu:2009xf} and we do not consider them further.

The interaction Lagrangians for scalar models $\rm S_I$ and $\rm S_{IX}$ are
\begin{align}
{\cal L}_{\rm I} &= \eta \,( \,S^0_{ij}  \bar{u}_{i\,L} u_{j \,R}   + S_{i j}^-  \bar{d}_{i\,L} u_{j\,R}  ) + h.c.\label{eq:SILag} ,\\
 {\cal L}_{\rm IX} &= \eta \,  \epsilon_{\alpha \beta\gamma} S_{ij}^{\alpha  }  d_{R}^{\,i\,\beta} \,u_{R}^{\,j\,\gamma} ~+~h.c.,  \label{eq:SIXLag} 
\end{align}
where $i,j$ and $\alpha,\beta,\gamma$ denote flavor and color indices, respectively. Explicit forms of the interaction Lagrangians for the remaining scalar models under discussion, $\rm S_{V, VI, X}$ are collected in Appendix \ref{AppendixA}. In the $\rm G_F$ symmetric limit 
the decay widths of the scalars are $\kappa \,\eta^2 \,m_S /16 \pi $,
where  $\eta$ is the coupling of the scalar field to the SM quarks, and $\kappa = 1,\,8,\,8,\,2,\,2$ in models $\rm S_I, S_V$, $\rm S_{VI}$, $\rm S_{IX}$, $\rm S_X$, 
respectively (assuming all quark decay channels are open and ignoring phase space effects). The  interaction Lagrangians  explicitly defining $\eta$ are collected in Appendix \ref{AppendixA}, while the relevant NP cross section formulae for top quark pair production can be found in Appendix B.

In Figs. \ref{Fig:ScalarVVI} and \ref{Fig:ScalarIXX} we collect predictions for the inclusive $A_{FB}^{t \bar{t}}$ and ${\sigma}_{t \, \bar{t}}$
as functions of the scalar masses in models $\rm S_{V}$, $\rm S_{VI}$, $\rm S_{IX}$, and $\rm S_X$, 
for several values of the couplings $\eta$ to quarks.  
These results do not depend on whether one is considering the $\rm G_F $ symmetric limits in Eqs. \eqref{eq:SIXLag}, \eqref{eq:SVandSVILag}, and \eqref{eq:SIXandSXLag}),
or the ${\rm H_F}$ symmetric limits (with the couplings of the light quarks to the top identified with the chosen values of $\eta$).
Large effects on $A_{FB}^{t \bar{t}}$ are possible, while remaining within the $2 \sigma$ bounds for ${\sigma}_{t \, \bar{t}}$.
In general  $\mathcal{O}(1)$  couplings $\eta$ are required.   Therefore, some $\rm G_F \rightarrow H_F$
flavor breaking will be necessary in order to satisfy the dijet constraints on the couplings to light quark pairs, 
particularly in models $\rm S_{VI}$ and $\rm S_X$ (we comment on this further below).  
For example, recent LHC dijet measurements imply that in model $\rm S_{VI}$ the couplings of light quarks to $\sim 1$ TeV mass scalars should be $\lsim 0.1$, see Section \ref{tevatron-constraints}.

\begin{figure}
\includegraphics[width=0.48\textwidth]{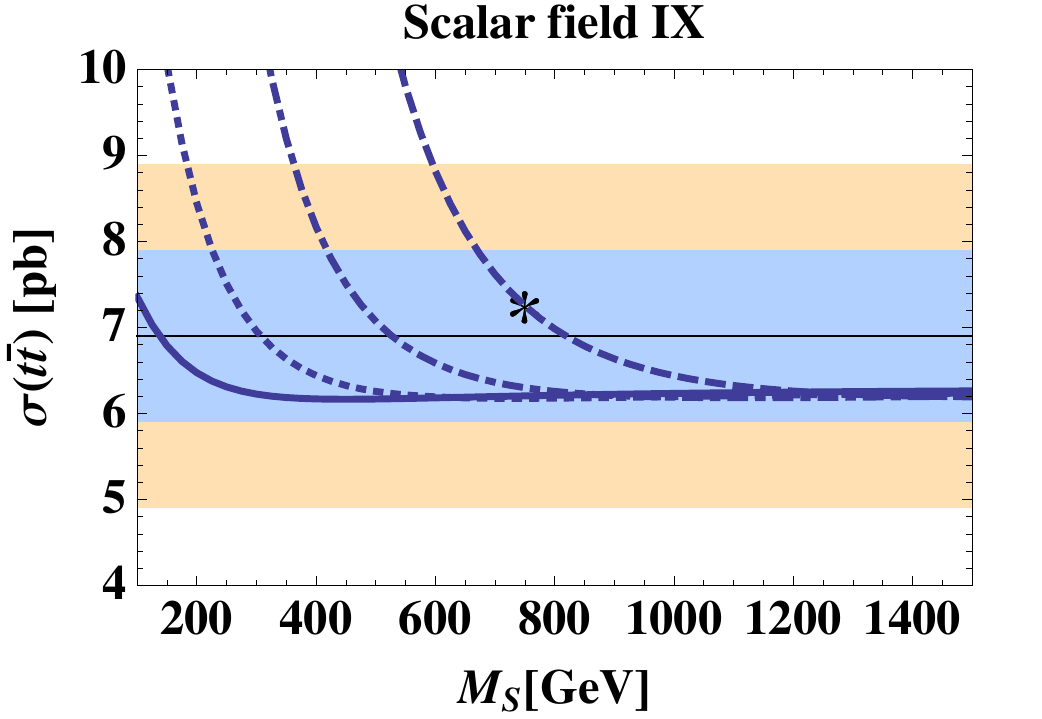}
\includegraphics[width=0.48\textwidth]{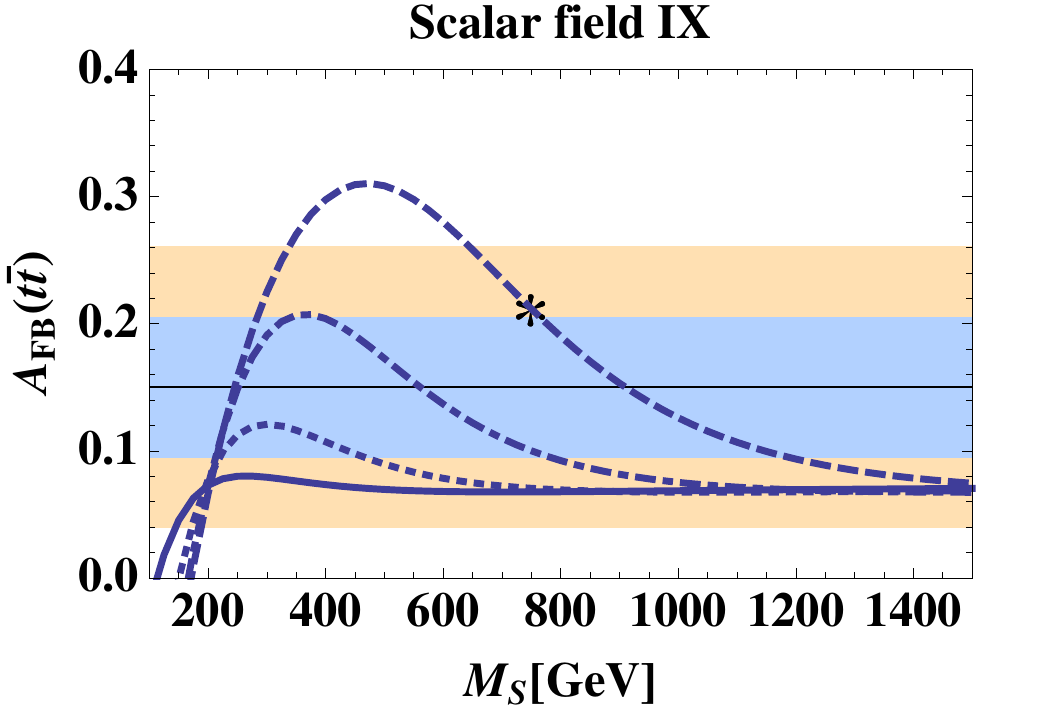}\\[2mm]
\includegraphics[width=0.48\textwidth]{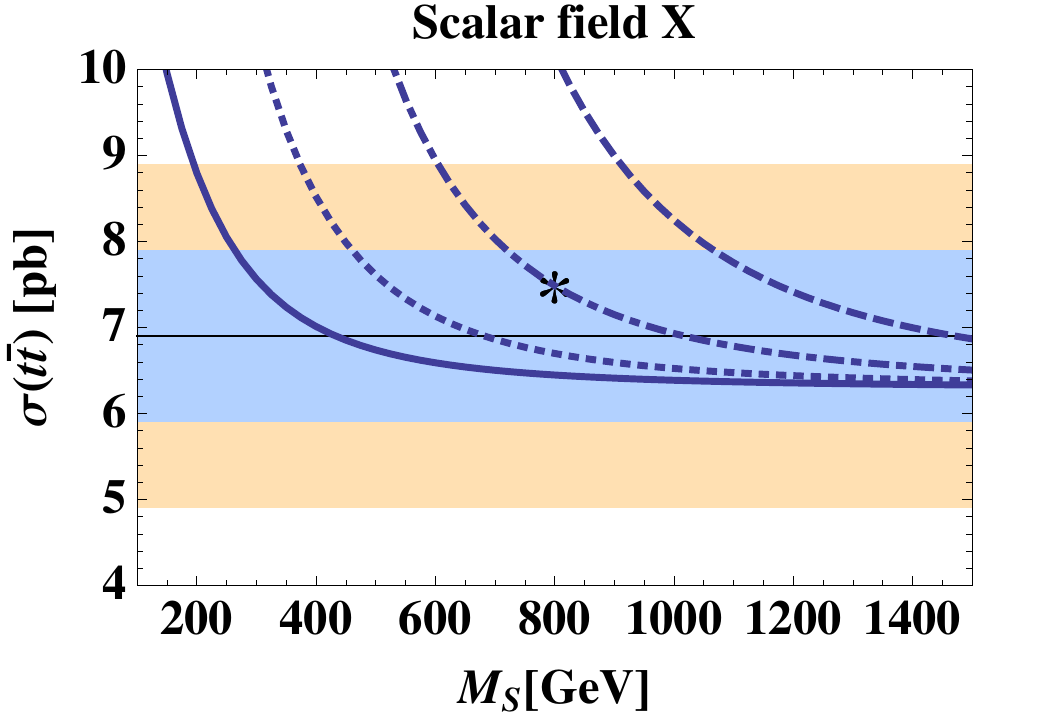}
\includegraphics[width=0.48\textwidth]{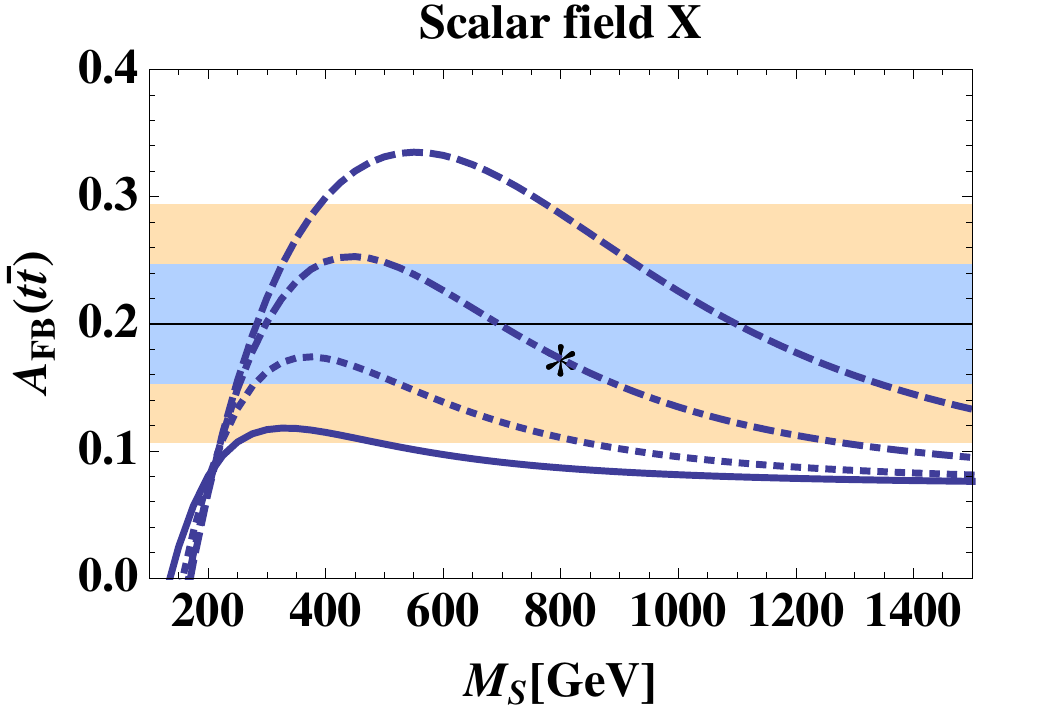}\\[2mm]
\includegraphics[width=0.48\textwidth]{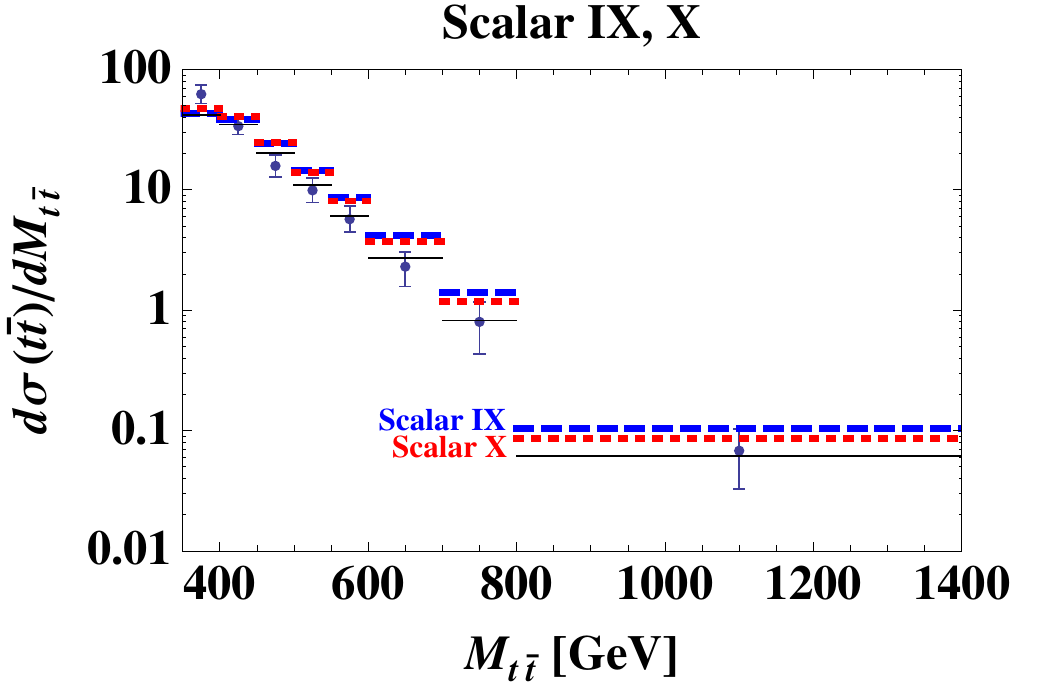}
\includegraphics[width=0.48\textwidth]{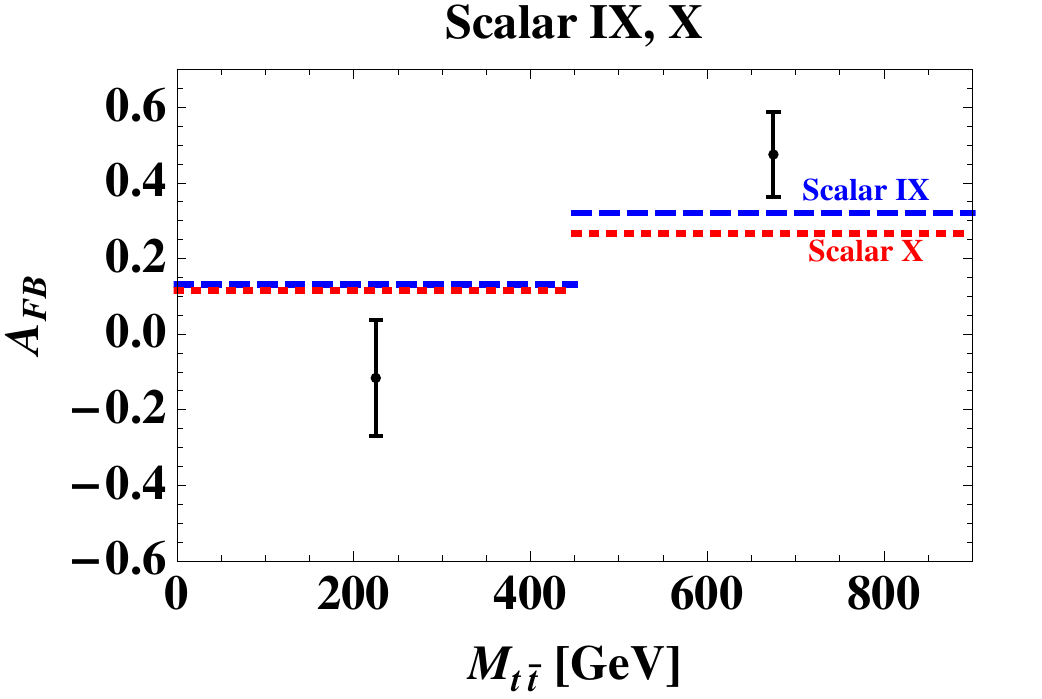}
\caption{The same as Fig. \ref{Fig:ScalarVVI}, but for models ${\rm S_{X,IX}}$ and couplings $\eta =1/\sqrt2 $ (solid line), $1$ (dotted), $\sqrt2$ (dot-dashed), $2$ (dashed). 
Benchmark masses are $m_S = 750$ GeV ($\rm S_{IX}$ ) and $m_S = 800$ GeV ($\rm S_X$ ).\label{Fig:ScalarIXX} 
  }
\end{figure}

\begin{figure}
\includegraphics[width=0.48\textwidth]{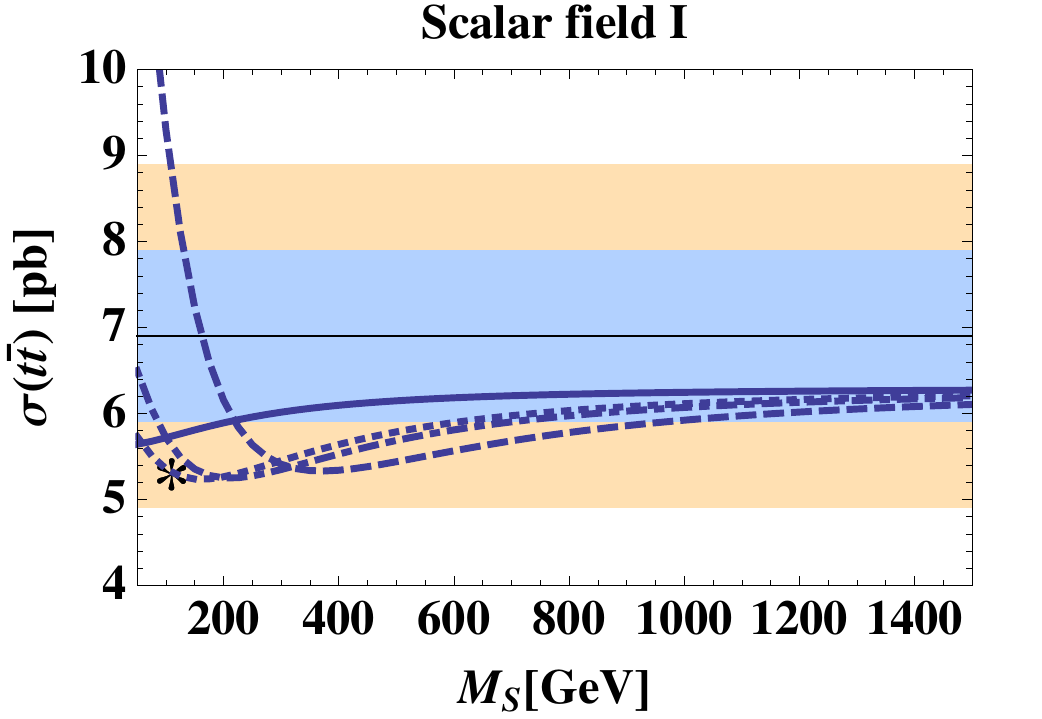}
\includegraphics[width=0.48\textwidth]{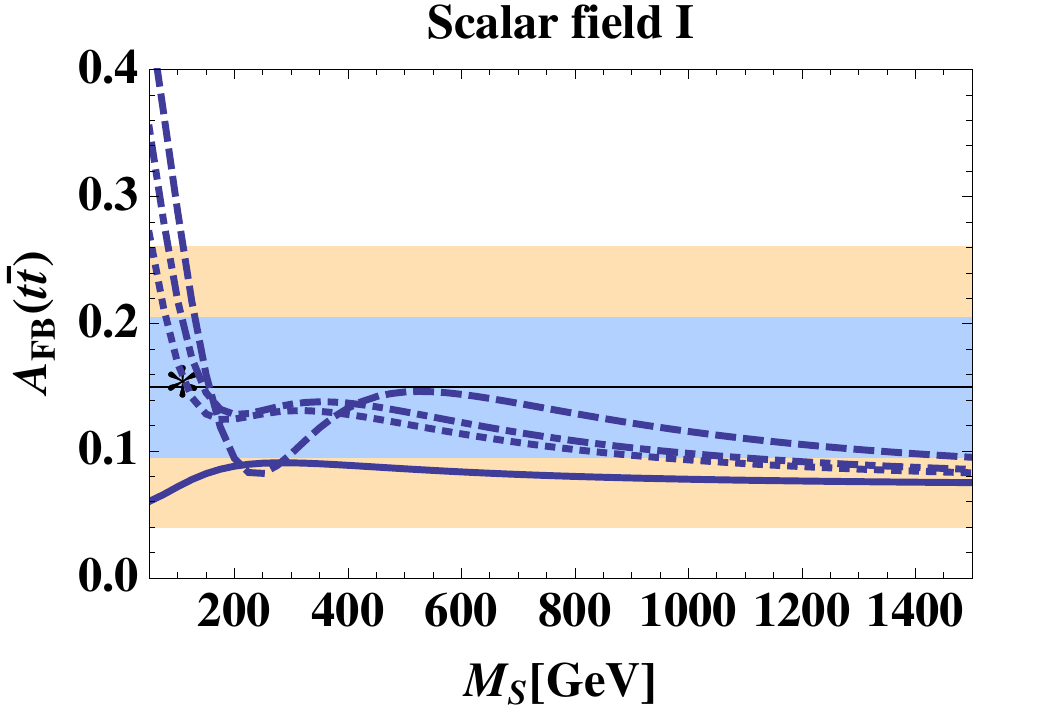}\\[2mm]
\includegraphics[width=0.48\textwidth]{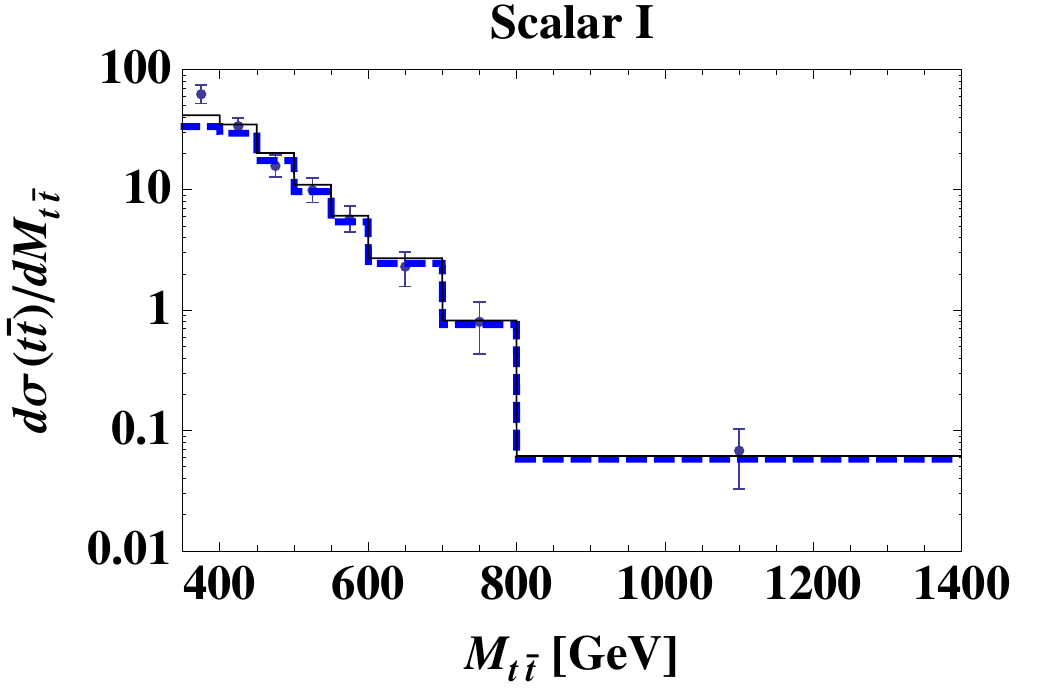}
\includegraphics[width=0.48\textwidth]{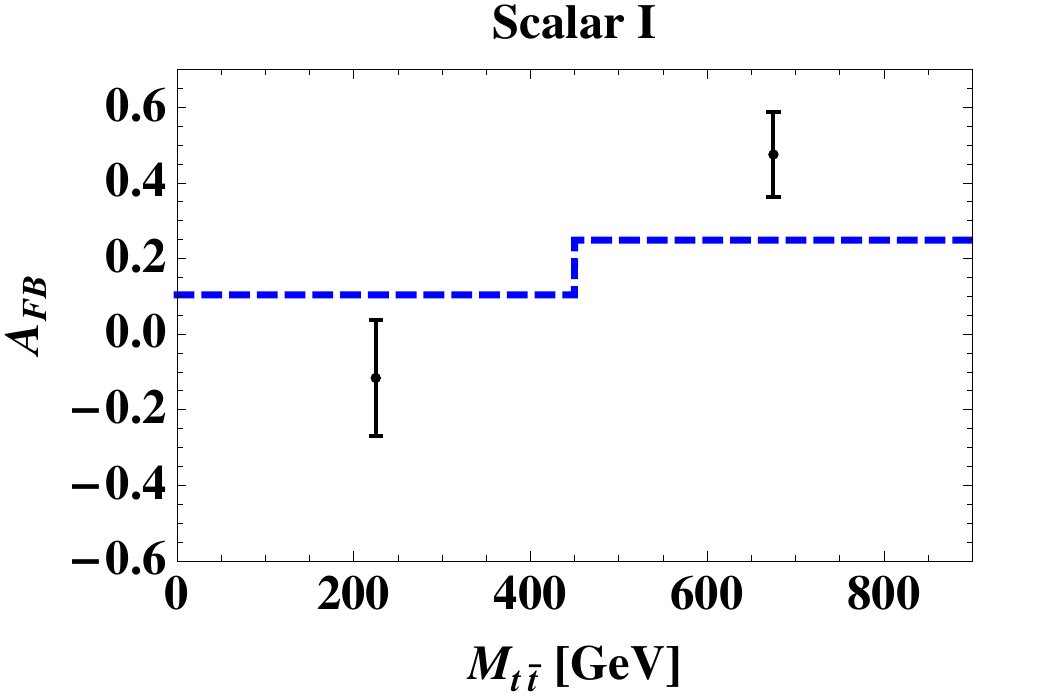}\\[-5mm]
\caption{Upper row: predictions for the cross section $\sigma(t\bar t)$ and inclusive forward-backward asymmetry $A_{FB}(t\bar t)$ as a function of scalar mass $m_S$ for model $\rm S_{I}$ and coupling $\eta =1/3$ (solid curve), $2/3$ (dotted), $3/4$ (dot-dashed), $1$ (dashed).  
Predictions for $A^{t\bar t}_{FB}$ with $M_{t\bar t}$ below and above 450 GeV and for $d \sigma_{t \, \bar{t}}/dM_{t \, \bar{t}}$ are shown in the last row for the benchmark point labeled with a $\star$ in the inclusive predictions ($m_S = 110$ GeV).}\label{Fig:FigSI}
\end{figure}

A strong constraint on these models is consistency with the measured $d \sigma(t \, \bar{t})/d M_{t \bar{t}}$ distribution.  
To demonstrate the potential of these models to explain the $A_{FB}^{t \bar{t}}$ anomaly and their impact on this spectrum we
pick particular benchmark values for the scalar masses and couplings,  denoted by a $\star$ in Figs.~\ref{Fig:ScalarVVI}, \ref{Fig:ScalarIXX}.  
The resulting $d \sigma(t \, \bar{t})/d M_{t \bar{t}}$ spectra and $ A_{FB}^{t \bar{t}}$ in the low and high $M_{t\bar t}$ bins are shown in the last rows of Figs. ~\ref{Fig:ScalarVVI}, \ref{Fig:ScalarIXX}.
We do not apply K-factors to the NP cross sections in any of the models we consider. A genuine $\rm NLO$ calculation for these models is beyond the scope of this work,
but is clearly required before precise conclusions can be drawn regarding the effects of the various models on the $d\sigma/d M_{t\bar t}$ distributions.
All chosen points predict a significantly enhanced $A_{FB}^{t\bar t}$ in the high invariant mass bin. 
In general, these models reduce the good agreement between the SM and measured $d \sigma(t \, \bar{t})/d M_{t \bar{t}}$ spectrum,
while improving the agreement in $A_{FB}^{t \bar{t}}$.
Clearly, there is significant tension between satisfying the constraints on the $M_{t \bar{t}}$ distribution while simultaneously creating a large $A_{FB}^{t \bar{t}}$ enhancement, 
as has recently been emphasized in \cite{Blum:2011fa,Gresham:2011fx}.
Recent D\O\, data \cite{Abazov:2011rq} shows a preference for a smaller $A^{t\bar t}_{FB}$ in the high $M_{t\bar t}$ bin.  While the $M_{t\bar t}$ dependence of $A_{FB}^{t\bar t}$ has not been unfolded by D\O\, , and thus cannot be directly compared to the results in Figs. \ref{Fig:ScalarVVI} and \ref{Fig:ScalarIXX}, some of the tension may be alleviated with the new data.

The flavor breaking naturally present in this framework can reduce the remaining tension with measurements associated with the light quark sector, e.g., from the dijet bounds mentioned above. 
For instance, keeping the leading insertions in~$y_t^2$ that break ${\rm G_F} \rightarrow {\rm H_F}$, 
the Lagrangian for scalar model $\rm S_{IX}$ in the mass eigenstate basis becomes,
 \bea
\mathcal{L}_{\rm IX} = \eta_1 \, (d_R')_{\alpha i} \, (u_R')_{\beta j} \,
S_\gamma^{' i,j} \, \epsilon^{\alpha \beta \gamma} + \left(\eta_1 + 2\, \eta_2\, y_t^2 \right)(d_R')_{\alpha i} \, (t_R')_{\beta} \, S_\gamma^{' i,3} \, \epsilon^{\alpha \beta \gamma}, 
\eea
with $ j = \{1,2\}$ and $i = \{1,2,3\}$. The mass spectrum is split, with $m_{i3}^2=m_{ij}^2+\tilde{\zeta}_1 m_S^2$.
Depending on the signs 
of $\eta_{1,2}$ and $\tilde \zeta_1$ one can have an enhanced coupling to top quarks (compared to couplings to light quarks) with a suppressed mass scale of the $m_{i3}^2$ components of the scalar. 
This can naturally lead to larger effects on top phenomenology, such as $A_{FB}^{t \bar{t}}$ and $\sigma_{t \, \bar{t}}$, while at the same time weakening the impact of the dijet constraints (see Section \ref{tevatron-constraints}).  This situation is similarly realized in the other scalar models.
Alternatively, one could consider minimal $\rm H_F$ symmetric versions of models $\rm S_{IX}$ and $\rm S_X$, which 
only contain the scalar $\rm G_F$ submultiplets transforming as $(1,2,1)$ under $\rm H_F$ (they only couple the light quarks directly to the top quark), 
while in the minimal $\rm H_F$ symmetric versions of models $\rm S_{V}$ and $\rm S_{VI}$ one would consider the $(2,1,1)$ representations of $\rm H_F$.
 
Finally, we discuss model $\rm S_I $.\footnote{Note that two studies have recently concluded \cite{AguilarSaavedra:2011hz,Blum:2011fa}  that 
a (color singlet) $SU(2)_L$ doublet 
would be the most promising, among the possible scalar representations, for explaining the $t\bar t$ asymmetry with minimal distortion of the Tevatron and LHC $M_{t\bar t}$ spectra.}
(Models with flavored scalar doublets have previously been considered in 
\cite{Nelson:2011us,Babu:2011yw}).
For illustration, we can define the left-handed (LH) quark fields in Eq. \eqref{eq:SILag} in either the up or down quark mass eigenstate basis (with the right-handed (RH) up quarks in their mass eigenstate basis).
In both cases, the recent D\O\, upper bounds \cite{Abazov:2011af} on anomalous resonant dijet production in $W$ + jets 
imply that the couplings of the scalars to light quarks should satisfy $\eta_{ij} \lsim 0.2$, $i,j=\{1,2\}$, given the light scalar doublet masses favored by 
the $A_{FB}^{t\bar t}$ anomaly (see below and Section \ref{tevatron-constraints}).
Significant breaking of $\rm G_F \to H_F$ would, therefore, be necessary in order to accommodate $O(1)$ scalar-top quark-light quark couplings.
However, we reiterate that $\rm H_F$ symmetry protects against NP contributions to $K-\bar K$ and $D-\bar D$ mixing, as well as to single top and same sign top production.

If the LH quarks in Eq.  \ref{eq:SILag} are defined in the up mass basis, then 
the $B \to K\pi$ branching ratio constraints \cite{Blum:2011fa,Zhu:2011ww} would also impose the bound $\eta_{3i} \lsim 0.2$, $i=\{1,2\}$, see Section \ref{FCNC}.
Therefore, in this case 
top phenomenology would be well approximated by
a minimal $\rm H_F$ symmetric model
consisting of scalar $\rm SU(2)_L$ doublets which transforms as $(1,1,2,a)$ under ${\rm H_F}$, where the $U(1)_3$ charge $a$ is opposite to that of the top and bottom quarks ($\rm U(2)_{Q_L}$ is defined with respect to the LH up quark mass eigenstate basis).
We can write the corresponding interaction Lagrangian, in the quark mass eigenstate bases, as 
\beq {\cal L}^{\rm min}_{\rm I(a)} = \eta \, ( S_{i \,3}^0 \, \bar{u}^\prime_{i\,L}\, t^\prime_{ R}   + S_{i\,3}^-\,  (V^\dagger_{\rm CKM})_{j \,i}\, \bar{d}^{\,\prime}_{j\,L} \,t^\prime_{R}  ) + h.c.,
\label{SImina}\eeq
where $i=\{1,2\}$, $j=\{1,2,3\}$ and $V_{\rm CKM}$ is the Cabbibo-Kobayashi-Maskawa (CKM) matrix.
If the LH quarks in Eq.  \eqref{eq:SILag} are defined in the down quark mass eigenstate basis, there are two minimal $\rm H_F$ symmetric and flavor safe alternatives
for $A_{FB}^{t\bar t}$ enhancement:
exchange of scalar $\rm SU(2)_L$ doublets which transform as $(2,1,1,a)$, or as $(1,1,2,a)$ (
$\rm U(2)_{Q_L}$ is now defined with respect to the LH down quark mass eigenstate basis). 
The relevant interaction Lagrangians would be
\beq {\cal L}^{\rm min}_{\rm I(b)} = \eta \, ( S_{3 \,i}^0 \, \bar{u}^\prime_{j\,L}\, (V_{\rm CKM})_{j\,3} \,u^\prime_{ i\,R}    + S_{3\,i}^-\,  \bar{b}^{\,\prime}_{L} \,u^\prime_{ i\,R}  ) + h.c.,
\label{SIminb}\eeq
or 
\beq {\cal L}^{\rm min}_{\rm I(c)} = \eta \, ( S_{i \,3}^0 \, \bar{u}^\prime_{j\,L}\, (V_{\rm CKM})_{j\,i} \,t^\prime_{R}    + S_{i\,3}^-\,  \bar{d}^{\,\prime}_{i\,L} \,t^\prime_{ R}  ) + h.c.,
\label{SIminc}\eeq
respectively,
where $i=\{1,2\}$, $j=\{1,2,3\}$.
For the light scalar masses ($\lsim 130$ GeV) and $O(1)$ couplings $\eta$ favored by $A_{FB}^{t\bar t}$, 
options I(a) and I(c) above (Eqs. \eqref{SImina}, \eqref{SIminc}) are disfavored at the $4\sigma$ level by constraints on non-oblique corrections to the $Z$ couplings to quarks, whereas option I(b) (Eq. \eqref{SIminb})  is consistent
at $2\sigma$, see Section \ref{EWPDobliqueandnonoblique}.  
In a generalization of option I(b) to complete $\rm G_F$ multiplets, non-oblique correctons would dictate that  the couplings $\eta_{i3}$ 
need to be roughly a factor of 2 smaller than $\eta_{3i}$ for $i=\{1,2\}$, while $Wjj$ constraints would dictate the bounds $\eta_{ij} \lsim 0.2$ for $i,j=\{1,2\}$. 

In Fig. \ref{Fig:FigSI} we show predictions for $A_{FB}^{t\bar t}$ (inclusive and $M_{t\bar t}$ bins) and $d\sigma /d M_{t\bar t}$ corresponding to Eq. \eqref{SIminb}.
Our results confirm the findings in \cite{Blum:2011fa} for $A_{FB}^{t\bar t} $ enhancement with scalar doublets: given CDF data the preferred scalar mass is small, below 130 GeV, while
the coupling $\eta$ is $O(1)$.
Also, the difference between the asymmetries in the low and high mass bins is not large when the latter is enhanced (which could be welcome in view of the recent D\O\, $A_{FB}^{t\bar t} $ measurements), and there is minimal impact on the $M_{t\bar t}$ spectrum.  
Finally, we note that for our benchmark points the bounds on the top quark decay width are not violated by new top quark decays to a scalar and a light quark (even though there are $2$ times as many such modes in I(b) than previously considered in \cite{Blum:2011fa}).

\begin{figure}
\includegraphics[width=0.48\textwidth]{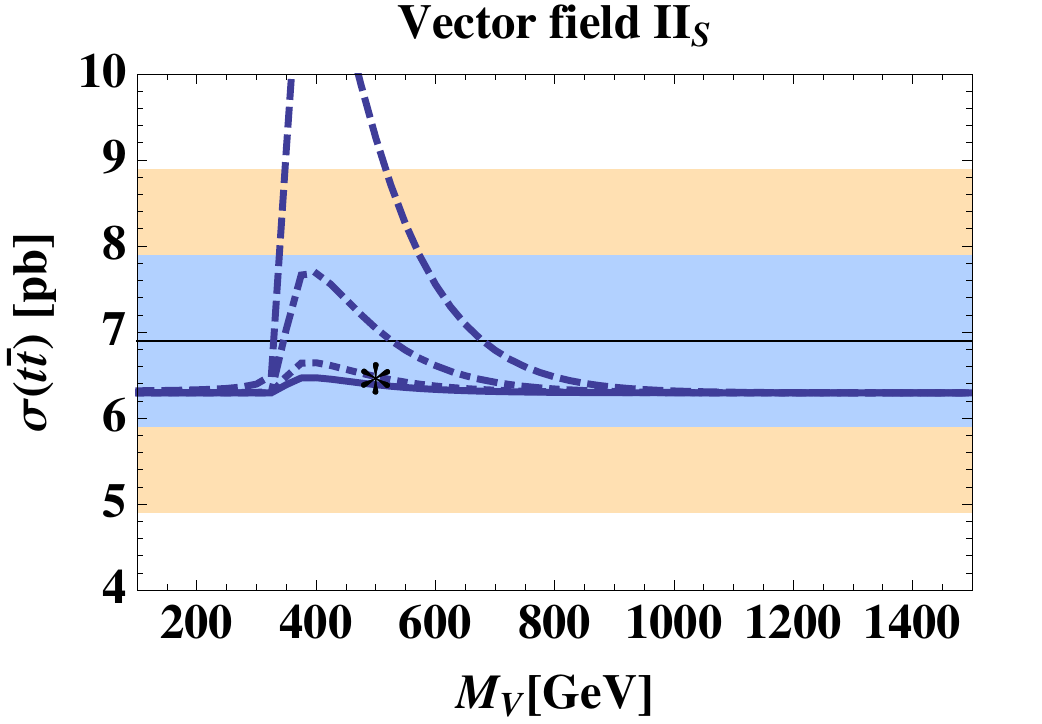}
\includegraphics[width=0.48\textwidth]{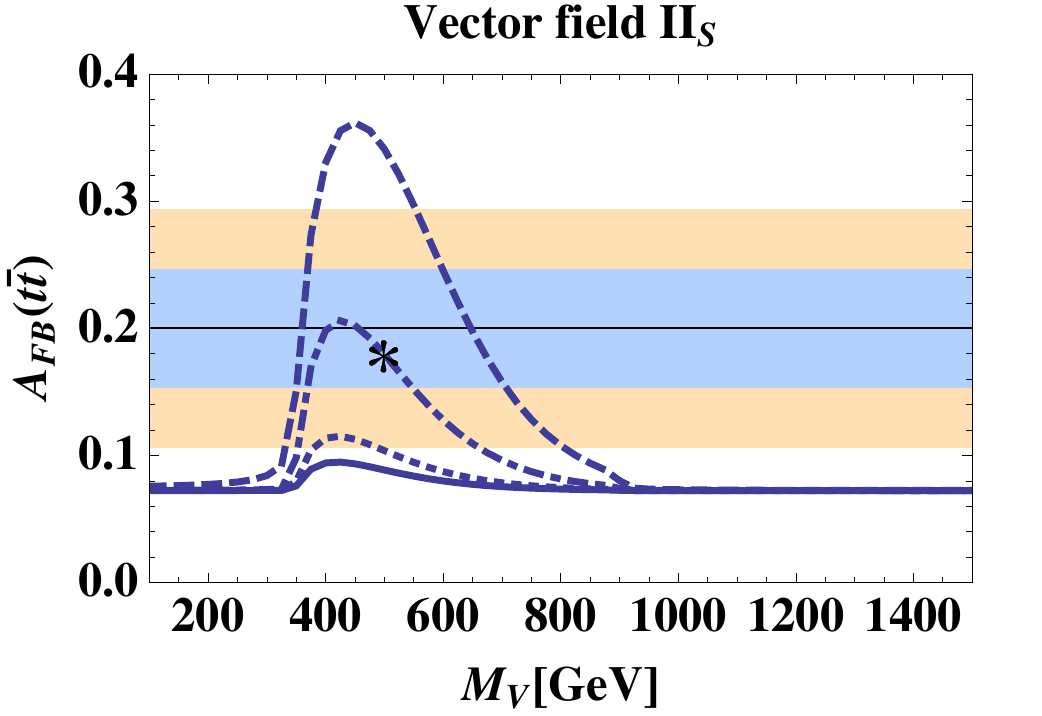}\\[2mm]
\includegraphics[width=0.48\textwidth]{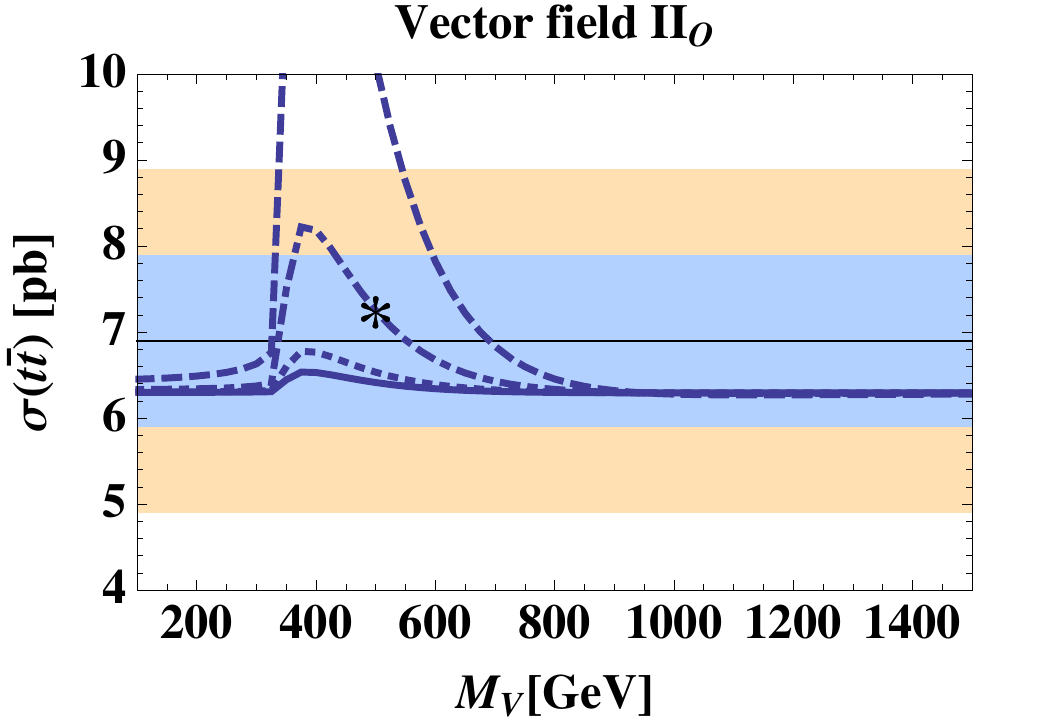}
\includegraphics[width=0.48\textwidth]{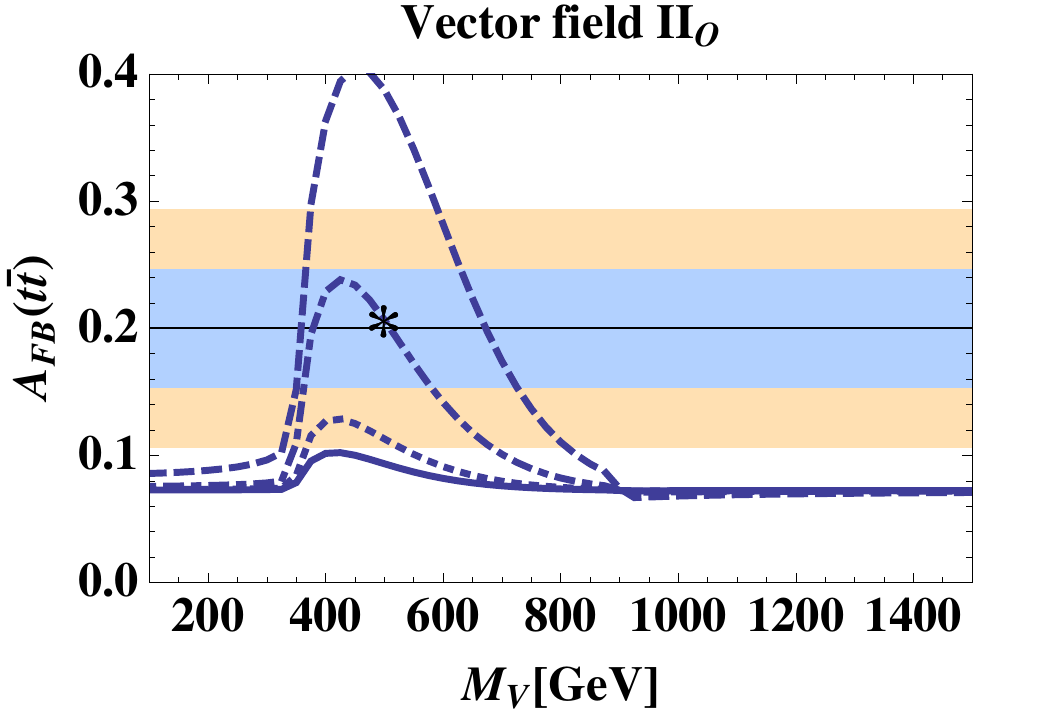}\\[2mm]
\includegraphics[width=0.48\textwidth]{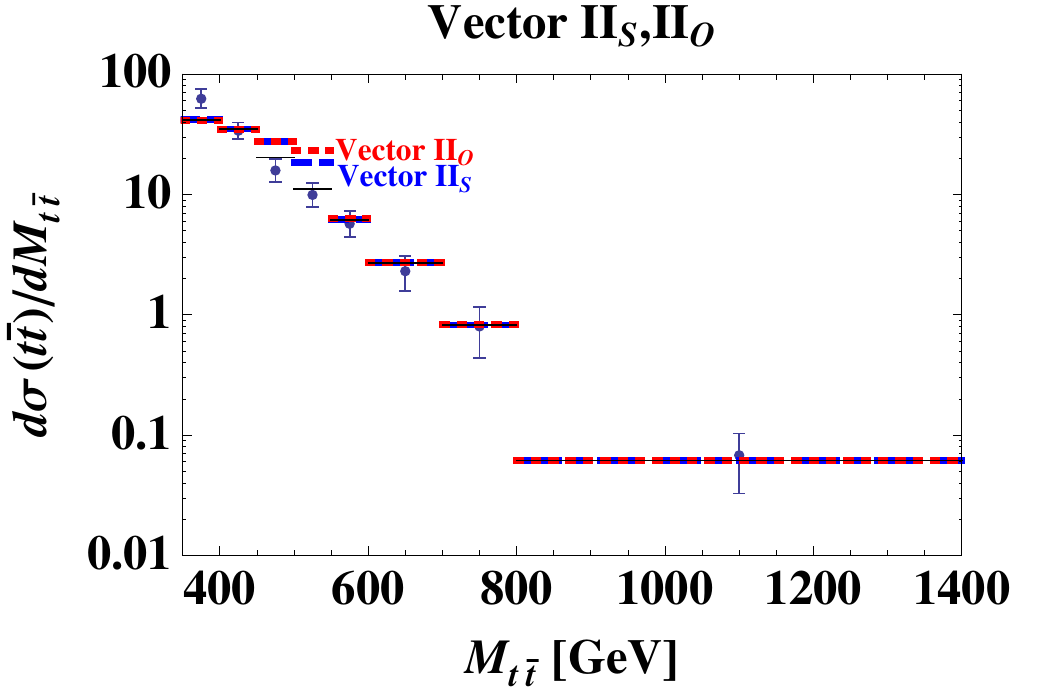}
\includegraphics[width=0.48\textwidth]{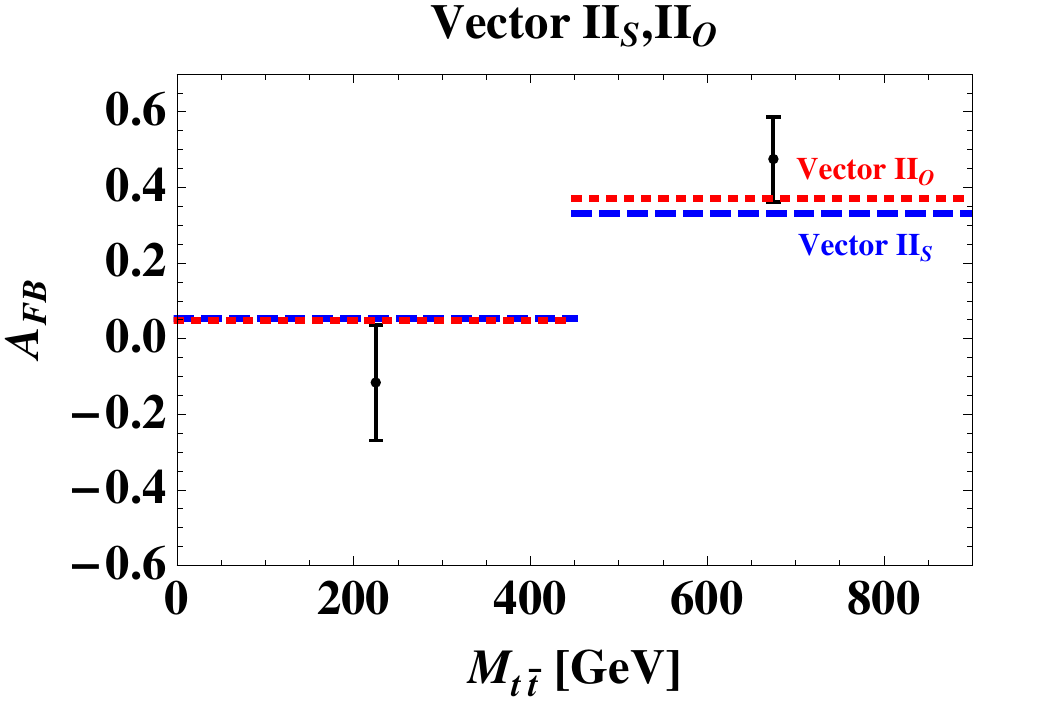}\\[-5mm]
\caption{Upper two rows: predictions for total cross sections $\sigma(t\bar t)$ and inclusive forward-backward asymmetry $A_{FB}(t\bar t)$ as a function of vector mass $m_V$ for models $\rm II_{s,o}$ and couplings $f_q f_t =1/256$ (solid curve), $1/128$ (dotted), $1/32$ (dot-dashed), $1/8$ (dashed).  The bands are the one and two $\sigma$ measurements of $A_{FB}^{t \bar{t}}$ and $\sigma_{t \, \bar{t}}$. The predictions for $A_{FB}$ with $M_{t\bar t}$ below and above 450 GeV and for $d \sigma_{t \, \bar{t}}/dM_{t \, \bar{t}}$ are shown in the last row for benchmark points labeled with a $\star$ in the inclusive predictions ($M_V = 500$ GeV).  The data are summarized in Table \ref{table:partondata}.  }\label{FigIIso}
\end{figure}

\begin{figure}
\includegraphics[width=0.48\textwidth]{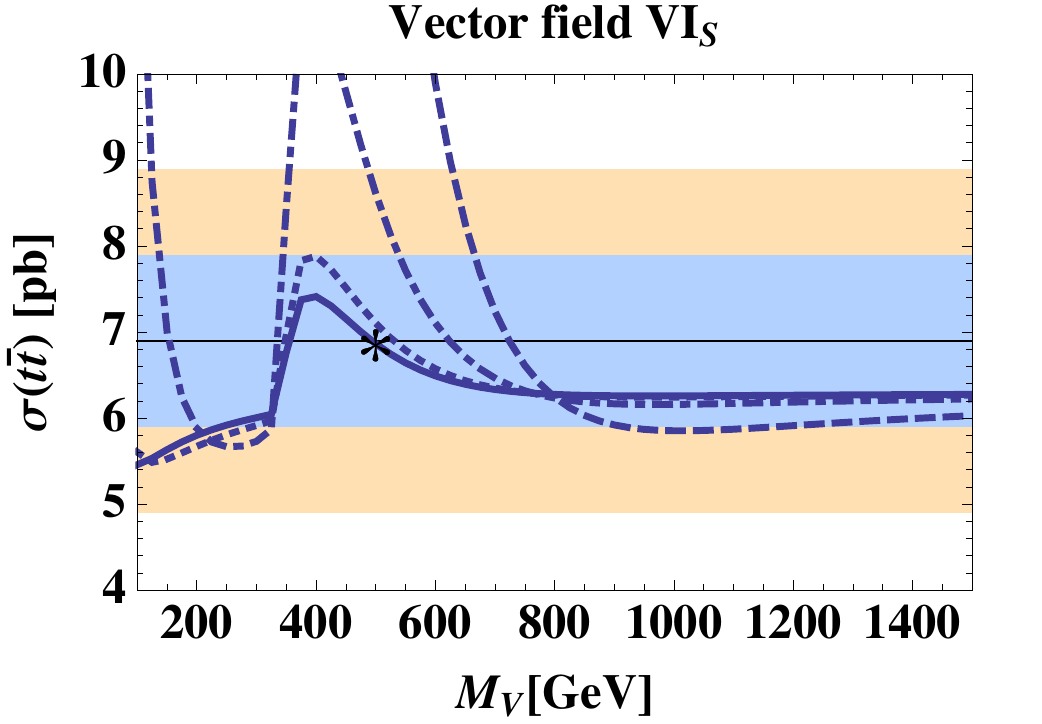}
\includegraphics[width=0.48\textwidth]{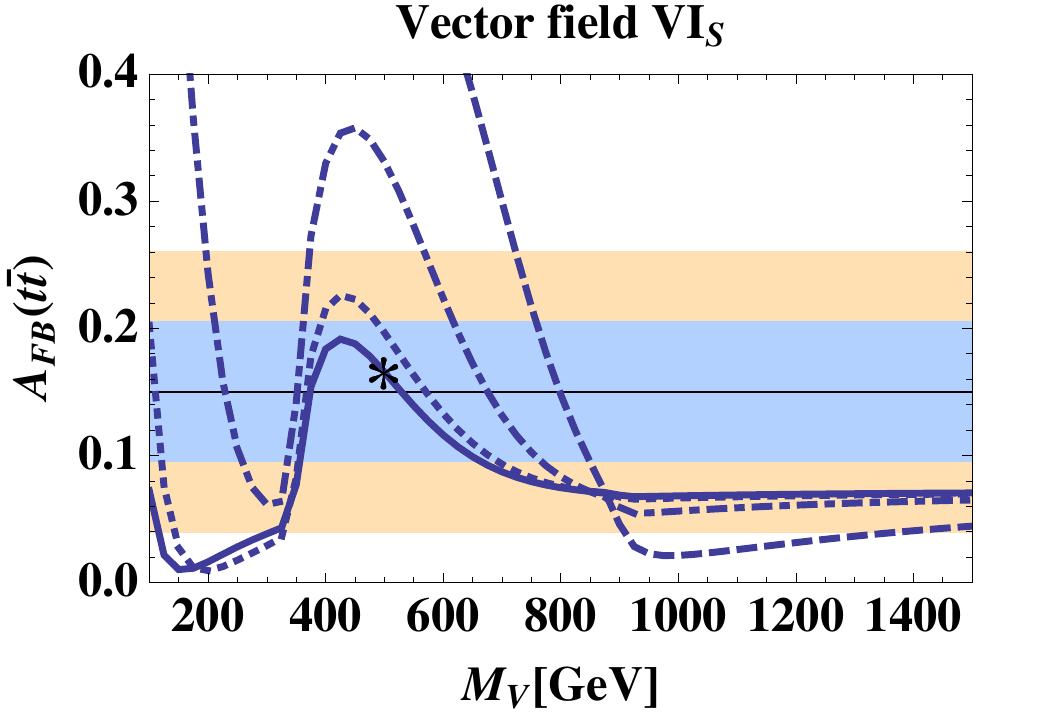}\\[2mm]
\includegraphics[width=0.48\textwidth]{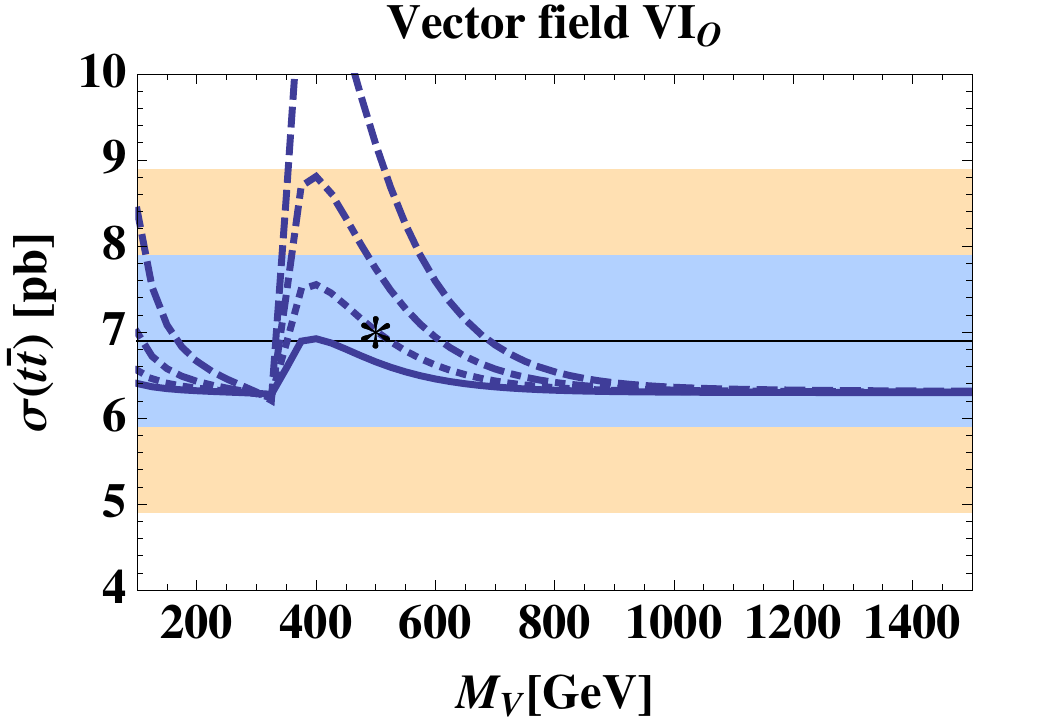}
\includegraphics[width=0.48\textwidth]{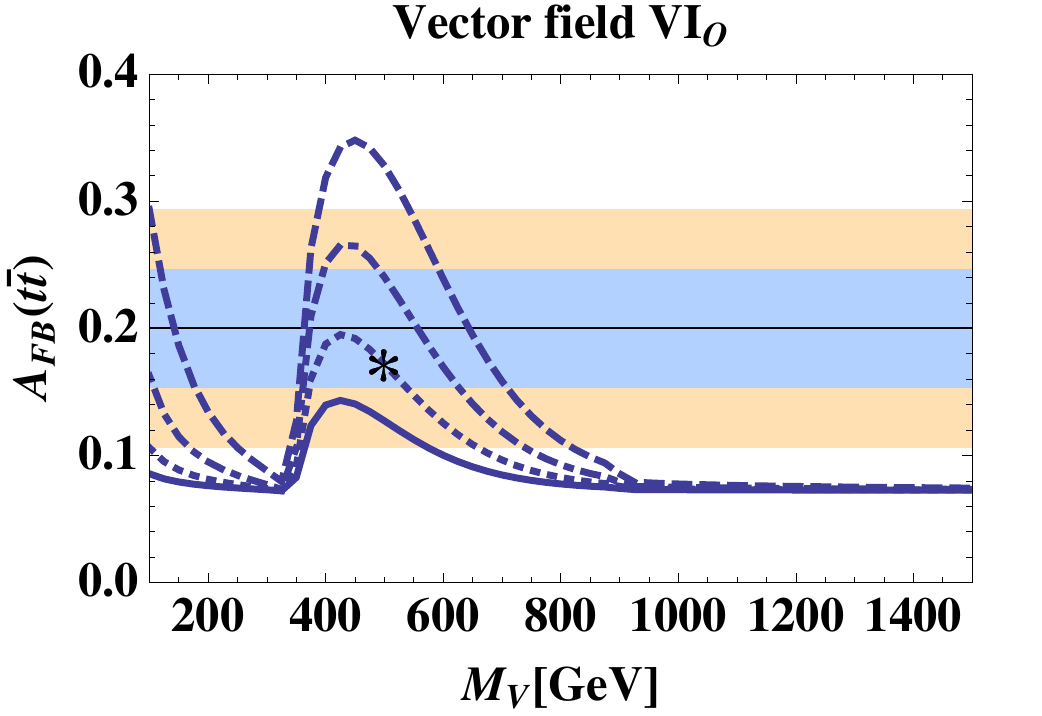}\\[2mm]
\includegraphics[width=0.48\textwidth]{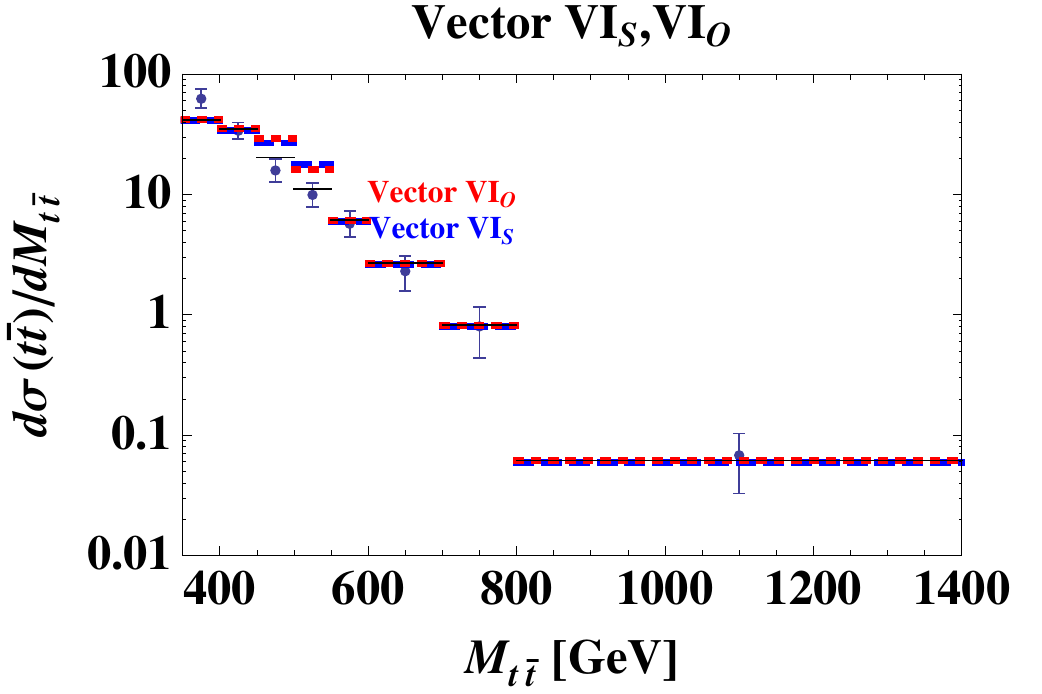}
\includegraphics[width=0.48\textwidth]{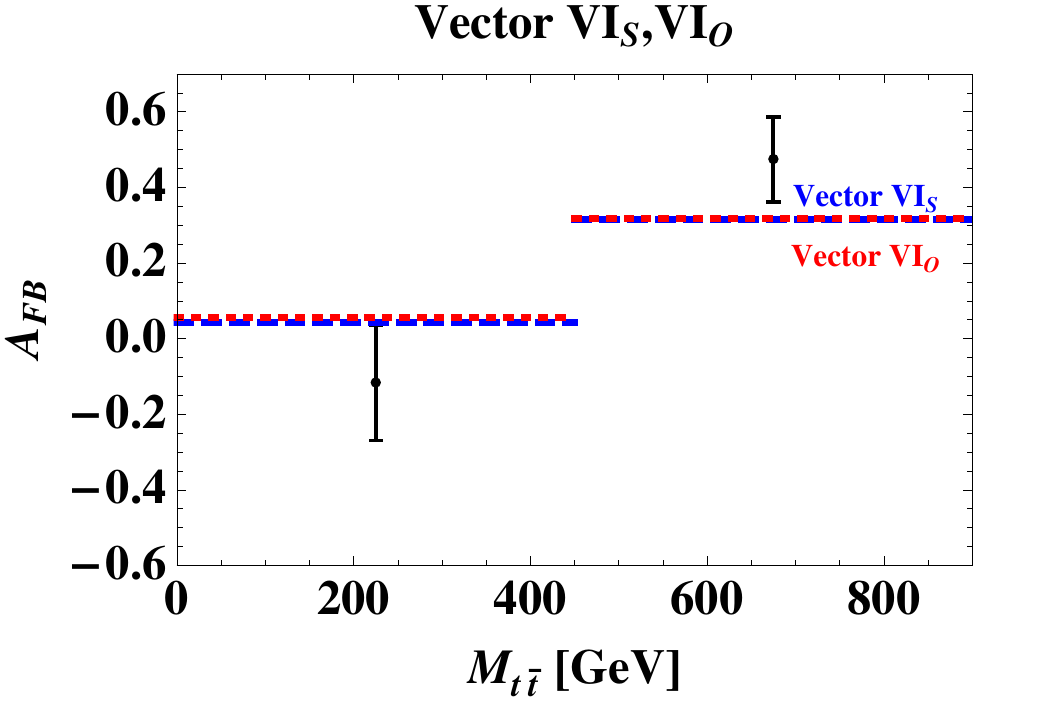}\\[-5mm]
\caption{Same as Fig. \ref{FigIIso} but for vector models  $\rm VI_{s,o}$ and couplings $f_q f_t =f_{qt}^2=1/4\sqrt{2}$ (solid curve), $1/4$ (dotted), $1/\sqrt{2}$ (dot-dashed), $2 \sqrt{2}$ (dashed) for model $\rm V_s$ and couplings $f_q f_t =f_{qt}^2=1/16$ (solid curve), $1/8$ (dotted), $1/4$ (dot-dashed), $1/2$ (dashed) for model $\rm V_o$. For the benchmark points, $m_V = 500$ GeV.  }\label{FigVIso}
\end{figure}

\begin{figure}
\includegraphics[width=0.48\textwidth]{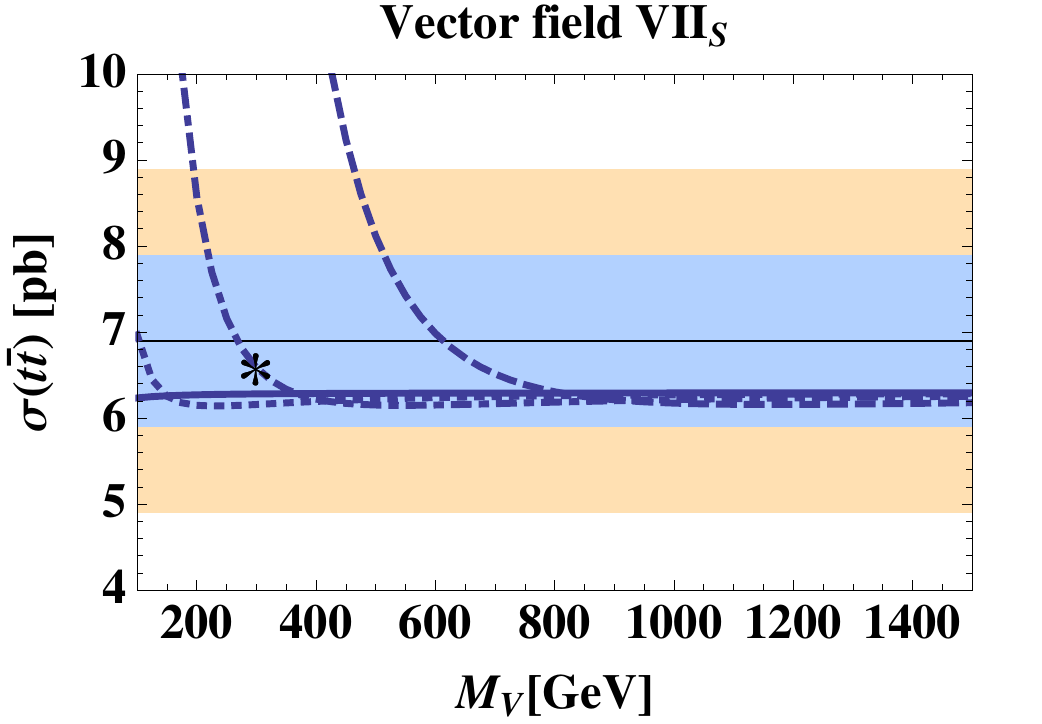}
\includegraphics[width=0.48\textwidth]{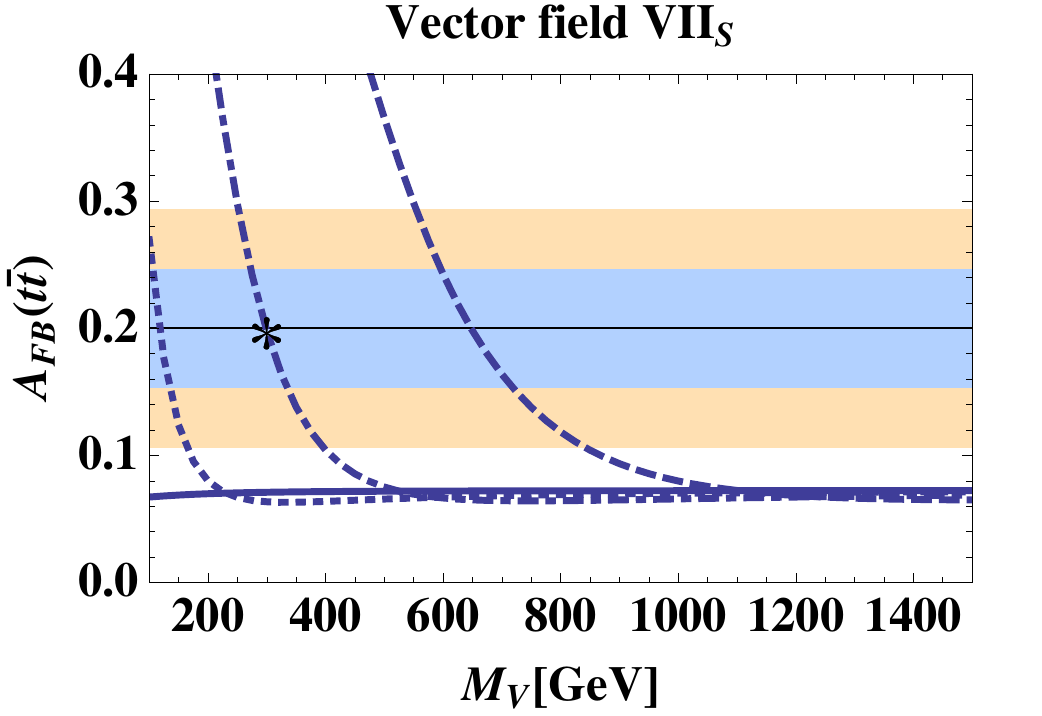}\\[2mm]
\includegraphics[width=0.48\textwidth]{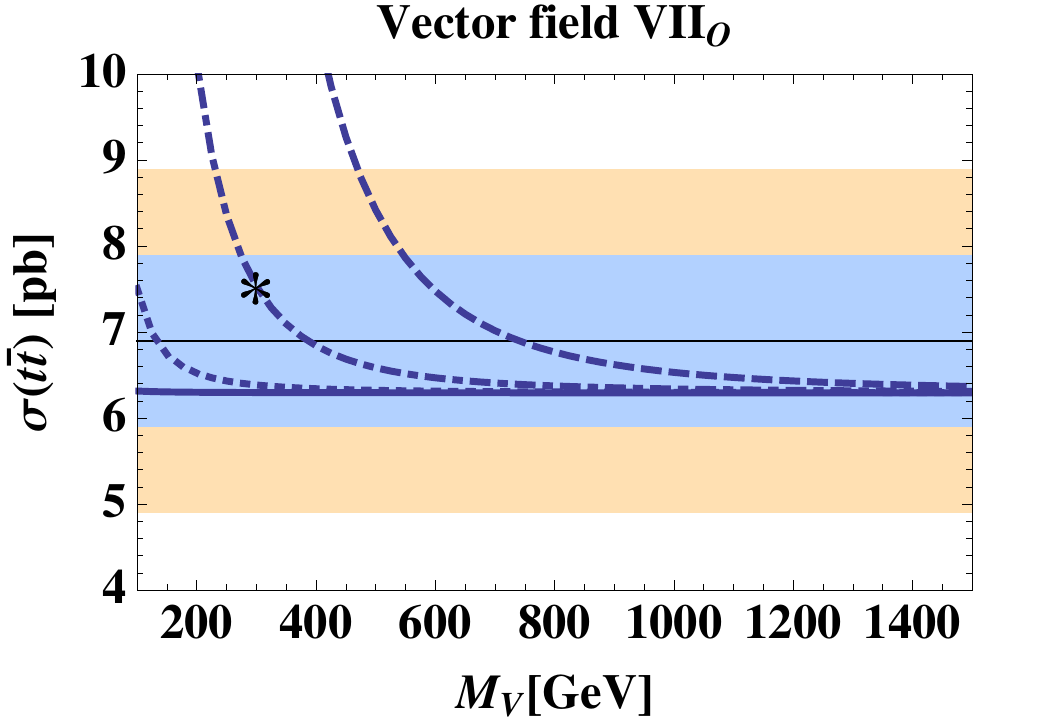}
\includegraphics[width=0.48\textwidth]{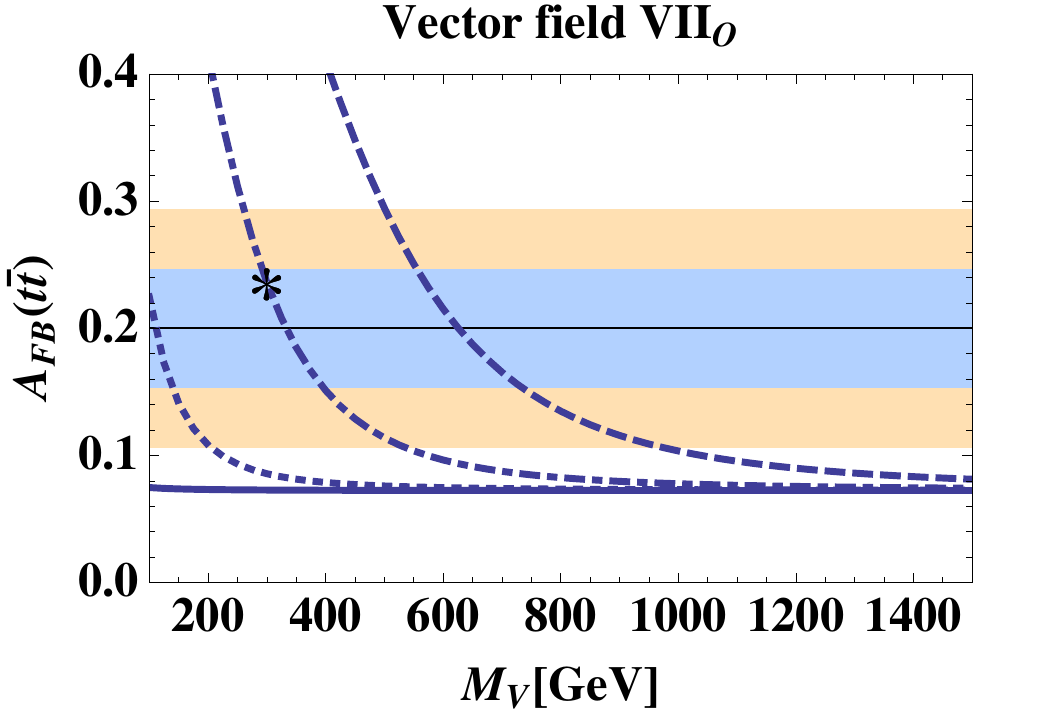}\\[2mm]
\includegraphics[width=0.48\textwidth]{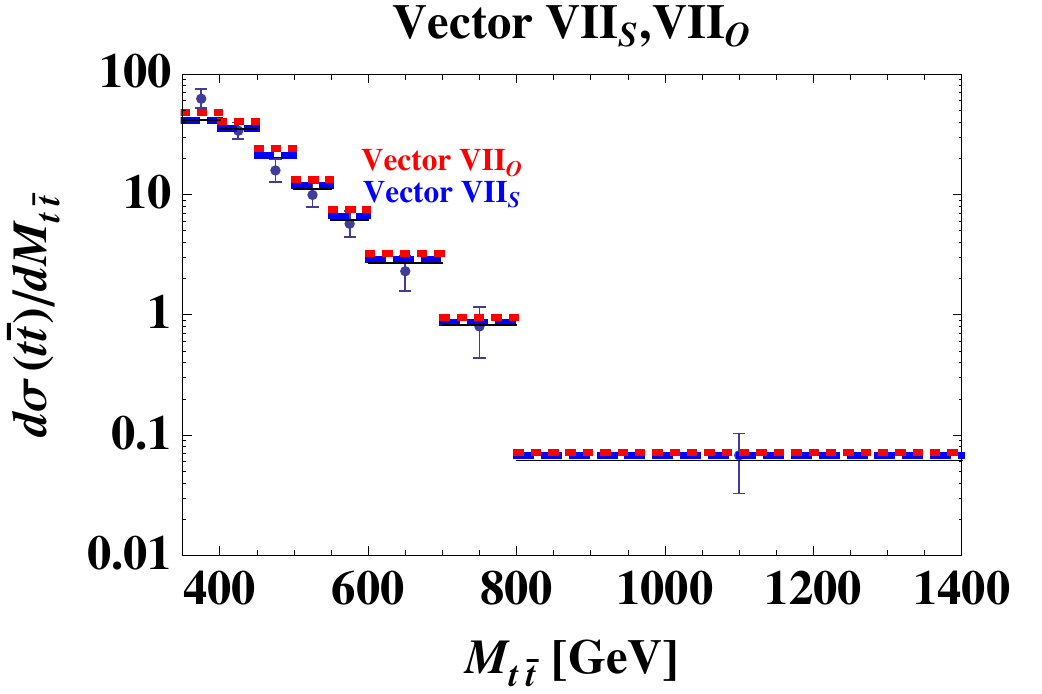}
\includegraphics[width=0.48\textwidth]{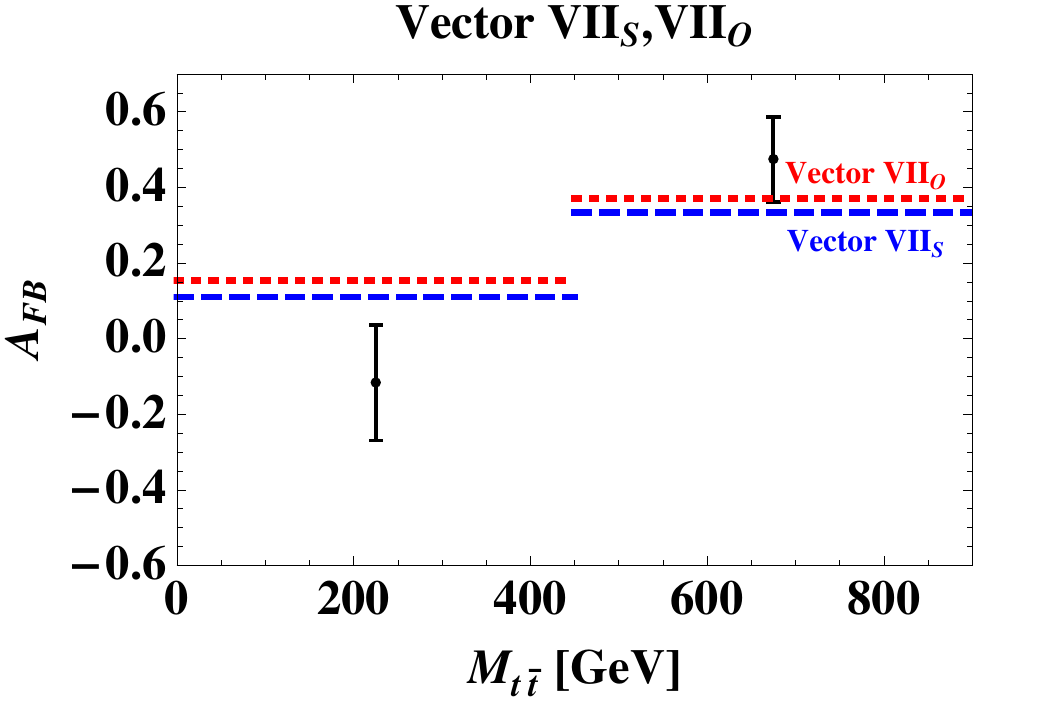}\\[-5mm]
\caption{Same as Fig. \ref{FigIIso} but for pure $t$ channel vector models  $\rm VII_{s,o}$ and coupling $f_{qt}=1/5$ (solid curve), $4/5$ (dotted), $8/5$ (dot-dashed), $16/5$ (dashed) for model $\rm V_s$  and couplings $f_{qt}=1/4$ (solid curve), $1$ (dotted), $9/4$ (dot-dashed), $4$ (dashed) for model  $\rm V_o$. For the benchmark points, $m_V = 300$ GeV.  }\label{FigVIIso}
\end{figure}

\begin{figure}
\includegraphics[width=0.46\textwidth]{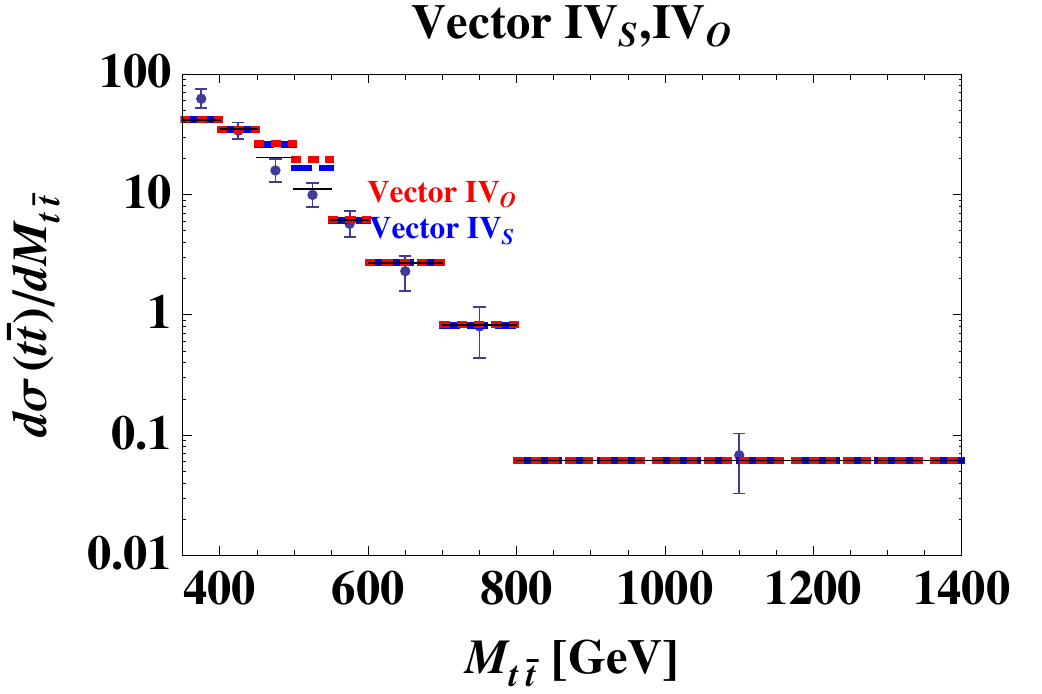}%
\includegraphics[width=0.46\textwidth]{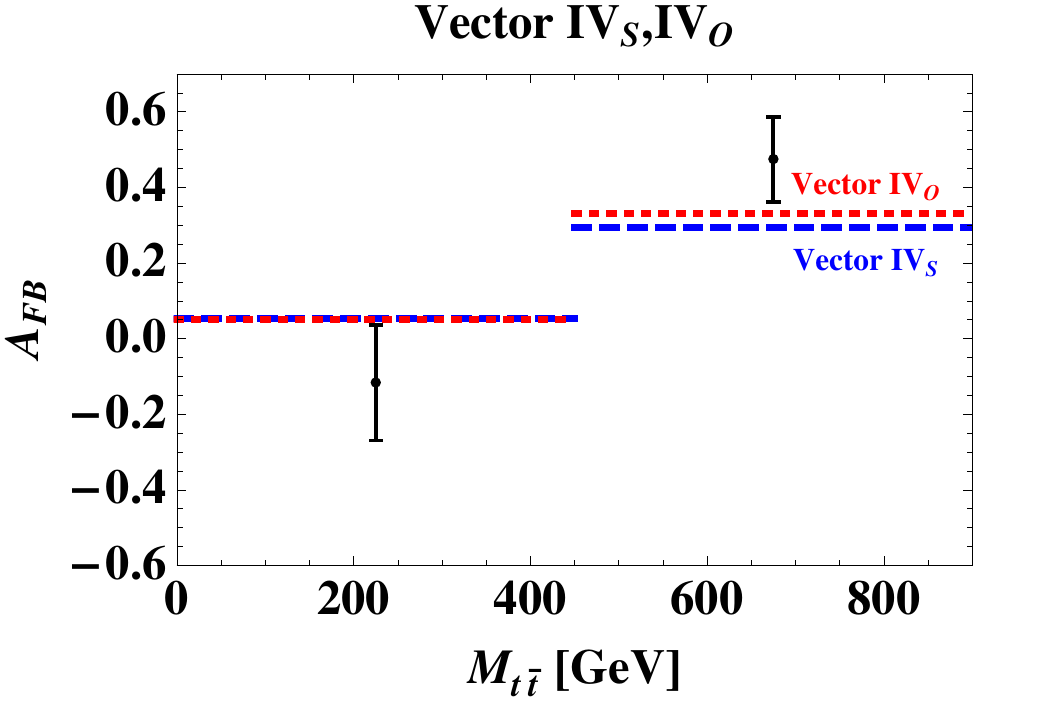}\\[2mm]
\includegraphics[width=0.46\textwidth]{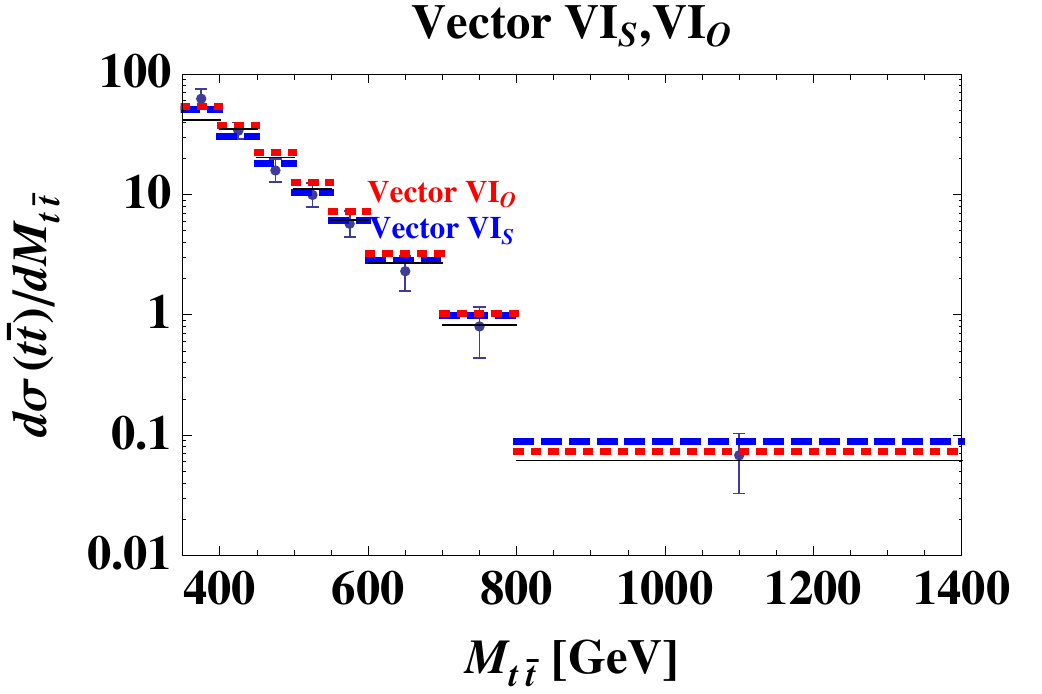}%
\includegraphics[width=0.46\textwidth]{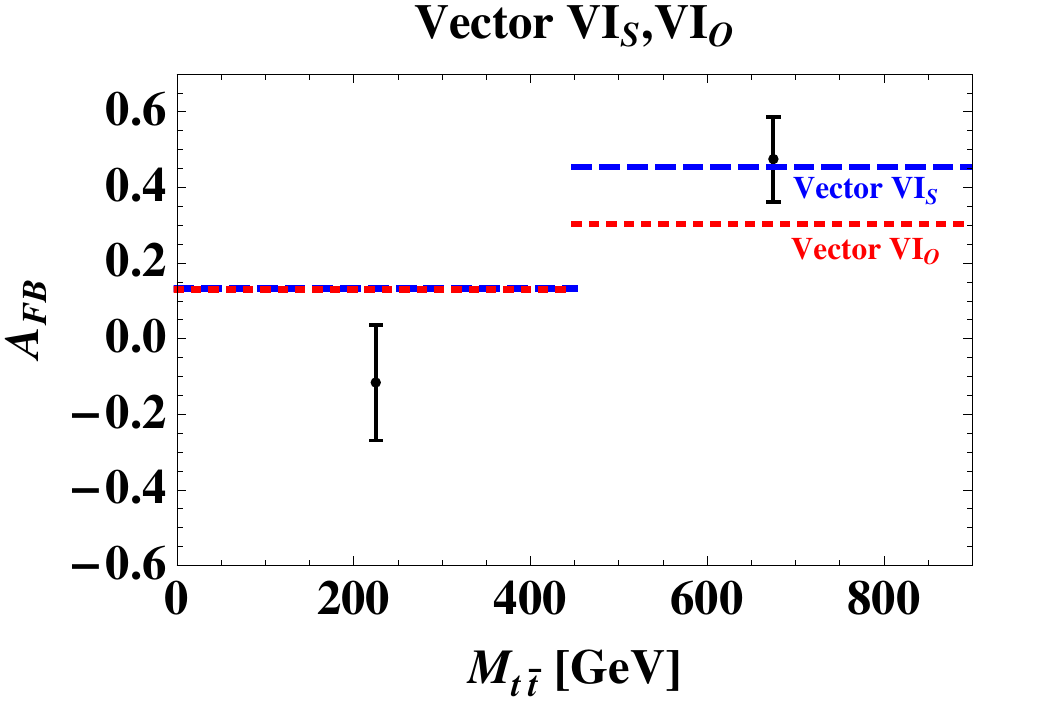}\\[2mm]
\includegraphics[width=0.46\textwidth]{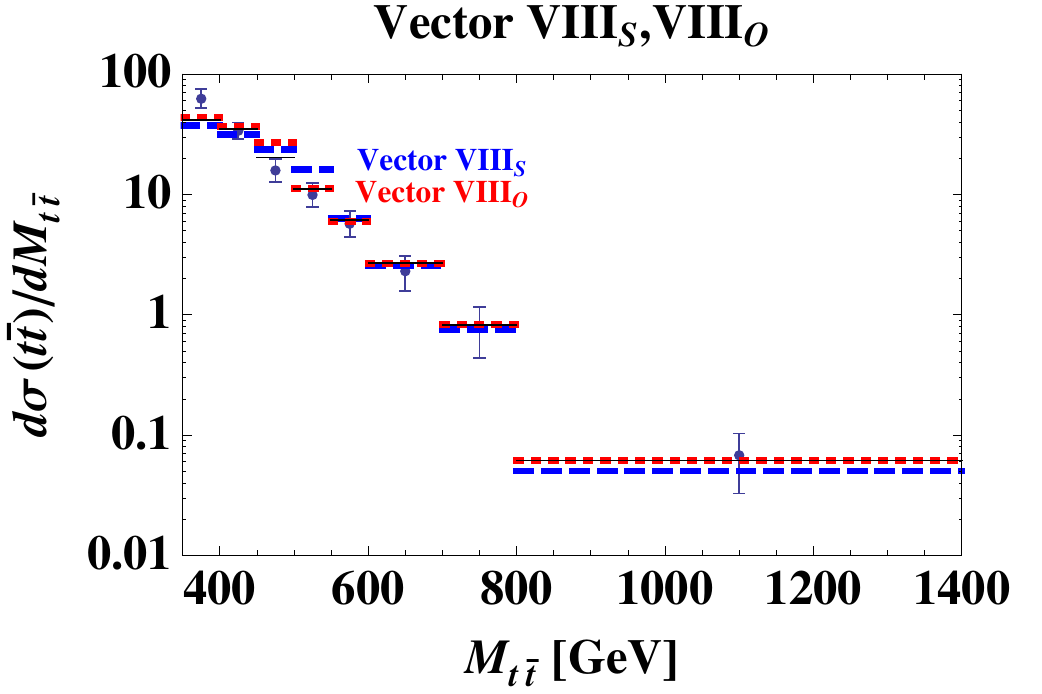}%
\includegraphics[width=0.46\textwidth]{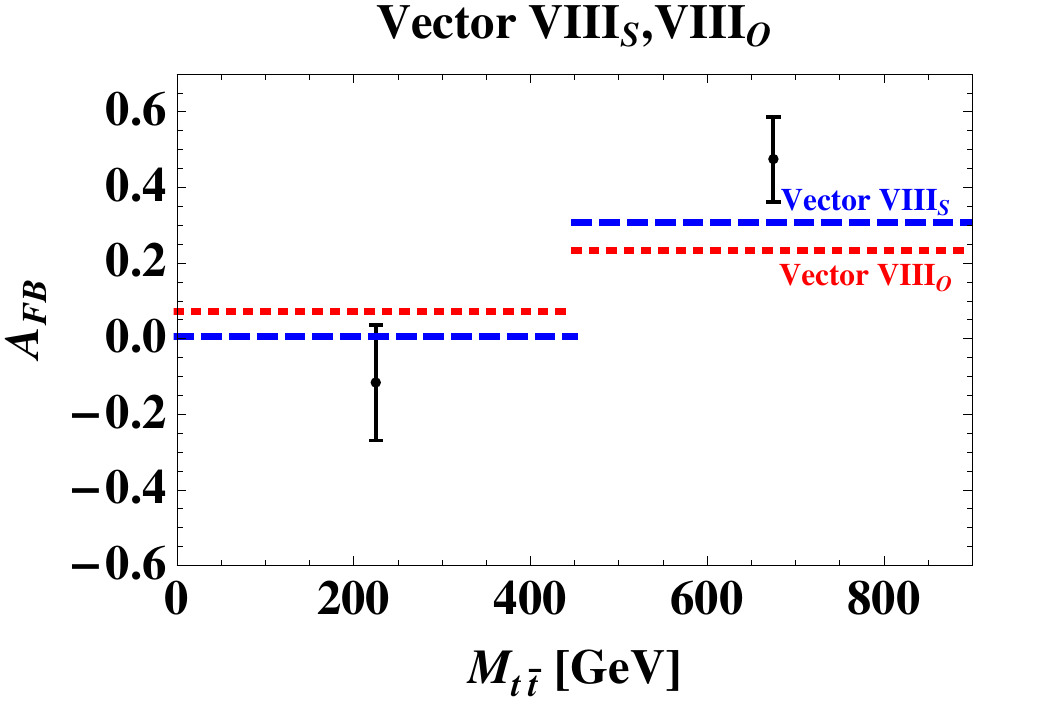}
\includegraphics[width=0.46\textwidth]{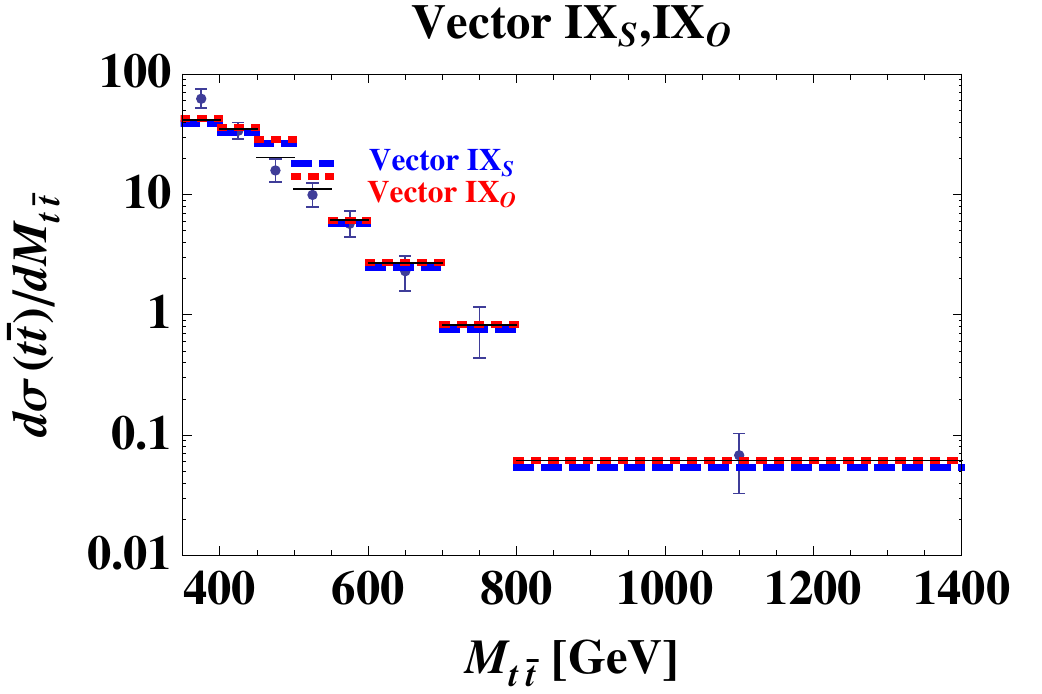}%
\includegraphics[width=0.46\textwidth]{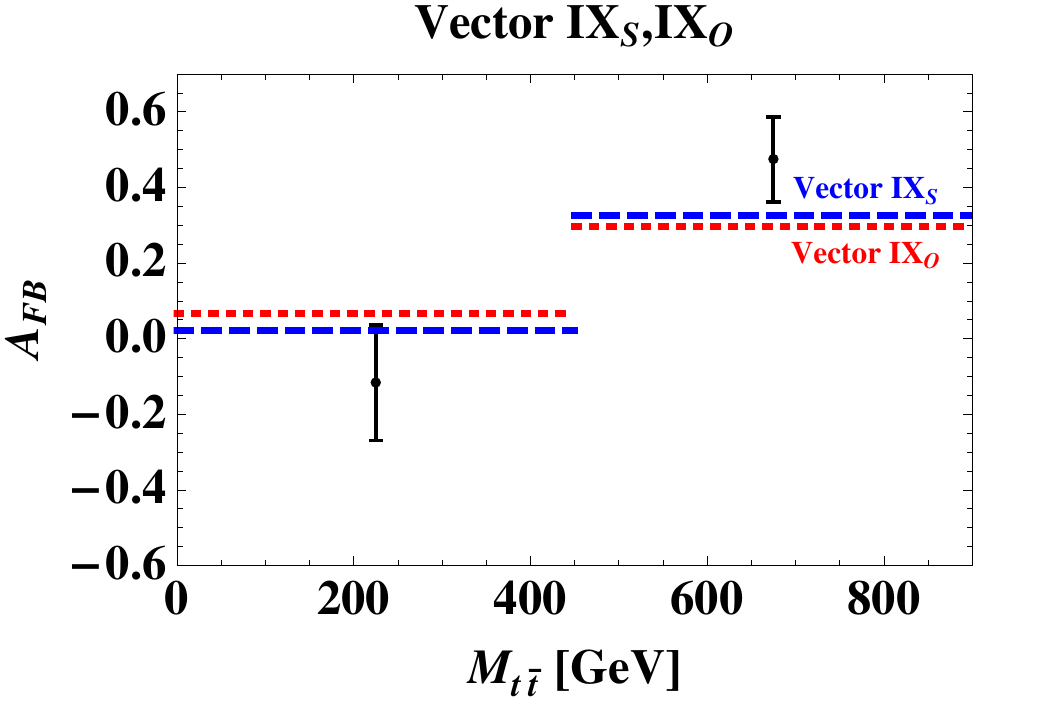}
\caption{Predictions for $d\sigma/dM_{t\bar t}$ and $A_{FB}$ for a set of vector models, ${\rm IV_{s,o}, VI_{s,o}, VIII_{s,o}}, IX_{s,o}$ fixing $m_V=$500 GeV (except for $\rm VI_{s,o}$ where  $m_V=$350 GeV, and using flavor breaking choices for couplings to quarks; $(\sqrt{f_q f_t},f_{qt})=(0.1,0.3) {\rm IV_s};$ $ (0.1,0.3) {\rm IV_o}; (0.55,1.3) {\rm VI_s};$ $ (0.55,1.3) {\rm VI_o}$ and $(\eta_1,\eta_2)=(1.4,-0.4) {\rm VIII_s};$ $ (1.3,-0.5) {\rm VIII_o};$ $(1.1,-0.4) {\rm IX_s}; $ $(1.1,-0.4) {\rm IX_o}.$
\label{fig:FV-VI}
}
\end{figure}

\subsubsection{The $t \, \bar{t}$ phenomenology of $\rm H_F$ symmetric vector fields}

In flavor symmetric scalar models, $t\bar t $ production can only proceed via the $u$ or $t$ channels.
Flavor symmetric vector models differ in this respect, as $t\bar t $ production can also proceed in the $s$ channel or both in $s$ and $t$ channels simultaneously.
Top phenomenology in these models is dictated to a large extent by which channel dominates.
The sizes of the couplings required in order to explain the top asymmetry are roughly fixed by whether the vectors couple to up or
down quarks in the initial states, as in the scalar models. 
The Lagrangians for the vector models 
are given in Section \ref{MFVreps} and Appendix \ref{AppendixA}.
Expressions for the $q \, \bar{q} \rightarrow t \, \bar{t}$ cross sections for models $\rm I- XI$ are collected in Appendix \ref{Apptevatron}.  
They are given in terms of effective couplings of the vectors to light quarks ($f_q $), to top quarks ($f_t $), and to top and light quarks ($f_{qt}$), defined in terms of $\eta_{1,2}$ etc, see Table \ref{table-int-vector}.  In this section we use these couplings as numerical inputs.
In the $\rm G_F$ symmetric limit ($y_t = 0$), one has $f_q = f_t = f_{qt}$.
Among the vector models, $\rm I_{s}$ and  $\rm V_{s,o}$ do not contribute to top production, and $\rm I_s $, $\rm II_s $ and $\rm III_s$ do not interfere with the 
SM amplitude at tree-level.

We begin with an examination of the representative models $\rm II_{s,o}$, which only contribute in the $s$-channel.
The vector decay widths are given by $ m_V \,( 2  f_q^2  + f_t^2 )/8 \pi $ ($\rm II_s$) and $m_V  \,( 2  f_q^2  + f_t^2 )/48  \pi $ ($\rm II_o$),
neglecting phase space suppression due to top quarks in the final state.  
In the first two rows of Fig.~\ref{FigIIso} we collect predictions for the inclusive $A_{FB}^{t \bar{t}}$ and ${\sigma}_{t \, \bar{t}}$
as functions of the vector masses,
for several values of  the product $f_q f_t $.  
We have taken vector decay widths appropriate for the 
$\rm G_F$ symmetric limit:  $ m_V 3 f_q f_t /8 \pi $ ($\rm II_s$) and $m_V f_q f_t  /16  \pi $ ($\rm II_o$).
In the last row, predictions for $d\sigma/dM_{t\bar t}$ and $A_{FB}^{t\bar t}$ in the low and high $M_{t\bar t}$ bins are shown for the benchmark points
(denoted by a $\star$).
Good fits to the inclusive cross section and the top asymmetries can be obtained.  However, agreement with $d\sigma/dM_{t\bar t}$ is much harder to achieve, as the effect of the 
$s$-channel resonance is clear.  The bounds from the Tevatron can be avoided if  $m_V \gtrsim 1 \, {\rm TeV}$.  Unfortunately, for $m_V \gtrsim 1 \, {\rm TeV}$ and $O(1)$ couplings the effect on $A_{FB}^{t \bar{t}}$ is strongly reduced.  The LHC $d\sigma/dM_{t\bar t}$ spectrum would also be problematic.
It is possible, though, to hide a light $s$ channel resonance via an increase of its decay width due to additional non-$q \bar q$ final states, as has recently been discussed  in the context of pure $s$-channel axial-vector models \cite{Tavares:2011zg}.  Another possibility for increasing the vector decay widths may be to consider the $\rm G_F$ breaking hierarchy $f_q^2 << f_t^2 $, keeping the product  $f_q f_t $ fixed.


The flavor octet models ${\rm VI_{s}}$ and $\rm VI_o$ contain admixtures of $s$ and $t$ channel contributions, while the flavor $(3,\bar 3 , 1)$ models ${\rm VII_{s}}$ and ${\rm VII_{o}}$ are purely $t$ channel.
The $s$ channel contributions in ${\rm VI_{s,o}}$ are due to the vectors associated with the generator $T^8$ of $\rm SU(3)_{U_R}$ and the coupling product $f_q f_t$.
The $t$ channel contributions are due to the vectors associated with the generators $T^{4,..,7}$ and the coupling product $f_{qt}^2 $.
While the generic problem of $s$ channel resonances persists, it is mitigated by their relative suppression compared to the  $t$ channel contributions
(the coefficients $C_{1,2}$ are significantly smaller than $C_3$ for these models, see Table \ref{table-int-vector}), by possible cancelations between the two channels,
and by the absence of interference between the SM and NP $s$ channel contribution in $\rm VI_s $  ($C_5 = 0$).
The $s$ channel effects can be further diminished, if the vector decay widths are enhanced by additional non-$q \bar q$ final states.\footnote{Given that these models are by themselves non-renormalizable, it is reasonable to expect that renormalizable UV completions could contain additional 
vector decay modes.}
This assumption was implicit in the model $\rm VI_o$ examples presented in~\cite{Grinstein:2011yv} and 
Fig.~\ref{Fig-eff2}.
In the following, we only consider vector decays to quark bilinears.  In the $\rm G_F $ symmetric limit ($f_q = f_t = f_{qt}$)
the decay widths can be written as $f_{qt}^2  m_V /16 \pi $ ($\rm VI_s $ and $\rm VII_s $) and 
$f_{qt}^2 m_V / 96 \pi $ ($\rm VI_o $ and $\rm VII_o $). 
Finally, $s$ channel effects in $\rm VI_{s,o}$ would be suppressed in the $\rm G_F$ breaking limit $f_{q} f_t << f_{qt}^2$ (this breaking is natural due to the large top yukawa), and would be absent entirely 
in the minimal $\rm H_F$ symmetric realizations which only contain the complex $\rm SU(2)_{U_R} $ doublets associated with the $SU(3)_{U_R} $ flavor generators $T^{4,..,7}$.

We show the effect of the models ${\rm VI_{s,o}}$ and $\rm VII_{s,o}$ on  $A_{FB}^{t \, \bar{t}}$ and
$\sigma_{t \, \bar{t}}$ in the $\rm G_F $ symmetric limit in Figs.~\ref{FigVIso} and ~\ref{FigVIIso}, respectively. In the purely $t$-channel models $\rm VII_{s,o}$, we see that it is possible to enhance $A_{FB}^{t \, \bar{t}}$, while at the same time not introducing any visible deviations in the $d\sigma/dm_{t\bar t}$ differential spectrum.
For models ${\rm VI_{s,o}}$, on the other hand, 
we see that it is difficult to enhance $A_{FB}^{t \, \bar{t}}$ sufficiently, without obtaining excessive $s$ channel peaks in the  $d\sigma/dM_{t\bar t}$ spectra, as in the model $\rm II_{s,o}$ examples above.
However, as already noted, much better agreement with the measured spectrum can be achieved in this case, if ${\rm G_F}$ is broken down to its subgroup ${\rm H_F}$\footnote{Whereas in models $\rm VII_{s,o}$ the $\rm G_F \to H_F$ breaking is not needed in order to obtain good agreement with the $d\sigma/dm_{t\bar t}$ spectrum, it may be required to evade dijet  and $B \to K\pi$ constraints.}.
In Fig. \ref{fig:FV-VI} we show such examples, with $f_{q} f_t << f_{qt}^2$, for various vector models.
The decay widths of the different vectors in a $\rm G_F$ multiplet will not be equal in this limit.  For simplicity, we
have identified all the widths in each multiplet with the quantities $ m_V f_q f_t /16 \pi  \times $   $(3,1/2,1,1/6,2,1/3,1/2,1/12)$ in models  $\rm IV_{s}$, $\rm IV_{o}$, $\rm VI_{s}$, $\rm VI_{o}$, $\rm VIII_{s}$, $\rm VIII_{o}$, $\rm IX_{s}$, $\rm IX_{o}$, respectively (which would hold in the $\rm G_F$ symmetric limit, neglecting phase space differences).
The rationale for this approximation is that the $s$ channel resonance widths are associated with the coupling products $f_q^2 $  and $f^2_t$, whereas 
the widths of the $t$ channel resonances, controlled by the product $f_{qt}^2$, have minimal impact on the $M_{t\bar t}$ spectrum.
In addition, as we will see in Section \ref{tevatron-constraints} (see Table \ref{table-vector-bounds}), dijet constraints on the light quark vector couplings $f_q$ are relatively mild,
so that the relation $f^2_q = f^2_t  = f_q f_t $ would be phenomenologically viable in the light vector examples of Fig. \ref{fig:FV-VI}.

{


\subsubsection{Limits from LHC measurements of the $t \, \bar{t}$ invariant mass spectrum.}

Recently, the first measurements of the $t\bar{t}$ invariant mass spectrum at the LHC have been presented by the ATLAS and CMS collaborations. With up to $0.7$ pb${}^{-1}$ the ATLAS collaboration obtains $\sigma_{t\bar t}=\rm 176\pm5(stat)^{+13}_{-10}(syst.)\pm7(lumi.)$ pb, 
while using $36 {\rm pb}^{-1}$ the CMS collaboration  obtains a combined cross section $\sigma_{t\bar t}=\rm 158\pm10(uncor.)\pm15(cor.)\pm6(lumi.)$ pb \cite{cmsnote}. This is to be compared with the NLO+NNLL predictions or the approximate NNLO predictions for the inclusive cross section that are listed in Ref. \cite{Ahrens:2011mw}, and range between $\sigma_{t\bar t}=157\pm 13$~pb and  $\sigma_{t\bar t}=145\pm 11$~pb, depending on the approximation used. ATLAS and CMS have also presented promising first measurements of the differential cross section at 200 pb${}^{-1}$ and 886 pb${}^{-1}$ \cite{CMS-Mulders-talk}, respectively. 

\begin{figure}
\includegraphics[width=0.55\textwidth]{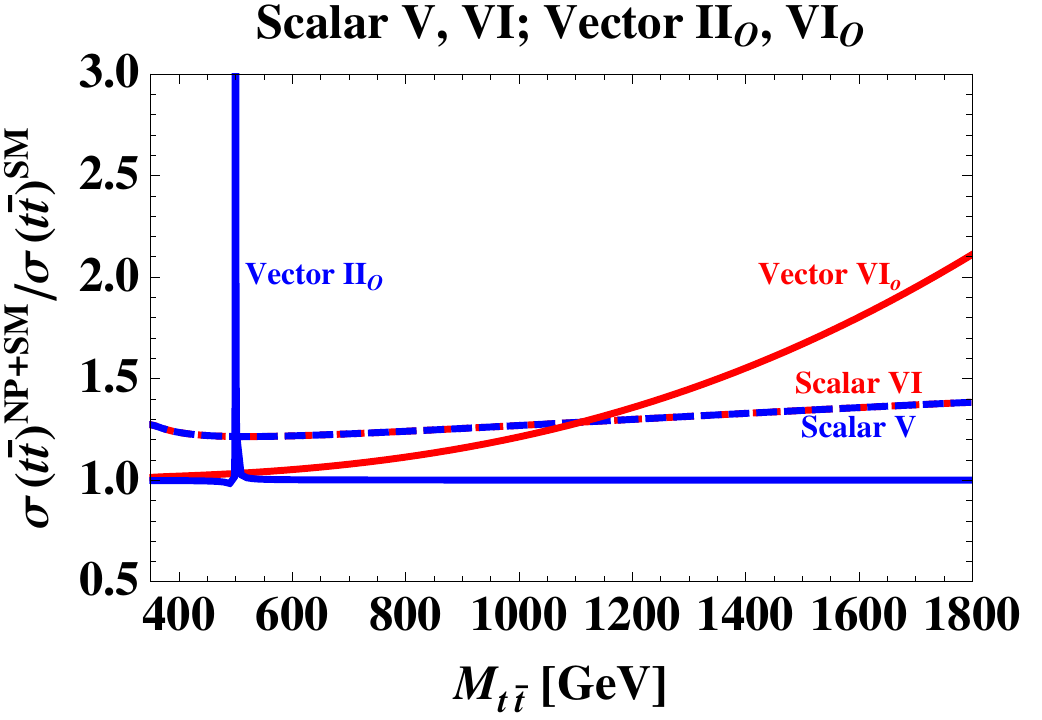}
\caption{Predictions for the ratio $(d \sigma(t \,\bar{t})_{NP+ SM}/d M_{t \bar{t}})/(d \sigma(t \,\bar{t})_{SM}/d M_{t \bar{t}})$ for scalar and vector models. The scalar model examples are for model $\rm S_{V}$ with $(\eta, m_S)=$ $(1, 400\, {\rm GeV})$ dashed blue, and for model 
$\rm S_{VI}$ with $(\eta, m_S)=$ $(1/\sqrt{2},800\, {\rm GeV})$ dotted red. The vector model 
examples are for $\rm II_{o}$  with $(f_q = f_t , M_V)=$ $(1/\sqrt{32}, 500 \, {\rm GeV})$ solid blue, and for  
 with 
$(f_q \, f_t , f_{qt} ,M_V )=$ $(0.5, 2, 500 \, {\rm GeV})$ solid red. 
}\label{LHCspecvector}
\end{figure}

In the near future the differential cross section measurements will place meaningful constraints on the models we have considered.
We therefore present below some representative examples of their impact   on the differential spectrum. In Fig. \ref{LHCspecvector} we show the ratio \beq
R_\sigma=\frac{d \sigma(t \,\bar{t})_{NP+ SM}/d M_{t \bar{t}}}{d \sigma(t \,\bar{t})_{SM}/d M_{t \bar{t}}}. 
\eeq
in models $\rm II_o, VI_o, S_V,$ and  $\rm S_{VI}$ for the benchmark points shown in Figs.  \ref{FigIIso}, \ref{FigVIso}, and \ref{fig:FV-VI}. In the ratio we use SM NLO+NNLL prediction  from Ref. \cite{Ahrens:2010zv}, while NP contribution including interference with the SM is calculated at LO, mirroring our procedure for Tevatron predictions used in previous sections. In Fig. \ref{LHCspecvector}  a sharp resonant peak is clearly visible for the disfavored pure $s$-channel model $\rm II_o$. Once convoluted with the experimental resolution the peak will be less prominent. For instance assuming 10 GeV resolution the peak leads to an ${\mathcal O}(30\%)$ enhancement of the cross section in the bin containing the peak.  The model $\rm VI_o$ example contains a rise in the tail region associated with the Rutherford scattering peak that is characteristic of $t$-channel models with ${\mathcal O}(1)$ couplings. The scalar models $\rm S_V$ and $\rm S_{VI}$, being $u$-channel models, display a relative enhancement throughout the entire differential spectrum \cite{Gresham:2011fx}, similarly to the case at Tevatron (see Fig. \ref{Fig:ScalarVVI}). Model $\rm S_I$ displays virtually no deviation from the SM spectrum, as pointed out in Refs. \cite{AguilarSaavedra:2011ug,Blum:2011fa}.

\subsection{Contributions to $B_{s}-\bar B_s$ mixing}

There is some evidence for NP contributions to the measurement of the $B_s$ mixing phase. A 3.9$\sigma$ deviation from the negligible SM prediction has been measured in the like-sign dimuon charge asymmetry by the D\O~collaboration \cite{Williams:2011nc,Abazov:2010hv}. This result is in agreement \cite{Ligeti:2010ia,Lenz:2010gu} with a hint for a nonzero weak phase in $B_s$ mixing (measured through flavor tagged decays \cite{D0-6098-CONF,CDF-10206}). The hints for NP in $B_s$ mixing have two preferred solutions  \cite{Ligeti:2010ia}, with 
\beq
h_s\sim0.5, \sigma_s\sim 130^\circ  {\rm ~and~} h_s\sim2, \sigma_s\sim 100^\circ.
\eeq
Here, $h_{s,d}$ and $2\sigma_{s,d}$ denote the magnitude and phase, respectively, of the NP contribution to the $B_{s,d}$ mixing amplitude, normalized to the SM one (see also \eqref{hsd-definitions}). Note that the above results have been obtained using the older measurement of the dimuon asymmetry \cite{Abazov:2010hv}. Since the two measurements are consistent, we do not expect a significant  change in the position of the two minima.
There is also a slight preference for $h_d\sim 0.2$, $\sigma_d\sim 100^\circ$, but at slightly more than $1\sigma$ $h_d$ is consistent with zero. At $3\sigma$ 
one finds $h_d<0.5$ for all $\sigma_d$  \cite{Ligeti:2010ia}. 

The flavor symmetric models can be grouped by whether or not they can give contributions to $B_{s}-\bar B_s$ mixing.  The models in which the new fields only couple to $u_R$ fall into the latter category
These are models ${\rm II}_{s,o}$, ${\rm VI_{s,o}}$ and ${\rm S_{V,VI}}$. Then there are models in which the NP contributions first arise at loop level: ${\rm VII_{s,o}}$ and ${\rm S_{IX,X}}$, which couple to $d_R$ and $u_R$, 
${\rm , XI_{\bar 3,6}}$ and ${\rm S_{I,II}}$ that couple to $u_R$ and $Q_L$. For the remaining models tree level contributions to $B_s$ mixing are possible. 

For a quantitative analysis one needs to specify the ${\rm H_F}$ breaking. We will adopt the MFV hypothesis, assuming that 
the breaking in the NP sector is only due to the SM Yukawa couplings.
(for an example with maximal breaking of $H_F$ see \cite{Shelton:2011hq}). This hypothesis ensures that the FCNCs generated from the exchange of NP fields have SM CKM suppression, making the FCNC constraints easier to satisfy. MFV scalar fields that can lead to the  dimuon anomaly have been discussed in the literature previously \cite{Blum:2010mj,Buras:2010pz,Buras:2010zm,Trott:2010iz}. We begin by classifying the flavor symmetric vector and scalar fields that can contribute to $B_s$ mixing at tree level.
Most of the details are relegated to Appendix \ref{AppFCNCs}.

The MFV models fall in two categories:``universal models" leading to class 1 $\Delta B=2$ mixing operators in the terminology of \cite{Kagan:2009bn}, and ``yukawa suppressed" models that give rise to class 2 operators and are additionally suppressed by light quark yukawa couplings, $y_{s,d}$. In universal models the NP contributions (normalized to the SM) are equal in the $B_s$ and $B_d$ mixing amplitudes. 
This is in some tension with experiment, where  there are indications for large effects in $B_s$ mixing, i.e.,
$50\%-200\%$ of the SM amplitude. Measurements in the $B_d$ system, on the other hand, can accommodate a NP contribution of up to  $\sim 20\%$. Universal models can thus explain the dimuon anomaly only if the real effect is on the lower end of the experimentally preferred range for $B_s$ mixing and on the upper range for $B_d$ mixing. 

The simplest examples of universal MFV models that can explain the dimuon anomaly are the two scalar doublet model, where in our notation the higgs linear combination without a vev is ${\rm S_H}$  \cite{Buras:2010pz,Buras:2010zm,Trott:2010iz}, and the color-octet, weak-doublet scalar model ${\rm S_8}$ \cite{Manohar:2006ga}. Effects are large even for $m_S=1$ TeV, if the couplings of the flavor breaking terms $\eta_i$ and the bottom Yukawa coupling $y_b$ are ${\mathcal O}(1)$. For the SM value $y_b\sim0.02$ the mass scale for the new uncharged scalars would have to be quite low. Such a low mass  scale, surprisingly, is consistent with present phenomenology for ${\rm S_8}$ \cite{Burgess:2009wm}. Similar comments apply to the other universal models, ${\rm III_{s,o}, IV_{s,o}, VIII_{s,o}, IX_{s,o}}$ and $\rm S_{XIV}$. In order to obtain the right size of $h_s$ for the dimuon anomaly one needs $\eta_i\sim 0.1$, if $m_{S,V}=1$ TeV.

The yukawa suppressed models can have mixing amplitudes suppressed either by one power or two powers of light yukawas. For $y_{s}^2$ suppressed models, ${\rm I_{s,o}, V_{s,o}}$ and $\rm S_{VIII}$, the $B_s$ mixing anomaly can be potentially explained only if $y_b, \eta_i \sim {\mathcal O}(1)$ and the masses of NP fields are no more that a few 100 GeV. The NP contributions to $B_d$ are predicted to be too small to be observed in this case, since they are additionally suppressed by $y_d^2/y_s^2$. The singly $y_{s}$ suppressed models, ${\rm X_{\bar 3,6}}$ and ${\rm S_I, S_{III}}$, give contributions to $h_s$ of the right order of magnitude for $\eta_i \sim {\mathcal O}(1)$ and $m_{S,V}\sim 1$ TeV. There is also an effect predicted for $B_d$ mixing with size $h_d\sim 0.1$.

A number of flavor symmetric models can potentially explain both the $B_s$ mixing and $A_{FB}^{t\bar t}$ anomalies simultaneously. If enhanced $B_s$ mixing is due to tree level exchange, the NP fields need to couple to left-handed doublets if they are to also contribute to $t\bar t$ production. The relevant models are ${\rm III_{s,o}, IV_{s,o}, VIII_{s,o}}$ and ${\rm IX}_{s,o}$, where for TeV masses, the flavor breaking couplings need to be $\eta_i\sim {\mathcal O}(0.1)$.
However, none of these models seem to fit  the observed data in top pair production particularly well (See Fig. \ref{fig:FV-VI} for models  $\rm IV_{s,o}, VIII_{s,o}$ and ${\rm IX}_{s,o}$. The models ${\rm III_{s,o}}$ are pure $s$-channel and thus lead to large deviations in the $t\bar t$ differential cross section.). 

There are also four models of interest that exhibit linear Yukawa suppression and can in principle contribute to $B_s$ mixing and to the $A_{FB}^{t\bar t}$ anomaly simultaneously. These are models ${\rm X}_{\bar 3,6}$ and models ${\rm S_{III}}$, ${\rm S_{IV}}$. However, due to their color representation $\rm S_{III}$, $\rm S_{IV}$, have limited potential to explain $A_{FB}^{t\bar t}$, while ${\rm X}_{\bar 3,6}$ reduce rather than increase $A_{FB}^{t\bar t}$. 

This leaves us with 
models where the contributions to $B_s$ mixing are loop suppressed. 
These are the vector models ${\rm VII_{s,o}}$, ${\rm XI_{\bar 3, 6}}$ and the scalar models ${\rm S_{I}, S_{II}}$,  ${\rm S_{IX, X}}$.  They lead to contributions to $B_s$ and $B_d$ mixing which are universal, if one assumes MFV. Among these, models ${\rm VII_{s,o}}$ and ${\rm S_{I}}$ can also explain the present top data.  The contributions to $B_s$ mixing can also carry a new weak phase, as required by the dimouon charge asymmetry. 

For instance, model ${\rm S_I}$, see Eq. \eqref{SImina},
contains contributions 
that are proportional to the product of the generally complex
quantity
$4 \eta_0^2\eta_3^2$ (where these coefficients are defined in Table \ref{kappa-table3} in
Appendix \ref{AppFCNCs}) with a loop function that is of the
same order of magnitude as the SM one, with its precise value
depending on the size of $m_S$.  The $B \to K\pi$ branching ratios require $\sqrt{\eta_0 \eta_3 } \lsim 1/4$ (see Section \ref{FCNC}),
while $A_{FB}^{t\bar t}$ favors $\eta_0 \sim 1$.
Thus, CPV contributions to $B_s$
mixing of ${\mathcal O}(20\%)$ are readily attainable for $\rm S_I$ scalar masses of order 100 GeV. Similar considerations apply to models $\rm VII_{s,o}$.

\section{Existing experimental constraints on a flavor symmetric sector}\label{lepEWPD}
We now determine some of the existing experimental constraints on flavor
symmetric extensions of the SM. We examine bounds on direct production at LEP, from
electroweak precision data (EWPD) and from collider searches at the Tevatron
and LHC. We also determine the (residual) FCNC constraints.

Our aim is characterize the general phenomenological bounds whenever possible and 
to check if some of the models that we have identified as promising candidates are
consistent with other phenomenology in some detail. Our results are not completely comprehensive,
as at times we have to focus on particular sample models. However, we believe our results are
valuable to gain some intuition as to the current phenomenological constraints on a flavor symmetric sector.

\subsection{LEP Constraints}

The LEP constraints depend on how the new states couple to electrons
and to the pair of final state fermions. In this section we will
discuss the constraints on the vector models from LEP in some detail.
The constraints for scalar models are very similar, see
\cite{Arnold:2009ay} for some details. Limits on anomalous four jet
events at LEP give a kinematic lower bound for pair production of the
scalar models of $105 \, {\rm GeV}$ \cite{Arnold:2009ay}. For the
vector and scalar models signals that involve couplings to quarks
and leptons are the most problematic at LEP.  For a $Z'$ vector boson
that couples to quarks (electrons) with couplings $g_z z_{fB}$ ($g_z
z_{eA}$), the bound is \cite{Carena:2004xs}
\bea M_{Z'}^2  \geq \, \frac{g_z^2}{4 \, \pi} \, | z_{eA} \, z_{fB}|
\, (\Lambda_{AB}^{f})^2\,.\label{bound}
\eea
with $A,B =\{L,R\}$ the chirality of the fermions. Typically $\Lambda_{AB}^{f} \sim 10~\rm TeV$, leading to
mass bounds on the order of $M_{Z'} \gtrsim 1~{\rm  TeV}$.  

The vectors in nontrivial color or flavor representations (${\rm
 I_o-IV_o}$ or ${\rm V -XI}$) are protected from having tree level
exchanges of this form. For models protected by flavor symmetry one
is forced to have the insertion of Yukawa matrices. The final
operating energy of LEP was $\sqrt{s} = 209 < 2m_t$ so the
flavor symmetry is sufficient to remove this bound for models $\rm V
-XI$.  These bounds do apply to $\rm I_s-IV_s$ unless the couplings to
electrons are somehow suppressed, {\it e.g.}, by extending flavor symmetry to the
lepton sector.

\begin{figure}[h]
\centerline{
\includegraphics[width=0.8\textwidth]{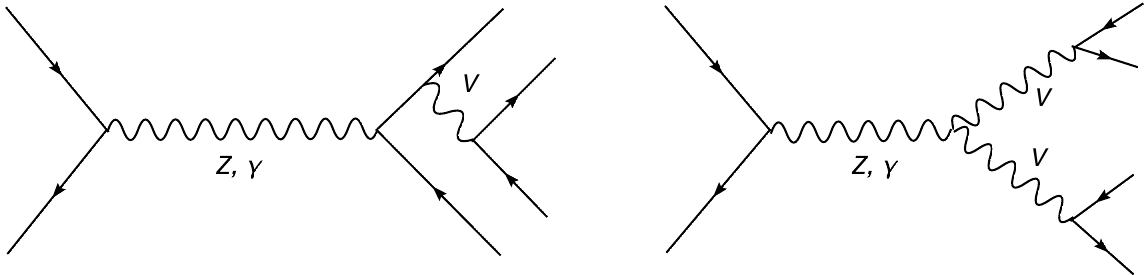}
} 
\caption{Tree level generation of multi-jet events at LEP due to flavor symmetric vectors. Similar diagrams (and bounds)
exist for scalars and the first diagram is representative of the emissions of the vectors and scalars off the final state fermions.}\label{LEPplot}
\end{figure}

For the models where the bound \eqref{bound} does not hold, there are
constraints from covariant derivative couplings to~$Z$ and~$\gamma$ in
the kinetic terms of the vectors (the factor $N$ is set by requiring
canonical normalization),
\bea \mathcal{L}_K = N \text{Tr}\left(D_{\mu} V^{\dagger}_{\nu} \, 
          D^{\mu} V^{\nu } - D_{\mu} V^{\dagger}_{\nu}
  \, D^{\nu} V^{\mu} \right)\,. 
\eea
Some tree level processes that would lead to anomalous four jet events
at LEP are shown in Fig.~\ref{LEPplot}. These processes do proceed
even in the absence of flavor violation. Furthermore, for vector pair
production (the second diagram) the couplings of the $Z$ or $\gamma$ to
the vectors are fixed by gauge invariance. This implies a kinematic
bound of $M_V \gtrsim 105 ~ {\rm GeV}$ on the flavor symmetric vectors.  The
analogous diagram for the scalar models is the origin of the
kinematic bound on the scalar masses discussed in
\cite{Arnold:2009ay}.

The first diagram(s) can lead to a stronger bound for a vector or a
scalar model as the vector or scalar is singly produced and has an order one
branching ratio to anomalous multi-jet final states due to the flavor
quantum numbers it carries.  We find that this production process gives a cross section of
$0.20|\eta|^2~\text{pb}^{-1}$ for
$M_V = 120 \, {\rm GeV}$, or, with an integrated luminosity of
$500$~pb above this mass scale, a total of  $100 |\eta|^2$ events. 
The number of expected events falls rapidly with increasing
$M_V$ and is  below $12|\eta|^2$ for $M_V \geq 140 \, {\rm
  GeV}$. Assuming an order one branching ratio to anomalous multi-jet
final states, this implies that the mass bound on the vectors from
LEP is $M_V \gtrsim 150 \, {\rm GeV}$ for ${\mathcal O}(1)$ couplings.
Colored vectors will have similar mass bounds from LEP from anomalous
multijet events. The calculation for the scalar models is very similar,
resulting in the same approximate bounds.

In summary, we consider the minimum LEP lower bound for all vector and
scalar models to be $\gtrsim 150 \, {\rm GeV}$ when order one
couplings exist for the fields to quarks.  
A dedicated study would be required for each model to obtain more precise bounds.
However, we note that while scalar (doublet) masses below 150 GeV were only considered in model $\rm S_I$, see Eq. \ref{SIminb}, for enhancing 
$A_{FB}^{t\bar t} $,  this bound could easily be evaded by making the charged components of the scalar doublets heavier.
Single exchange of the neutral components (the ones involved in $t\bar t$ production), as in the scalar analog of the first diagram of Fig.~\ref{LEPplot}, would 
lead to kinematically suppressed virtual $t\bar t  $ pair production. 
The LEP bound can be
generically stronger for a color singlet that does not transform under the
flavor group: for vectors $\rm I_s-IV_s$ the lower mass bound
is $\gtrsim 1 \, {\rm TeV}$ for order one couplings to leptons and
quarks.

\subsection{Electroweak precision tests}
\label{EWPDobliqueandnonoblique}

We consider the models for massive $\rm H_F$ symmetric vector and scalar fields to be
effective field theories.  Up to this point we have assumed that the
cut-off of the theory, $\Lambda$, is high enough that we can neglect
operators suppressed by powers of $\Lambda$.  For the scalar models,
in principle, we do not have to consider effects of higher dimensional
operators.  In this case, for scalar models that are $\rm SU_L(2)$
singlets, the results of \cite{Arnold:2009ay} show that the generic
lower bound from fitting to EWPD is $m_s \gtrsim 100 ~{\rm GeV}$.  The
LEP bounds we have just derived are stronger than these bounds for
scalar models that do not break custodial symmetry.

Below, we summarize our findings for model $\rm S_I$, for which particularly light scalar doublets were considered in the context 
of $\rm A_{FB}^{\bar t t}$.
We find that it 
is easily consistent with $S$ parameter bounds,
even in the presence of nine light doublets, as in the $\rm G_F$ symmetric realization,
rather than only two, as in the minimal $\rm H_F$ symmetric versions I(a) - I(c) (Eqs. \eqref{SImina} -- \eqref{SIminc}) considered in the previous section (we have generalized the analysis of \cite{Blum:2011fa} for a single scalar doublet).  The $T$ parameter bounds are easily satisfied in the minimal versions, with neutral scalar doublet component masses as low as 110 GeV
and heavier charged component masses of 150 GeV, a mass splitting that may be required in order to simultaneously evade the LEP bounds discussed above
and enhance $A_{FB}^{t\bar t} $ to observed levels.
Mass splitting between the charged and neutral scalar doublet components can be obtained via $\rm G_F $ symmetric 
couplings to the SM Higgs vacuum expectation value.
We have also checked the sizes of non-oblique corrections in the minimal versions of $\rm S_I$.  
In options I(a) and I(c), 
loop graphs containing the top quark lead to problematic shifts 
in the LH down and strange quark couplings to the $Z$.  For instance, taking
charged scalar masses of 150 GeV, neutral scalar masses of 110 GeV, and couplings $\eta \approx 2/3$
yields $ \delta g_{s_L} / g_{s_L},\, \delta g_{d_L} / g_{d_L} \approx -0.0046$, roughly $2\sigma $ in excess of the 2 $\sigma$ upper bound obtained, for equal shifts in the two couplings, from the $Z$ hadronic width. 
In option I(b), loop graphs containing the charm and up quarks lead to a shift in the LH bottom quark coupling to the $Z$.
For the same parameter choices (see Fig. \ref{Fig:FigSI}) the shift is $\delta g_{b_L} / g_{b_L} \approx -0.0028$, which is 
consistent with $R_b $ at $2\sigma$.  We therefore conclude that whereas I(b) is a viable option for enhancing $A_{FB}^{t\bar t}$, this is unlikely for cases I(a) and I(c).

For the massive vector fields we do have to consider the effects of
higher dimensional operators suppressed by the cut off scale as the theory is nonrenormalizable. This is
particularly the case when considering electroweak precision data
(EWPD) for vectors. To the effective fields we have considered
at the scale $\mu\sim M_V$ we should add dimension six
operators. These arise from integrating out heavier modes of the UV
complete theory. At the electroweak scale $\mu\sim m_Z$ we integrate
out also the vectors and can match onto an effective field theory
constructed from only the SM fields, appropriate for examining
EWPD. The dimension six operators then receive contributions from
vectors and from heavier fields from the UV completion.

For cases $\rm I_s - IV_s$, the EFT will in general have a dangerous
dimension 4 operator $\mathcal{O}_1 = V^\mu \, (H^\dagger \, D_\mu \,
H)$.
Integrating out the vectors gives the dimension six operator
$\mathcal{O}_T = |H^\dagger \, D_\mu \, H|^2/M_V^2$, well constrained
by EWPD. For positive (negative) Wilson coefficient $C_{T}$ the bound
on the vector mass is $M_V \ge 5.6 \, (4.6) \, {\rm TeV} \times
\sqrt{|C_{T}|}$ (adding only this operator to the SM EWPD fit, and
taking $m_h = 115 \, {\rm GeV}$)~\cite{Barbieri:2000gf}.

\begin{figure}[t]
\centerline{
\includegraphics[width=0.4\textwidth]{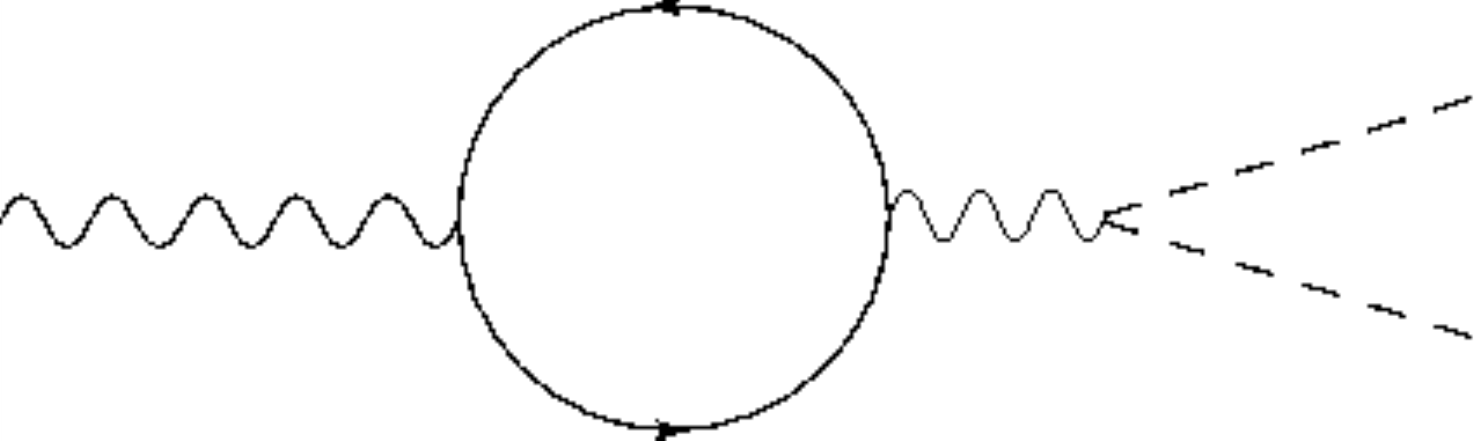}
} 
\caption{Contribution to the operator
  $\mathcal{O}_1$ generated by the SM through mixing with the SM $B$ field.}\label{LEPplot-2}
\end{figure}

The operator $\mathcal{O}_1$ could be suppressed by the unknown
dynamics in the UV theory.  However, it is also generated via the quarks
that the vectors couple to and SM interactions, see Fig.~\ref{LEPplot-2}. Due to mixing with the SM $B$ field (with coupling g) as shown
an effective wilson coefficient of  ${\rm C}_1 \simeq \Sigma_i \, y_i^2 \,
\eta^2 \, g^3 \, v^4/(16 \pi^2 \, M_V^2)^2$ is generated, where $y_i$ is the quark Yukawa coupling and the sum
is over all the allowed flavors (we neglect small $\rm G_F$
breaking). The bound is very weak as this is effectively a two loop effect due to mass mixing with the 
$B$ field. This gives a bound $M_V/\eta^{2/5} \gtrsim 50 \, {\rm GeV}$ for models $\rm
II_s, III_s, IV_s$ generated through SM interactions, weaker than the LEP bound discussed. 
The bound is even weaker if the vector only couples to
down quarks and $y_b$ is not enhanced from its SM value. Models that transform under flavor can
still generate this operator through SM interactions but require the insertion of the appropriate number of Yukawas in $\mathcal{O}_1$.
If the vector transforms under $\rm SU(3)_c$ the operator $\mathcal{O}_1$ is forbidden. 
In general, if this operator is not present due to matching onto the underlying theory, the bound through
custodial symmetry violating loop effects is weaker than the direct production LEP bound.

Note that a key assumption of this analysis --- fitting to EWPD with
the SM supplemented with a single operator such as $\mathcal{O}_T$ ---
is not well justified. We leave a more complete analysis of EWPD to a future publication
but note that non oblique observables, such as $R_b$ and the decay width of the SM $\rm Z$ boson
potentially can place stronger bounds on the $\rm H_F$ symmetric fields with a large number of degrees of freedom.
This is particularly the case when flavor off diagonal couplings are present and the top quark mass scale is 
present in large numbers of loop corrections to  $R_b$  and $\Gamma_Z$.

\subsection{Tevatron and LHC dijet Constraints}
\label{tevatron-constraints}

At the LHC and the Tevatron the most relevant constraints on flavor
universal couplings come from dijet resonance searches and dijet
angular distribution studies. These constraints limit the couplings of
the $\rm H_F$ symmetric fields to the light quarks. In this section we give a brief
account of these dijet bounds on scalar models $\rm S_{V}, S_{VI}$ and
vector models ${\rm II_o}$, ${\rm VI_o}$, and ${\rm VI_s}$, relegating details of the
calculations to Appendix \ref{Apptevatron}.   These models are of interest for the tevatron's top quark
forward-backward asymmetry, $A_{FB}^{t \bar{t}}$. 
The vector examples contain
prominent dijet contributions in both the $s$ and $t$-channels at the tevatron and LHC.
Scalar model $\rm S_{V}$ is $u$-channel dominated, 
due to the flavor antisymmetry of its quark couplings
($s$-channel effects are present in the $uc\to uc$, $\bar u c \to \bar u c$ channels);
model $\rm S_{VI}$
contains $s$-channel ($uu \to uu$) and $u$-channel contributions.
Related bounds on some of these models were
also discussed recently in
\cite{Grinstein:2011yv,Ligeti:2011vt,Giudice:2011ak}.  

The relevant scalar interaction Lagrangians in the $\rm G_F$ symmetric limit are given in Eq. \eqref{eq:SVandSVILag}. The
relevant vector Lagrangians are listed in Eqs. \eqref{symmetricV}--\eqref{XosEq} in Section \ref{section:MFVrepresentations}, and in Appendix \ref{AppendixA}
(with $\rm G_F $ breaking corrections also discussed).  Expressions for the dijet cross sections in these models 
are 
given in Appendix \ref{Apptevatron}. The new scalar or vector fields can have ${\mathcal O}(1)$
couplings to first generation quarks, a necessary condition for obtaining large $A_{FB}^{t \bar{t}}$.  Thus, they can mediate large contributions to
$2\rightarrow 2$ quark scattering, potentially spoiling the good agreement
between measurements and SM theory predictions for dijets. Tevatron and LHC bounds on their couplings to
light quarks are collected in Table \ref{table-scalar-bounds}.  
They are derived using partonic cross sections evaluated at LO in $\alpha_s$, with the renormalization 
and PDF factorization scales identified with the transverse momenta of the two outgoing partons.
Rapidity related experimental cuts on the two leading jets (the dijet measurements are inclusive) 
are emulated by imposing these cuts on the outgoing partons.
Clearly, a parton level treatment introduces considerable uncertainties.
Nevertheless, our bounds should still be indicative of the 
constraints one would obtain with more precise inclusive Monte Carlo simulations, carried out at NLO and including showering and
full detector response.  

\begin{table}
\begin{tabular}{c|cccc|c|cccc}
\hline\hline
\text{$\rm S_{V}^{3}$ Mass} &  \text{TeV  $M_{jj}$} &  \text{LHC  $M_{jj}$} &  \text{TeV $\chi$}  &  \text{LHC $\chi$} & \text{$\rm S_{VI}$ Mass} & \text{TeV  $M_{jj}$} &  \text{LHC  $M_{jj}$} &  \text{TeV $\chi$}  &  \text{LHC $\chi$} \\
\hline
300 \, & 1.0 & - & 1.2 & 1.1 & 300 &  0.3 & - & 0.4 & 0.5  \\
500 \,  & 1.2 &  $ \rm n.b.$  &  0.5 & 0.9 &  500 & 0.3 & 2.2 & 0.2 & 0.5  \\
700 \, & 2.0 &  0.7 & 0.7 & 0.6 & 700 &  0.6 & 0.2 & 0.2 & 0.2  \\
900 \,  & 2.5 &  0.3  & 0.6 & 0.5 &  900 & 0.7 & 0.1 & 0.2 & 0.2  \\
1100 \,  & 2.8 &  0.4  & 0.5 & 0.6 &  1100 & 1.4 & 0.1 & 0.2 & 0.1 \\
1300 \,  & 4.0 & 0.5 & 1.3 & 0.6 &  1300 & 1.6 & 0.1 & 0.7 & 0.1 \\
1500 \,  & 6.0 &  0.6  & 1.6 &  0.3 &  1500 & 1.8 & 0.1 & 0.8 & 0.1 \\
1700 \,  & $ \rm n.b.$ &  0.6  & 1.8 & 0.5 & 1700 & 2.0 & 0.1 & 0.8 & 0.1 \\
1900 \,  & $ \rm n.b.$ &  0.6  & 2.0 & 0.4 &  1900 & 2.6 & 0.1 & 0.9 & 0.1 \\
2100 \,  & $ \rm n.b.$ &  0.7  & 2.1 &  0.6 &  2100 & 3.0 & 0.1 & 1.0 & 0.1\\
\hline\hline
\end{tabular}
\caption{Approximate upper bounds on the couplings of the scalars $\rm S_{V}$, $\rm S_{VI}$ to light quarks due to the measured dijet invariant mass spectra (labeled $M_{jj}$) and angular distributions (labeled $\chi$) at the  tevatron and LHC, as explained in the text.
The masses correspond to the scalar field flavors $\rm S_{V}^{3}$ and $\rm S_{VI}^{11=22=12}$.
If no bound is determined we denote this with ``n.b.''.}\label{table-scalar-bounds}
\end{table}

In Tables \ref{table-scalar-bounds} and \ref{table-vector-bounds} 
we quote upper bounds on the couplings $\eta$ and $\eta_1$ (or $f_q$) of the light quarks to the scalar and vector fields, respectively,
for a range of masses.  
There are four sets of bounds, obtained from searches for dijet resonances in the dijet invariant mass ($M_{jj}$) spectra at CDF \cite{Aaltonen:2008dn} and LHC \cite{Khachatryan:2010jd,Aad:2011aj,ATLASnote095,CMSCollaboration:2011ns}, and from dijet angular distribution measurements at \dzero~\cite{:2009mh} and CMS~\cite{Khachatryan:2011as}.
 The CDF dijet resonance search \cite{Aaltonen:2008dn} (corresponding to a luminosity of 1.3 fb$^{-1}$) imposes a rapidity cut of $|y| < 1$
on the two leading jets.  For the LHC dijet resonance searches we use the CMS data \cite{Khachatryan:2010jd} (2.9 pb$^{-1}$) 
for 500 GeV scalar or vector masses, the ATLAS data \cite{Aad:2011aj}  (36 pb$^{-1}$) for 700 GeV masses, and the more recent ATLAS  \cite{ATLASnote095} (0.81 fb$^{-1}$) 
and CMS \cite{CMSCollaboration:2011ns} (1 fb$^{-1}$) data for heavier masses.  These studies impose a rapidity cut of $|y| < 2.5$ (2.8 in \cite{ATLASnote095}) on the two leading jets,
with a rapidity separation for these jets satisfying $|\Delta y | < 1.3$  (1.2 in \cite{ATLASnote095}).
Rapidity cuts are imposed in order to eliminate a large fraction of the dominantly $t$-channel, hence forward, QCD dijet background at large $p_T$;
$s$-channel NP effects will be more isotropic.  As already mentioned, we implement these cuts on the two outgoing partons.

The dijet angular distribution measurements quote normalized differential
cross sections $1/\sigma_{\rm dijet} \, d \sigma/ d \chi$, where 
the angular variable is $\chi \equiv (1 + |\cos \theta|)/(1 -|
\cos \theta|)$ and $\theta $ is the scattering angle for the $2 \to 2$ parton scattering process in the parton CM frame.
The differential cross sections are integrated over dijet mass intervals of a few hundred 
GeV in size, and results are presented in bins of $\chi$ for $0
< \chi < 16$.
The normalization, $\sigma_{\rm dijet}$, is the
measured cross section integrated over the dijet mass interval and over $0
< \chi < 16$.   Whereas the $\chi$ distributions for the QCD dijet background are relatively flat (due to
$t$-channel dominance), more central NP effects, e.g., from $s$-channel resonances, will peak at low $\chi$. 
The \dzero~  angular measurements~\cite{:2009mh} (0.7 fb$^{-1}$) include a cut on
$y_{\rm boost} \equiv 0.5 \, |y_1 + y_2  | <  1$, where $y_{1,2}$ are the rapidities of the two leading jets.
The CMS angular measurements~\cite{Khachatryan:2011as} (36 pb$^{-1}$ ) employ a similar cut of $y_{\rm boost} < 1.11$.
Again, we implement these cuts on the two outgoing partons.

In general, the dijet invariant mass spectrum of a NP model can exhibit both an $s$-channel peak in the vicinity of the mass of the mediating field, 
and a monotonic rise relative to the SM prediction at larger $M_{jj}$.
At the Tevatron, our scalar examples would possess
the latter feature, due to $u$-channel effects, but an $s$-channel peak would be less prominent, particularly for model $\rm S_{V}$.
Therefore, for these models the bounds on the light quark couplings quoted in the Tevatron ``$M_{jj}$'' columns of Table \ref{table-scalar-bounds} are obtained by requiring that the ratios of the predicted to SM $d \sigma/d M_{jj}$ spectra (with both evaluated at LO) lie within the PDF uncertainty band in Fig.~1b of
the CDF study \cite{Aaltonen:2008dn}.  In particular, 
we do not make use of the bump hunter bounds on the new particle production cross sections.
The widths of the intermediate scalar fields are identified with the two-body decay widths to quarks neglecting phase space,
$ \Gamma=\eta^2 \, m_s /(2 \, \pi)$ for $\rm S_{V}$ and $\rm S_{VI}$.  
The vector examples can produce prominent $s$-channel peaks at the Tevatron.  
We therefore obtain the Tevatron $M_{jj} $ bounds quoted in Table \ref{table-vector-bounds} from the CDF bump hunter
95\% CL upper limits on the product of a $Z^\prime$ production cross section $\times$ its branching ratio to dijets (Br) $\times$ acceptance ($A$), setting $\rm Br \times A=1$ (see Table I of  \cite{Aaltonen:2008dn}).  
For vector masses in excess of the last measured invariant mass bin, $M_{jj} \in [1225,1350]$ GeV,
we require that the ratios of the predicted to SM dijet cross sections in this bin lie below the 95\% CL upper bound, obtained 
from the systematic and PDF uncertainty errors added in quadrature (the corresponding ratios in the lower bins would be much smaller than such bounds, given the widths of the vector fields).  
The intermediate vector two body decay widths to quarks are  $\eta^2 \, m_V /(16 \, \pi)$ for IIo and VIs, and 
 $\eta^2 \, m_V /(96 \, \pi)$ for VIo.   
To obtain bounds on the scalar and vector couplings from the ATLAS and CMS dijet resonance searches we again use bump hunter 95\% CL upper limits on new particle production cross sections,
setting $\rm Br \times A=1$.   Specifically, for the CMS studies \cite{Khachatryan:2010jd,CMSCollaboration:2011ns} we make use of the bounds provided 
for $qq$ final states (see Table 1 of  \cite{Khachatryan:2010jd} for a 500 GeV resonance mass, and Table 1 of \cite{CMSCollaboration:2011ns});
for the ATLAS studies in \cite{ATLASnote095} we use the upper limits in Table 3, and in \cite{Aad:2011aj} we use the bound in Figure 3 (for a 700 GeV 
resonance mass).

\begin{table}
\begin{tabular}{c|cccc|cccc|cccc}
\hline\hline
 &  &  \text{$\rm  II_{o}$} &   &  &  &  \text{$\rm  VI_{o}$} &  &   &  &  \text{$\rm  VI_{s}$} & & \\
\hline
\text{Mass} &  \text{TeV  $M_{jj}$} &  \text{LHC  $M_{jj}$ } &  \text{TeV $\chi$}  &  \text{LHC \, $\chi$} & \text{TeV  $M_{jj}$} &  \text{LHC  $M_{jj}$} &  \text{TeV $\chi$}  &  \text{LHC \, $\chi$}  &  \text{TeV  $M_{jj}$} &  \text{LHC  $M_{jj}$} &  \text{TeV $\chi$}  &  \text{LHC \, $\chi$}\\
\hline 
300 \, & 0.4 & - & 0.9 &1.7 & 0.6 & - & 1.4 &2.2 & 0.6 & -     & 1.7     &  1.4     \\
500 \,  & 0.4 &0.8 & 0.3 & 1.3 & 0.4 & 0.9 & 0.4 &1.5 &0.5 &    1.0  & 0.5     &    1.4      \\
700 \, & 0.4 & 0.8 & 0.3 &0.9& 0.4 & 1.0 & 0.5 & 1.0 & 0.5  &  1.1   &   0.6   &  1.2    \\
900 \,  & 0.3 & 0.3 & 0.2 & 0.7 & 0.5 &0.3 &0.3 & 0.9 & 0.6 &  0.3  &  0.4   &  1.0     \\
1100 \,  & 0.4 & 0.3 & 0.4 & 0.8 & 1.1 &0.4& 0.6 &  1.0 & 1.3  &  0.5  &   0.8  &   1.2    \\
1300 \, & 0.7 & 0.5& 1.1 &0.9 & 1.0 & 0.6 & 1.6 & 1.2 & 1.2 &  0.6   &  1.6     & 1.2      \\
1500 \, & 2.7 & 0.5 &2.6& 0.9& 4.8 &0.6& 4.3 & 1.1 &3.4     &  0.7   &  3.1   &  1.2    \\
1700 \, & 4.0 & 0.5 & 3.5& 1.2& 6.4 &0.7 & 5.7 & 1.6 & 4.8 &  0.7   &   4.1  &   1.6    \\
1900 \,  & 5.3 & 0.7 & 4.4 & 1.0& 7.7 & 0.9 & 6.5 & 1.3 &6.1 &  0.8   & 4.8    &   1.4    \\
2100 \, & 6.5 &  0.8 & 5.2 & 1.4& 8.8 & 1.0 & 7.2 & 1.8 & 7.8  & 0.8   & 5.6     &   1.8    \\
\hline\hline
\end{tabular}
\caption{Approximate upper bounds on the couplings $f_q$ of the vectors $\rm  II_{o}, VI_{o}$ and $VI_s$ to light quarks, due to the measured dijet invariant mass spectra (labeled $M_{jj}$) and angular distributions (labeled $\chi$) at the Tevatron and LHC, as explained in the text.} \label{table-vector-bounds}
\end{table}

To set the bounds on the light quark couplings from the dijet angular distribution measurements,
we conservatively require that in each bin of $\chi$ and in each $M_{jj}$ interval, the addition of NP does
not result in a shift of $1/\sigma_{\rm dijet} \, d \sigma/ d \chi$ 
that is larger than the $1\sigma$ error.  In the case of the \dzero~measurement~\cite{:2009mh},
the errors are statistics dominated (see the ``EPAPS'' document cited therein),
and we can neglect the systematic and SM theory errors.
For the CMS study~\cite{Khachatryan:2011as} we add the theory and experimental errors in quadrature.
In all of our examples, the largest deviations occur at small $\chi$, and they tend to concentrate in the invariant mass intervals 
closest to the new particle masses, as would be expected for $s$-channel resonances.  
We can see from a comparison of all the bounds in Tables \ref{table-scalar-bounds} and \ref{table-vector-bounds} that
the LHC bounds have superseded the Tevatron bounds for new scalar or vector masses 
$\ge 1$ TeV, and that they are particularly restrictive for model $\rm S_{VI}$.

It is important to note that the constraints determined above are on the couplings of the $\rm H_F$ symmetric couplings of the new fields to light quarks. Only in the
$y_t \rightarrow 0$ limit, when the full $\rm G_F$ symmetry is restored, are these couplings the same as the couplings involving
heavy quarks. Thus, these bounds
cannot be directly translated into limits on the allowed $A_{FB}^{t\bar{t}}$, in general.  
Rather, these bounds 
dictate to what degree $\rm G_F \rightarrow H_F$ flavor breaking is required, in order to have
couplings that are suitable for explaining the $A_{FB}^{t \bar{t}}$ anomaly while being consistent with dijet constraints.
Considering the benchmark scalar examples in Fig. \ref{Fig:ScalarVVI},
flavor symmetry breaking is probably not required in model $\rm S_{V}$, while moderate breaking is required in model ${\rm S_{VI}}$.
In the case of vector models $\rm VI_{s,o}$, we can see that in examples in which simultaneous agreement with all of the top data is obtained with the help of the 
$\rm G_F \to H_F$ symmetry breaking hierarchy  $f_q f_t << f_{qt} $, see Fig. \ref{fig:FV-VI} some flavor symmetry breaking would probably also be required  in order accommodate the dijet bounds ($f_q < f_{qt} $);
however, in models in which agreement with the top data is achieved with the help of enhanced vector decay widths, see Fig. \ref{fig:vecXacc}}, 
little or no flavor symmetry breaking appears to be required by the dijet bounds.

We close this Section with a brief discussion of collider constraints on scalar model $\rm S_I$.
In particular, we consider the recent D\O\ bounds \cite{Abazov:2011af} on anomalous dijet production in $Wjj$ final states, prompted by 
a potential signal at CDF \cite{Aaltonen:2011mk},
and constraints from UA2 dijet measurements \cite{UA2}.
The scalar doublet fields $S_{ij}$,  $i,j=\{1,2\}$ couple to light quarks with equal $\rm H_F$ symmetric couplings $\eta_{ij}$, see Eq.  \ref{eq:SILag}, which would lead to
dijet peaks in $Wjj$ final states via associated $W + S_{ij}$ production.  
D\O\  has presented 95\% CL upper bounds on the cross section for such processes, ranging from 2.57 pb to 1.28 pb for dijet resonance masses of
110 - 170 GeV, i.e., in the preferred range for enhancing $A_{FB}^{t\bar t} $ in model $\rm S_I$.  We apply these bounds to the sum over cross sections for 
the four scalar doublet fields (charged and neutral components).  The resulting 95\% CL upper bounds on the couplings $\eta_{ij}$ are $\approx 0.2$
for these masses.
Finally, we find that the UA2 dijet measurements do not provide meaningful constraints on these couplings once finite width effects for the $S_{ij}$ and interference with the SM are taken into account.

\subsection{Residual Constraints from FCNCs} \label{FCNC} The fact that there
is a global flavor symmetry ${\rm H_F}$ in the models we consider
helps to avoid low energy constraints from Flavor Changing Neutral
Currents (FCNCs) even though the NP has weak scale masses. In this
section, as always when considering flavor bounds, we adopt a strict interpretation of the MFV hypothesis and
assume that ${\rm G_F}$ is only broken by the SM Yukawas. The detailed
analysis of FCNC constraints from $B_{d,s}, D$ and $K$ mixing is given
in Appendix \ref{AppFCNCs} while here we only collect the main
results.

We discuss the models in the two categories, universal and Yukawa
suppressed, that we have introduced. (In the notation of
Ref. \cite{Kagan:2009bn} these are the models that receive only
contributions from class-1 or class-2 operators). The universal models
give roughly the same contributions to $B_d$ and $B_s$ mixing amplitudes, when
normalized to the SM. For vector or scalar masses about 1 TeV, the
coupling constants have to be less than about 0.1. The
models of the universal type are vector models ${\rm III_{s,o},
IV_{s,o}, VIII_{s,o}, IX_{s,o}}$ and scalar model $\rm S_{XIV}$.

The Yukawa suppressed models give contributions to the mixing amplitudes
that are CKM suppressed (with the same CKM structure as the leading
short distance contributions in the SM). They are, in addition, Yukawa
suppressed. The suppression can start at linear or  quadratic order in light quark Yukawas. 
The FCNC bounds are satisfied for $\eta$ couplings of
${\mathcal O}(1)$ for vector or scalar masses above $\sim1$ TeV for
linear Yukawa suppressed models. For quadratically Yukawa suppressed
models the FCNC constraints are even weaker and are already satisfied 
for masses above $\sim 300$~GeV.

Finally, meaningful constraints can be obtained for certain quark couplings in model
$\rm S_I$ from  the $B \to K\pi$ branching ratios  \cite{Blum:2011fa,Zhu:2011ww}.
For example, if the LH quarks in the ${\rm S_I}$ Lagrangian of Eq. \eqref{eq:SILag} are defined in the 
up quark mass eigenstate basis, the couplings of the charged scalars in the quark mass eigenstate bases would 
include the terms
\beq 
\eta_{k\, 1}\, S^-_{k \,1}   \,(V_{\rm CKM}^\dagger )_{j \,k}  \,\bar{d}^{\,\prime}_{j\, L } \,u^\prime_{R}  ~~+~~ h.c. \,,
\eeq
where $j,k=\{1,2,3\}$.
In the $\rm G_F$ symmetric limit, the contributions to $B \to K\pi ,\,\pi\pi $ from 
tree-level exchanges of the scalars $S^-_{k1}$ would exactly cancel.
However, in the limit of large $\rm G_F  \to H_F$ symmetry breaking (as would be required in order to maintain consistency 
with the $Wjj$ bounds while 
accounting for enhanced $A_{FB}^{t\bar t}$)
the couplings would be constrained.
The tightest bounds come from $B \to K^+ \pi^-$.  Borrowing from the analysis of \cite{Blum:2011fa}, we find 
$\eta_{ k \,1 } < 0.23  \,(m_S / 125~\rm GeV)$, when considering the exchange of $S^-_{31}$ and $S^-_{21} $ separately.  
Thus, tree-level exchange of the scalar doublet fields $S_{3i}$ could not account for $A_{FB}^{t\bar t}$ 
in this case.  However, if the couplings of the LH quarks in  Eq. \eqref{eq:SILag} are defined in the LH down quark mass eigenstate basis,
the scalars $S^-_{3\, i}  $ would only couple to 
$b_L$ in the ${\rm H_F }$ symmetric limit, as in Eq. \ref{SIminb}, and there would  be no obstruction to $A_{FB}^{t\bar t} $ enhancement from $B \to K \pi$.

\section{LHC Signals}\label{LHC}
We next turn to 
signals at the LHC from a flavor symmetric sector. 
The NP fields couple to light quark bilinears so that generically deviations
in the phenomenology of dijet observables at LHC and the Tevatron are expected. No statistically
significant deviation in these observables has been measured to date, thus placing bounds 
on the couplings of NP fields to light quark bilinears. For a set of sample models we derived the bounds 
in Section \ref{tevatron-constraints} 
showing that $\mathcal{O}(1)$ couplings of these fields to quark bilinears are generally still allowed.
Deviations from the QCD prediction
can arise from an $s$-channel resonance effect, a $t$-channel driven rise in the dijet invariant mass spectra at large
$M_{jj}$, or a combination of both signatures. 

\begin{figure}
\includegraphics[width=0.5\textwidth]{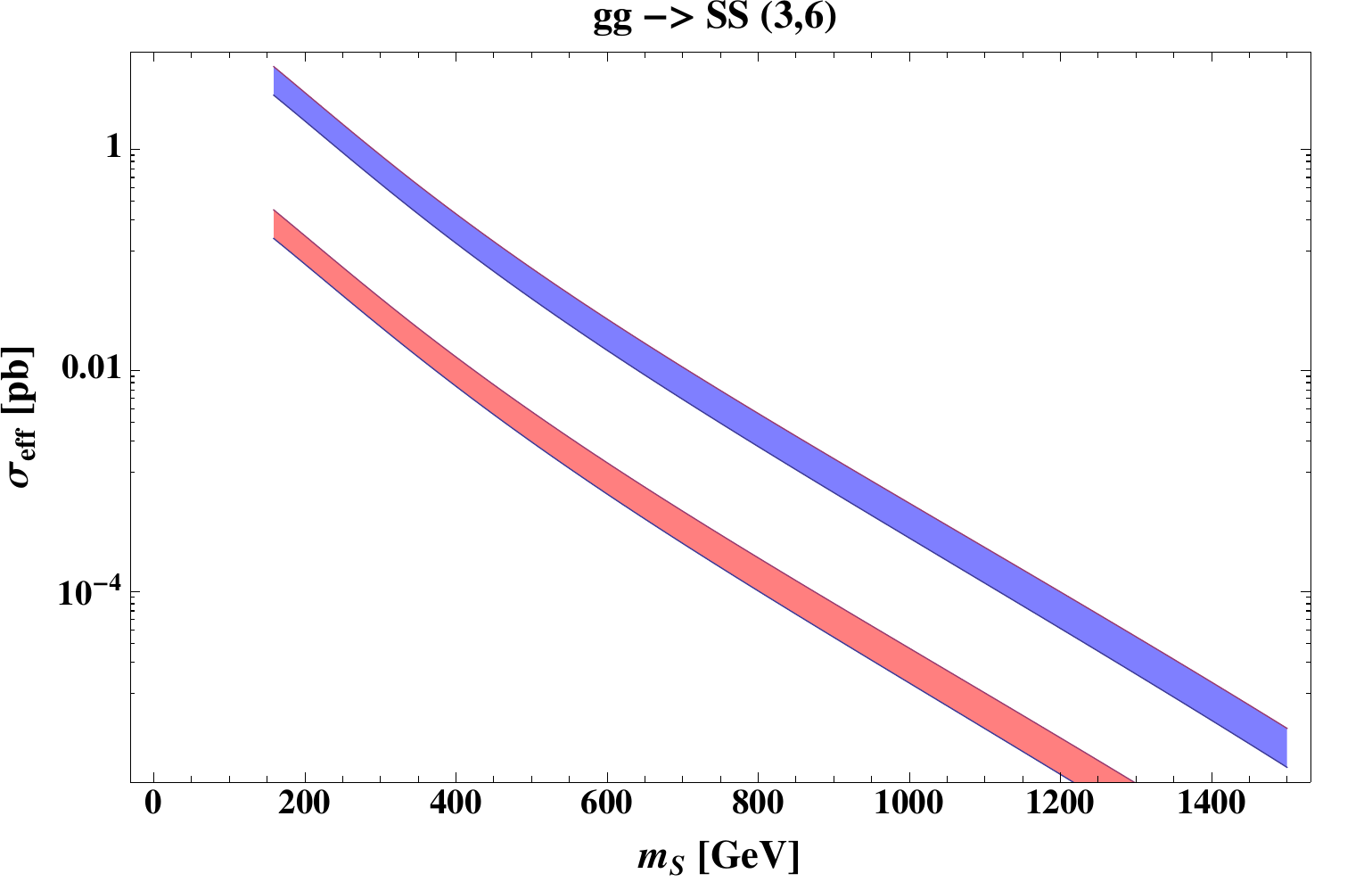}%
\includegraphics[width=0.5\textwidth]{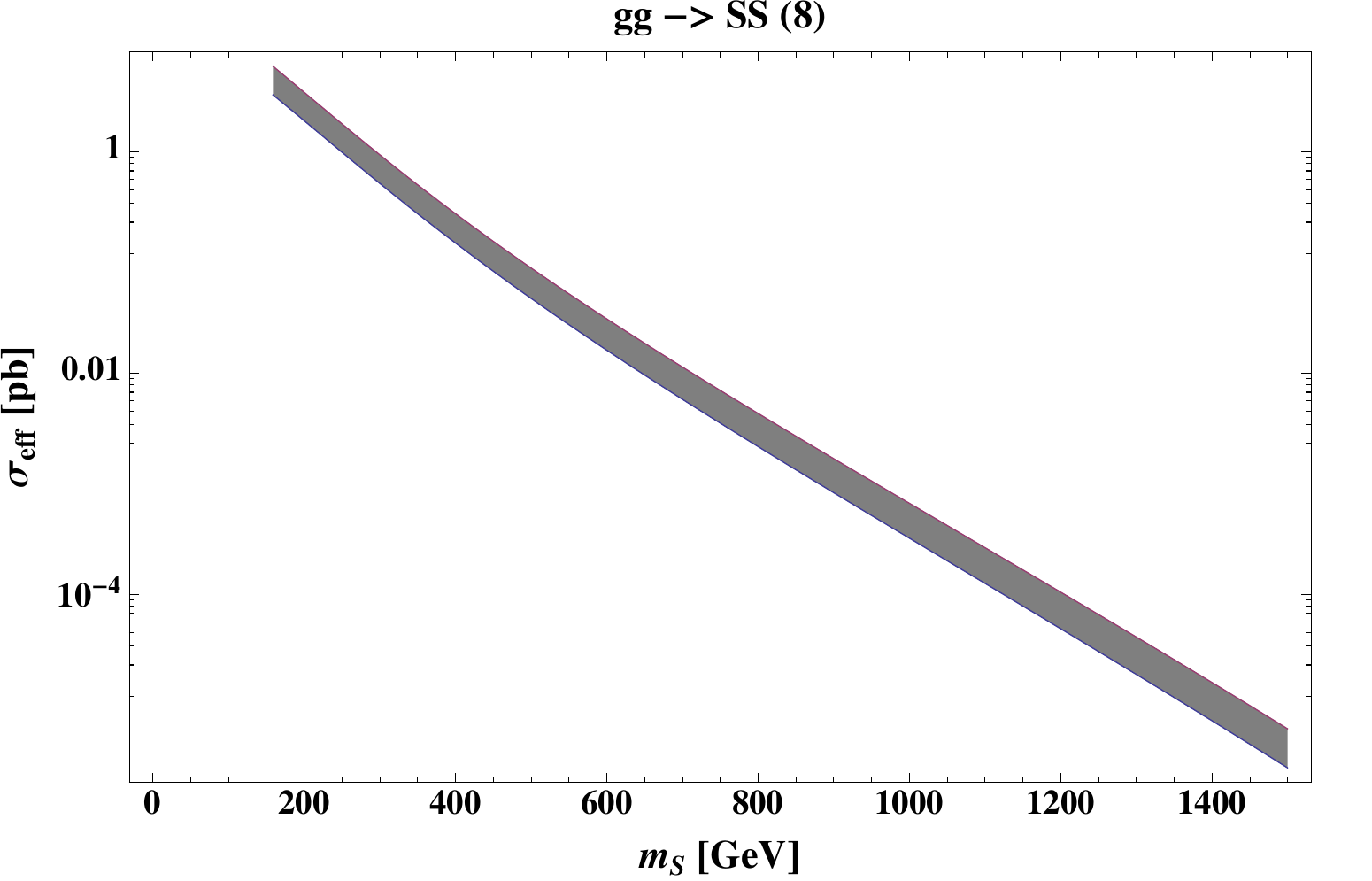}
\caption{Production cross sections of $g \, g \rightarrow S\,S$ for color triplets (red left), sextets (blue left) and octet (grey right) scalar field S. We show $\sigma_{eff} = \sigma/{\rm dim}[F]$ for a single flavor.
We use MSTW208 pdfs and the PDF error bands shown are defined by taking $\mu = [2 (2 m_S), (2 m_S)/2 ]$. Details of the calculation are in the Appendix. 
} \label{Fig-xSS}
\end{figure}

A prototypical signal of colored $\rm G_F$ symmetric fields is pair production, $g \, g \rightarrow V\, V$ and  $g \, g \rightarrow S\,S$.
This flavor diagonal scattering comes about due to the kinetic terms of the new scalar and vector fields, and the rate is driven by a 
coupling fixed by gauge invariance. 
The $gg\to SS$ production cross section in the  $\rm G_F$ symmetric limit as a function of scalar meson mass is shown in Fig. \ref{Fig-xSS}. 
As this signal requires the pair production of the NP states, there is a significant fall in the cross section
with the mass of the scalars.  
A simple approximation to the fall in the pair production cross section
as a function of $m_S$ is given by $\sigma_{\rm eff} \sim (1,16,20)  {\rm ~pb} \times  \exp(-m_s/95 \, {\rm GeV})$ for the color representations $({\bf 3,6,8})$.
This approximates the full result within the PDF uncertainty band over the mass region $0.15 -1.5 \, {\rm TeV}$.
We expect the behavior of the $g \, g \rightarrow V\, V$ cross sections to be of a similar order of magnitude.
Note that the non renormalizable nature of the effective vector field will lead to corrections to the cross section that grow with $\sqrt{s}/m_V$.
These effects, however, are accompanied with the fall in the cross section with $\sqrt{s}$ due to PDF suppression.

Dedicated four quark jet studies where one searches for pairs of equal mass resonances  with no missing energy can place strong bounds on models of this form
in principle. The cross section for 4-jets at the 7 TeV LHC with all four jets having at least $p_T$ of 300 GeV (500 GeV) is $1.6\cdot 10^{-3}$ pb ($1.6\cdot 10^{-4}$ pb), obtained using {\tt ALPGEN}. Compared to the pair production cross section the QCD background is thus of moderate size. The condition that the four jets reconstruct resonances of approximately equal mass can aid in reducing the QCD four jet background. 
We leave a detailed study of the discovery prospects for a future study. 

There are a number of other LHC signal features that are common to $\rm G_F$ symmetric models.
When a signal requires flavor violation, such as the production of same sign tops, the rate
is suppressed by the required CKM suppression and Yukawas to generate the tops from the initial state - as in the SM.
The $t$-channel models aiming to explain the $A_{FB}^{t \, \bar{t}}$ anomaly without such a flavor symmetry 
can be strongly constrained by bounds on same sign top signal
at LHC \cite{Chatrchyan:2011dk}. The $\rm G_F$ symmetric models we have studied are essentially unbounded by this measurement.

Other signatures that are linked to explanations of the $A_{FB}^{t \, \bar{t}}$ anomaly do offer the prospect of constraining
these models, such as a resonance decaying to $t+j$. 
This can be seen in $t\bar t+j$ production, and is a generic feature of $t$-channel models that can give enhanced $A_{FB}^{t\bar t}$ 
not just $\rm G_F$ symmetric models. Prospects for early LHC discovery were explored in Ref. \cite{Gresham:2011dg,Gresham:2011fx}.
It is also feasible that for $t$-channel dominated models, the enhanced very forward $t\bar t$ cross section may be observable at LHCb \cite{Kagan:2011yx}.
We have shown that generically the new physics effect on the $t \, \bar{t}$ invariant mass spectra in the models we have considered are measurable in the near future.
 The cuts can be optimized at LHC to increase the prospects of measuring an excess of $t \, \bar{t}$ events due to new physics of this form
while suppressing the SM background - see \cite{Arguin:2011xm} for a recent study.

\section{Conclusions}\label{conc}
In this paper, we have explored the collider physics of new physics
sectors symmetric under the flavor group  $\rm G_F= U(3)_{U_R}\times U(3)_{D_R}\times U(3)_{Q_L}$ and its subgroup $\rm H_F= U(2)_{U_R}\times U(2)_{D_R}\times U(2)_{Q_L}\times U(1)_3$. These are the flavor symmetry groups of the SM quark sector in the absence of all yukawa couplings, and in the limit that only the top and bottom yukawa couplings are kept, respectively.  
Flavor symmetric fields cataloged in Tables I and II can have EW scale masses and $\mathcal{O}(1)$ couplings to quark bi-linears.
This is because the representations of these fields under the SM gauge group and the flavor group 
are motivated to address a generic experimental conflict  -- the flavor problem of NP.  Moreover, the flavor symmetries eliminate NP contributions to single top and same sign top production, which otherwise would be tightly constrained.
For fields that transform under $\rm G_F$ and have ${\mathcal O}(1)$ couplings to light quarks we find that a lower bound on their masses of roughly $150 \, {\rm GeV}$ can be obtained from the absence of anomalous multi-jet events at LEP2. 
More detailed studies of EWPD, however, have the potential to improve these constraints.

Upper bounds on the couplings of $\rm G_F$ symmetric fields to light quarks can be set through studies of
dijet resonance searches and angular distributions at the Tevatron and LHC. Based on the partonic LO study of dijet production we have determined these bounds in 
a number of sample models. Our results indicate that the LHC 
data sets are starting to constrain
the allowed couplings to be $\lesssim 0.1$ in some of these models, in particular if they lead to resonant $uu$ quark pair production as in models with color sextet scalars (e.g., model ${\rm S_{VI}}$ in Table \ref{int-scalar}).

FCNC bounds on flavor symmetric models depend on how $G_F$ is broken. In the $\rm H_F$ symmetric limit there are no NP contributions to meson mixing.  We have analyzed NP contributions to meson mixing assuming MFV breaking of the $\rm H_F$ symmetry. The contributions to meson mixing are of second order in the product of MFV  flavor breaking coefficients and vector or scalar couplings to quarks. We find that for models that contribute to $B_d$ and $B_s$ mixing with the same product of CKM elements as the SM box diagrams (and no additional light quark yukawa coupling factors) these coefficients have to be $\lesssim 0.1$ for vector or scalar masses of 1 TeV (these are
vector models ${\rm III_{s,o},
IV_{s,o}, VIII_{s,o}, IX_{s,o}}$ and the scalar model $\rm S_{XIV}$). The bounds on couplings for the remaining models are much looser.

The fields we have introduced and studied are of particular interest in the LHC discovery era. As an application we have examined the potential impact of such field content on two current experimental anomalies,
the top quark forward-backward asymmetry, $A_{FB}^{t \, \bar{t}}$, and the $B_{s,d}$ dimuon charge asymmetry. Many of the flavor symmetric models can yield large $A_{FB}^{t \, \bar{t}}$, while maintaining agreement with the total top pair production cross section. However, simultaneous agreement with the  $m_{t\bar t}$ differential cross section is more difficult to achieve. Generically, scalar models have difficulties accommodating all of the top quark data. The scalar models can either give $u$-channel contributions to $t\bar t$ production (models ${\rm S_{V}-S_{XIV}}$) or $t$-channel contributions (models ${\rm S_{I}-S_{IV}}$). The $u$-channel models exhibit significant tension with the differential spectrum. Of the $t$-channel models we have checked that model $\rm S_{I}$ shows good agreement with the differential spectrum, for masses of order 100 GeV.
While it has difficulty reproducing the CDF $m_{t\bar t}$ dependent $A_{FB}$ at the one sigma level, it should be noted that the recent D\O\ $A_{FB}^{t\bar t}$ measurement appears to exhibit a smaller $m_{t\bar t}$ dependence.

Vector models can contribute to $t\bar t$ production in the $u$-channel, $s$-channel, $t$-channel, or the $s$-channel and $t$-channel simultaneously. Phenomenologically preferred $t$-channel dominance can be a feature of the flavor structure (models $\rm VII_{s,o}$), or can be attained either through some flavor breaking or additional decay channels to broaden the $s$-channel decay widths. This leads to good agreement with all the top quark data, with a preference for lighter vector meson masses (see  models ${\rm VI_{s,o}}$ in Fig. \ref{fig:FV-VI}, or Fig. \ref{fig:vecXacc} and Ref. \cite{Grinstein:2011yv}, and models ${\rm VII_{s,o}}$ in Fig. \ref{FigVIIso} ). Note that flavor breaking or  broadening of the widths (via non-$q\bar q$ final states) also  insures consistency with resonance searches in the dijet data.

When studying the potential of new physics explanations of the $A_{FB}^{t \, \bar{t}}$ anomaly we have 
determined model independent acceptance corrections. These are appropriate for comparing against 
Tevatron 
data that can be applied to any model to good approximation at the partonic level- these results are given in Section \ref{Acceptance}.

Assuming MFV for $\rm H_F$ symmetry breaking, we have systematically examined the effects of scalar and vector fields on the dimuon charge asymmetry anomaly. Flavor symmetric models that can easily give relatively larger contributions to $B_s$ mixing than to $B_d$ mixing are preferred. Several flavor symmetric models can
accommodate this pattern with tree level exchanges: vector models $\rm I_{s,o}, V_{s,o}$ and scalar models $\rm S_{IV}, S_{VIII}$, with ${\mathcal O}(1)$ MFV flavor breaking coefficients and NP masses of no more than a few 100 GeV,  and models $\rm X_{\bar{3},6}$,  $\rm S_{III}, $ for  masses around 1 TeV. None of these, however, can also explain the large $A_{FB}^{t\bar t}$, while simultaneously obeying all constraints. The dimuon charge asymmetry anomaly can also be explained by the experimentally less favored possibility - equal contributions to $B_s$ and $B_d$ mixing amplitudes (once normalized to the SM contributions). There are a number of models that can contribute  to $t\bar t$ production as well as $B_{s,d}$ meson mixing. Of these, vector models $\rm VII_{s,o}$ and scalar model $\rm S_I$ are preferred, leading to both very small deviations in the differential $t\bar t$ spectrum and enhancement of $A_{FB}^{t\bar t}$. The contributions to $B_{s,d}$ arise at 1-loop and can be of the right size.

The models we have introduced and motivated can be discovered at the LHC through a number of processes: $t\bar t+j$ with $t+j$ forming a vector or scalar resonance peak;  deviations in the $m_{t\bar t}$  differential cross section; pair production via $gg \rightarrow SS,VV$, where two resonance peaks are reconstructed in four jet final state; and deviations in dijet observables.

In conclusion, flavor symmetric fields by virtue of their flavor safe properties can have masses near the electroweak scale. They are therefore candidates for discovery at the LHC, as the lowest lying states of more complete models, which may or may not be directly linked to electroweak symmetry breaking.  Flavor symmetric sectors have a rich, predictive, and mostly unexplored phenomenology.
It is interesting to note that the list of fields that are $\rm G_F$ symmetric and can directly couple to quark bilinears is relatively short.
Systematically exploring the phenomenology of flavor symmetric sectors is therefore feasible.

\section*{Acknowledgments}
 The work of 
B.G.\ was supported in part by the U.S.\ Department of Energy under 
contract DE-FG03-97ER40546. A. K. is supported by DOE grant 
FG02-84-ER40153.  AK thanks the Weizmann Institute Phenomenology group for their hospitality while this work was being carried out. MT thanks Mark Wise, Maxim Pospelov and Jonathan Arnold for past collaboration
related to some of the material discussed in this paper and the Triumf theory group for hospitality
while this paper was being finalized. We thank K. Blum, O. Gedalia, Y. Hochberg, J. Kamenik, I.-W. Kim, S. J. Lee, Z. Ligeti, Y. Nir, G.Perez, J. Shu,   and Y. Soreq for discussions regarding NP models in $t\bar t$ production.  
We use this opportunity to congratulate Jernej Kamenik on the birth of his first son Tian.

\pagebreak

\appendix
\section{Flavor Symmetric Lagrangians}\label{AppendixA}
In this appendix we collect the Lagrangians of the vector fields in nontrivial ${\rm G_F}$ representations. The form of the scalar Lagrangians is given in \cite{Arnold:2009ay} and we use that notation in the main part of the paper
and in the Appendix for ease of comparison (for concreteness we also show explicit form of the relevant scalar Lagrangians at the end of this Appendix). For parameterizing the breaking of ${\rm G_F}$ we use the language of MFV, where the breaking is due to SM Yukawas only. For the unitary transformations that diagonalize the quark mass matrices we use the following notation
 \begin{equation}
 \begin{split}\label{transform}
 {\cal U}(u,R)^{\dagger} &\, Y_U \, {\cal U}(u,L)={\sqrt{2}\mathcal{M}_u}/{v}\equiv Y_U^{\rm diag}\simeq {\rm diag}(0,0,y_t),\\
{\cal U}(d,R)^{\dagger} &\, Y_D \, {\cal U}(d,L)={\sqrt{2}\mathcal{M}_d}/{v}\equiv Y_D^{\rm diag}\simeq {\rm diag}(0,0,y_b).
 \end{split}
 \end{equation}
These relate the flavor ($q_i$) and mass eigenstates ($q_i'$)
 \beq\label{rot}
 \begin{split}
u_L=&{\cal U}(u,L) u'_L,~~~~~u_R={\cal U}(u,R) u'_R, \nonumber \\
d_L=&{\cal U}(d,L) d'_L,~~~~~d_R={\cal U}(d,R) d'_R.
\end{split}
 \eeq
In  \eqref{transform} $\mathcal{M}_u$ and $\mathcal{M}_d$ are the diagonal up-type and down-type quark mass matrices and $v\simeq 246$ GeV is the SM Higgs vev. 
 The Cabbibo-Kobayashi-Maskawa (CKM) matrix $ V_{\rm CKM}$ is then given by
 \begin{equation}
 V_{\rm CKM}={\cal U}(u,L)^{\dagger} \, {\cal U}(d,L).
 \end{equation}
 
 Two basis choices that we will frequently use are the ``up'' and
 ``down'' quark mass bases. In the ``up'' quark basis
$\yudyu \to (Y_U^{\rm diag})^2, 
\yddyd \to V_{\rm CKM} (Y_D^{\rm diag})^2 V_{\rm CKM}^\dagger$,
so that down-quarks are rotated from the mass eigenstates to keep interactions with the $W$ bosons diagonal. Similarly in the down quark mass basis, the up quarks are rotated by $V_{\rm CKM}^\dagger$, so that 
$\yudyu \to V_{\rm CKM}^\dagger (Y_U^{\rm diag})^2 V_{\rm CKM}, 
\yddyd \to (Y_D^{\rm diag})^2$.

\subsection{Vector fields $\rm I-IV$}
The vector fields $\rm I_{s,o}-IV_{s,o}$ are flavor singlets, but can be either singlets or in adjoint representations of color and/or weak $SU(2)_L$. The interaction Lagrangian with quarks then has the same forms as the interactions with corresponding gauge fields would have. For instance, for ${\rm I_{s,o}}$, it is
\beq
{\cal L}_{I}=\eta \bar d_R \slashed V d_R,
\eeq
where $V_\mu=V_\mu^A T^A$ for $\rm I_o$ fields, with $T^A$ the Gell-Mann matrices normalized to $\Tr(T^A T^B)=\delta^{AB}/2$. The interaction Lagrangians for ${\rm II_{s,o}, III_{s,o}}$ are obtained from the above through replacements, $d_R\to u_R$ and $d_R\to Q_L$, respectively. In the interaction Lagrangians $\rm IV_{s,o}$ the short-hand notation for the sum of vector field multiplets is $V_\mu=V_\mu^{a} \sigma^a/2$ for ${\rm IV_s}$ model and  $V_\mu=V_\mu^{A,a} T^A \sigma^a/2$  for ${\rm IV_o}$ model, with $\sigma^a$ the Pauli matrices. 

\subsection{Fields $\rm V_{s,o}$}
These fields are color singlets (octets) and ${\rm SU(3)_D}$ octets. The interaction and mass terms in the Lagrangian are 
similar to the case of the ${\rm SU(3)_U}$ octet fields $\rm VI_{s,o}$ given in Eqs. \eqref{Vmus}-\eqref{Vimass} and Eqs. \eqref{interactionsVI}-\eqref{VIsEq}, but with the obvious $u_R\to d_R$ and $y_t\to y_b$, $Y_U\to Y_D$ replacements. 
In the SM, $y_b\sim 0.02$ so that all the
vectors in the multiplet are degenerate to a good
approximation. The degeneracy can be lifted in theories with more than one Higgs where $y_b$ can be $O(1)$, as occurs for example in the large $\tan \beta$
limit of the MSSM. 

\subsection{Fields $\rm VI_{s,o}$}
These real fields are $\bf 1, 8$ under color and $\bf 8$ under
flavor. We denote them by $V_\mu^B$ and $V_\mu^{A,B}$, with color (flavor) labels $A(B)=1,\dots,8$. The quark-vector interaction Lagrangian and the mass terms were already given in Section \ref{section:MFVrepresentations} using the short-hand notation
 $V^s_\mu= T^B V_\mu^B ,  V^o_\mu={\cal T}^A T^B V_\mu^{A,B}$. Here we write out the terms keeping all the indices explicit. The ${\rm G_F}$ symmetric quark-vector interactions \eqref{symmetricV} are 
\begin{equation}\label{interactionsVI}
 \mathcal{L}_{\rm VI_s} =  \eta_1^s \,  \, (\bar{u}_R)^{\alpha,i} \,
 \,(T^{B})_i^j \slashed V^{B}(u_R)_{\alpha, j} ,
\qquad
\mathcal{L}_{\rm VI_o} =  \eta_1^o \, (\bar{u}_R)^{\alpha, i} \, (\mathcal{T}^A)_\alpha^\beta \, (T^{B})_i^j \, \slashed  V^{A,B} (u_R)_{\beta, j},
\end{equation}
where  $\alpha, \beta, \gamma ...$ ($i,j,k ...$) are color (flavor) indices. The couplings $\eta_1^{s,o}$ are real because the interaction terms are hermitian. The  ${\rm G_F} \rightarrow {\rm H_F}$ breaking interaction terms  due to  $\yuyud $ insertions \eqref{breakVI} have the following  explicit form in the mass eigenstate basis (keeping only the terms proportional to $y_t$)
\begin{align}
\Delta \mathcal{L}_{\rm VI_o}  &=    y_t^2  \left[(\eta_2^o)^\star  (\bar{u}'_R)^{\alpha} \, \slashed \tilde{V}_\mu^{A,B} \, (\mathcal{T}^A)_\alpha^\beta \, (T^{B})_3^i (u'_R)_{\beta,i} + h.c.\right] -\frac{y_t^4 \,\eta_3^o }{2\sqrt3}(\bar{u}'_R)^{\alpha, i}\, \slashed  \tilde{V}_\mu^{A,8} \, (\mathcal{T}^A)_\alpha^\beta (u'_R)_{\beta}, 
\label{VIoEq}
\\
\Delta \mathcal{L}_{\rm VI_s}  &= y_t^2 \left[ (\eta_2^s)^\star  \, (\bar{u}'_R)^{\alpha} \, \slashed \tilde{V}_\mu^{B}   \, (T^{B})_3^i (u'_R)_{\alpha, i} + h.c.\right] - \frac{ y_t^4 \,\eta_3^s }{\sqrt3}
(\bar{u}'_R)^{\alpha, i} \, \slashed \tilde{V}_\mu^{8}   \, (T^{8})_3^3 (u'_R)_{\alpha}. \label{VIsEq}
\end{align}

\subsection{Fields $\rm VII_{s,o}$}
These fields are in the $(\bar 3, 3,1)$ representation of flavor. As for fields V, VI we use the short-hand notation
\begin{equation}
V^s_\mu=T^B V_\mu^B , \qquad V^o_\mu={\cal T}^A T^B V_\mu^{A,B},
\end{equation}
with ${\cal T}^A$ the color  Gell-Mann matrices.  The flavor matrices $T^B$ now also include an identity matrix, so that for  $B=1,\dots,8$,  $T^B$ are the Gell-Mann matrices, and 
$T^9=\, \mathds{1}/\sqrt{6}$. Note that $V_\mu^\dagger\ne V_\mu^{\phantom{\dagger}} $, because the $V_\mu$ fields are in the $(\bar 3, 3,1)$ representation of $\rm G_F$. The transformation to the mass eigenstate basis is
\beq
V_\mu^a= {\cal U}(d,R) \tilde{V}_\mu^a \,{\cal U}(u,R)^{\dagger},\qquad a=o,s,
\eeq
explicitly showing that $V_\mu^{o,s}$ are not hermitian (note also that $V_\mu^B$, $V_\mu^{A,B}$ carry nonzero hyper-charge). The tree level quark coupling Lagrangian terms are (no summation over $a=o,s$)
\beq
\mathcal{L}_{\rm VII_a} =  \eta_1^a  \bar{d}_R \slashed V_\mu^a u_R + h.c.,
\eeq
with $\eta_1^{o,s}$ complex in general. Explicitly the two terms are
\bea
\mathcal{L}_{\rm VII_o} &=&  \eta_1^o \, (V_\mu^{A,B} T^B)^{i}_j \, (\bar{d}_R)^{\alpha, j} \, \gamma^\mu \, (\mathcal{T}^A)_\alpha^\beta \,  (u_R)_{\beta, i}, + {\rm h.c.}, \\
 \mathcal{L}_{\rm VII_s} &=&  \eta_1^s \, (V_\mu^B T^B)^{i}_j \,(\bar{d}_R)^{\alpha, j} \, \gamma^\mu \,  (u_R)_{\alpha, i} + {\rm h.c.} .
\eea
The flavor breaking introduces corrections to these interaction terms from insertions of the $\yuyud$ and $\ydydd$. Keeping only the terms proportional to $y_t^2$ for $\rm G_F \rightarrow H_F$, the leading contribution is given by
$\eta_2^a  \bar{d}_R \slashed{V}_\mu^a Y_U^{\phantom{\dagger}}  Y_U^\dagger u_R$, which gives in the mass eigenstate basis the following interactions 
\bea
\Delta \mathcal{L}_{\rm VII_o}  &=&   y_t^2 \left[\eta_2  (\bar{d'}_R)^{\alpha, i} \, \gamma^\mu \, (\mathcal{T}^A)_\alpha^\beta \,  (t'_R)_{\beta},\right] \, (\tilde{V}_\mu^{A,B} T^B)^{3}_i + h.c., \\
\Delta \mathcal{L}_{\rm VII_s}  &=&  y_t^2 \, \left[ \eta_2  (\bar{d'}_R)^{\alpha, i} \, \gamma^\mu \,  (t'_R)_{\alpha}  \right] \, (\tilde{V}_\mu^B T^B)^{3}_i + h.c.
\eea
The mass terms are 
\bea\label{mass1-VII}
\mathcal{L}_{\rm VII_{a}}  = 2(1+\delta_{a,o})\left\{m_{V_a}^2 \, {\rm Tr} \left[ \tilde{V}_\mu^{a} \,  \tilde{V}^{\mu a\dagger} \right] + \lambda \, (H^\dagger \, H) \, {\rm Tr} \left[ \tilde{V}_\mu^a \,  \tilde{V}^{\mu a\dagger} \right]\right\}, \qquad a=o,s,
\eea
while the top quark Yukawa splits the mass spectrum through corrections to the mass Lagrangian of the form 
$2(1+\delta_{a,o}) \, \zeta_1 Tr[ \tilde{V}_\mu^{a} Y_U^{\phantom{\dagger}}  Y_U^\dagger \,  \tilde{V}^{\mu a\dagger}]$. The mass of the field $V_\mu^8$ gets
split from the rest by the term $2(1+\delta_{a,o}) \, \zeta_2 Tr[ \tilde{V}_\mu^{a} Y_U^{\phantom{\dagger}}  Y_U^\dagger \,  \tilde{V}^{\mu a\dagger} Y_D^{\phantom{\dagger}}  Y_D^\dagger]$.
These splittings result in the following mass spectrum
\bea
m_1^2 =  m_2^2  = m_3^2  &=& m_V^2 \left(1 +  \frac{\lambda}{2} \frac{v^2}{m_V^2} \right), \\
m_4^2 =  m_5^2  = m_6^2 = m_7^2  &=& m_V^2\left(1 +  \frac{\lambda}{2}\frac{ v^2}{m_V^2} + \frac{{\zeta}_1}{2} \, y_t^2 \right), \\
m_8^2 &=& m_V^2 \left(1 +  \frac{\lambda}{2} \frac{v^2}{m_V^2}+ \frac{2{\zeta}_1}{3} y_t^2 + \frac{2{\zeta}_2}{3} \, y_b^2 \, y_t^2\right),\\
m_9^2 &=& m_V^2 \left(1 +  \frac{\lambda}{2} \frac{v^2}{m_V^2}+ \frac{{\zeta}_1}{3} y_t^2 + \frac{{\zeta}_2}{3} \, y_b^2 \, y_t^2\right), 
\eea
where we neglect terms suppressed by off-diagonal CKM elements or first two generation Yukawas. 

\subsection{Fields $\rm VIII_{s,o}$}
The scalar matrix of fields $V^s_\mu= T^B V_\mu^B$, and the octet matrix of fields $V^o_\mu={\cal T}^A T^B V_\mu^{A,B}$, now transform as a bi-fundamental of $\rm U(3)_Q$, i.e. $V_\mu^{o,s} \to V_Q   {V}_\mu^{o,s} V_Q^{\dagger}$ for a transformation that takes the left-handed quark fields to $Q_L \to V_Q Q_L$ (suppressing the flavor indices). Note that the fields $V_\mu^{o,s}$ are hermitian, $V_\mu^{o,s\dagger}=V_\mu^{o,s}$. The interaction with quarks is then given by
\bea
\mathcal{L}_{\rm VIII_a} &=&  \eta_1^a \bar{Q}_L \slashed V_\mu^a Q_L, 
\eea
or writing out the color and flavor indices explicitly
\bea
\mathcal{L}_{\rm VIII_o} &=&  \eta_1^o \, V_\mu^{A,B} \, (\bar{Q}_L)^{\alpha, i} \, \gamma^\mu \, (\mathcal{T}^A)_\alpha^\beta \, (T^{B})_i^j (Q_L)_{\beta, j}, \\
 \mathcal{L}_{\rm VIII_s} &=&  \eta_1^s \, V_\mu^{B} \, (\bar{Q}_L)^{\alpha, i} \, \gamma^\mu  \, (T^{B})_i^j (Q_L)_{\alpha, j}.
\eea
The flavor symmetric mass Lagrangian for the vectors is
\bea\label{mass1}
\mathcal{L}_{\rm VIII_{o,s}}^{\rm mass}  = (1+\delta_{a,o})\left\{ m_V^2 \, {\rm Tr} \left[ \tilde{V}_\mu \,  \tilde{V}^{\mu} \right] + \lambda \, (H^\dagger \, H) \, {\rm Tr} \left[ \tilde{V}_\mu \,  \tilde{V}^{\mu} \right]\right\},
\eea
where the color and flavor indices are suppressed. The masses are split by the presence of the SM Yukawas breaking the flavor group. In contrast to cases V and VII here both insertions of $\yudyu$ and $\yddyd$ are possible. The flavor breaking terms are 
\bea\label{mass1VIII}
\Delta \mathcal{L}_{\rm VIII_{o,s}}^{\rm mass}/m_V^2= \zeta_1 {\rm Tr} \left[ \tilde{V}_\mu \, \yudyu \tilde{V}^{ \mu} \right]  +\zeta_1' {\rm Tr} \left[ \tilde{V}_\mu \, \yddyd \tilde{V}^{ \mu} \right]+ \zeta_2 {\rm Tr} \left[ \yudyu \tilde{V}_\mu \, \yudyu \tilde{V}^{ \mu} \right] +\cdots,
\eea
with $\zeta_i$ all real (complex couplings are possible at higher orders in Yukawa insertions). In general, the masses for $V_\mu$ are not diagonalized by the ${\cal U}(u,L)$ or ${\cal U}(d,L)$ unitary transformation, but by a unitary transformation that differs by $\sim V_{\rm CKM}$ from the two. 
For simplicity, we work in the limit where the contributions from $y_b$ are much smaller than from $y_t$. In this limit the vector field mass matrix is diagonalized using  ${\cal U}(u,L)$, giving a spectrum
\bea
m_1^2 =  m_2^2  = m_3^2  &=& m_V^2 \left(1 +  \frac{\lambda}{2} \frac{v^2}{m_V^2} \right), \\
m_4^2 =  m_5^2  = m_6^2 = m_7^2  &=& m_V^2\left(1 +  \frac{\lambda}{2}\frac{ v^2}{m_V^2} + \frac{{\zeta}_1}{2} \, y_t^2 \right), \\
m_8^2 &=& m_V^2 \left(1 +  \frac{\lambda}{2} \frac{v^2}{m_V^2}+ \frac{{\zeta}_1}{3} y_t^2 + \frac{{\zeta}_2}{3}  y_t^4\right). 
\eea
The mass degeneracy is lifted at $y_b^2$ order from a contribution of
the form $y_b^2 (V_{\rm CKM}^{\phantom{\dagger}})_{i3}(V_{\rm
  CKM}^\dagger)_{j3}$ ({\it i.e.,} from the term proportional to $\zeta_1'$ in \eqref{mass1VIII}).

We are now ready to write down the flavor breaking couplings between the vector and the quark fields. These are given by 
\beq
\begin{split}
\Delta \mathcal{L}_{\rm VIII_{a}}&=  \big[\eta_2^a \bar{Q}_L Y_U^\dagger Y_U^{\phantom{\dagger}}  \slashed V_\mu^a Q_L+h.c.\big]+ \big[\eta_3^a \bar{Q}_L Y_D^\dagger Y_D^{\phantom{\dagger}}  \slashed V_\mu^a Q_L+h.c.\big] \\
&+ \big[\eta_4^a \bar{Q}_L Y_U^\dagger Y_U^{\phantom{\dagger}}  \slashed V_\mu^a  Y_D^\dagger Y_D^{\phantom{\dagger}} Q_L+h.c.\big]+\cdots, 
\end{split}
\eeq
where we only display one of the possible terms with four insertions of Yukawa matrices. Note that $\eta_{2,3,4}$ can all be complex. We first display the couplings to the up-quarks in the ``up'' basis 
\beq
\begin{split}
\Delta \mathcal{L}_{\rm VIII_o}  &=   y_t^2 \eta_2^o \left[ (\bar{t}'_L)^{\alpha}  (\mathcal{T}^A)_\alpha^\beta(T^{B})_3^i \,\, \slashed \tilde{V}_\mu^{A,B} (u'_L)_{i \beta}\right] \\
& + y_b^2 \eta_3^o  (V_{\rm CKM})^3_i \, (V_{\rm CKM}^\ast)_3^j  \left[  (\bar{u}'_L)^{i\alpha} (\mathcal{T}^A)_\alpha^\beta (T^{B})_j^k  \slashed  \, \tilde{V}_\mu^{A,B} (u'_L)_{k \beta}  \right] \\
 &+ y_t^2 y_b^2 \eta_4^o  \, (V_{\rm CKM})_i^3 (V_{\rm CKM}^\ast)_3^j \left[ (\bar{t}'_L)^{\alpha} (\mathcal{T}^A)_\alpha^\beta \, (T^{B})_3^i \slashed  \tilde{V}_\mu^{A,B} (u'_L)_{j\beta}  \right] . 
 \end{split}
 \eeq  
The couplings to the bottom quarks we display in the ``down'' basis
\beq
\begin{split}
\Delta \mathcal{L}_{\rm VIII_o}  &=   y_t^2 \eta_2^o  (V_{\rm CKM}^\ast)^3_i \, (V_{\rm CKM})_3^j  \left[  (\bar{d}'_L)^{i\alpha} (\mathcal{T}^A)_\alpha^\beta (T^{B})_j^k  \slashed  \, {V}_\mu^{A,B} (d'_L)_{k \beta}  \right] \\
& +y_b^2 \eta_3^o \left[ (\bar{b}'_L)^{\alpha}  (\mathcal{T}^A)_\alpha^\beta(T^{B})_3^i \,\, \slashed {V}_\mu^{A,B} (d'_L)_{i \beta}\right]  \\
 &+ y_t^2 y_b^2 \eta_4^o  \, (V_{\rm CKM}^\ast)_i^3 (V_{\rm CKM})_3^j \left[ (\bar{d}'_L)^{i \alpha} (\mathcal{T}^A)_\alpha^\beta \, (T^{B})_j^3 \slashed  {V}_\mu^{A,B} (b'_L)_{\beta}  \right] . 
 \end{split}
 \eeq 
For each of these two cases, the vector fields have to be put in the ``up'' or ``down'' basis respectively. In the limit where we neglect $y_b$ insertions, the ``up'' basis coincides with the mass basis for the vector fields, while the vector fields
are not mass diagonal in the ``down'' basis. The couplings for the singlet case are obtained from the above expressions by replacing $(\mathcal{T}^A)_\alpha^\beta\to \delta_\alpha^\beta$. 

\subsection{Fields $\rm IX_{s,o}$}
 The discussion for this case is very similar to the previous one.  The fields are given by $V_\mu^{B,i}$ for the color singlet and $V_\mu^{A,B,i}$ for the color octet case, where $i=1,2,3$ is the weak $\rm SU(2)_L$ index. All basis independent expressions apply also in this case, if the matrices of fields are defined as  $V^s_\mu= T^B \sigma^i V_\mu^{B,i}/2$, and the octet matrix of fields $V^o_\mu={\cal T}^A T^B \sigma^i V_\mu^{A,B,i}/4$, where $\sigma^i$ are the Pauli matrices. Phenomenologically the most important difference is that there are now charged currents.
 
\subsection{Fields $\rm X_{\bar{3},6}$}
These fields are weak doublets  and belong to the bi-fundamental
representation of the flavor group $(1,3,3)$. They are either
anti-triplets of color, in which case we have the fields $(V_\mu)^{r
  \gamma}_{i,j}$, or they are sextets of color, in which case the
fields are
$(V_\mu)^{r}_{i,j,\alpha,\beta}=(V_\mu)^{r}_{i,j,\beta,\alpha}$. Here
$r$ is the weak $\rm SU(2)_L$ index, $\alpha, \beta$ are the flavor
indices, while $i$ and $j$ are the indices of the $(1,3,1)$ and $(1,1,3)$ representations respectively. In Section \ref{MFVreps} we have written the Lagrangian in short-hand notation, while here we show the terms keeping the indices explicit. 
The tree level quark coupling Lagrangian terms are (suppressing the  $\rm SU(2)_L$ indices)
\bea
\mathcal{L}_{X_{\bar{3}}} &=&  \eta_1 \, \epsilon_{\alpha \beta \gamma} \, (V_\mu)_{i,j}^{\gamma} \, (\bar{d}_R)^{\alpha i} \, \gamma^\mu \, (Q^c_L)^{\beta j} + h.c., \label{X3:Eq}\\
\mathcal{L}_{X_{6}} &=&  \eta_1 \, (V_\mu)_{i,j, \alpha, \beta} \, (\bar{d}_R)^{\alpha i} \, \gamma^\mu \, (Q^c_L)^{\beta j}+ h.c.,\label{X6:Eq}
\eea
which in the mass eigenstate basis are
\bea
\mathcal{L}_{X_{\bar{3}}} &=&  \eta_1 \, \epsilon_{\alpha \beta \gamma} \, (\tilde{V}^1_\mu)_{i,j}^{\gamma} \, (\bar{d'}_R)^{\alpha i} \, \gamma^\mu \, (u'{}^{c}_L)^{\beta j}
+ \eta_1 \, \epsilon_{\alpha \beta \gamma} \,  (V_{\rm CKM})^k_j \, (\tilde{V}^2_\mu)_{i,k}^{\gamma} \, (\bar{d'}_R)^{\alpha i} \, \gamma^\mu \, (d'{}^{c}_L)^{\beta j} + h.c., \label{X3:EqBreak}\\
\mathcal{L}_{X_{6}} &=&  \eta_1 \, (V^1_\mu)_{i,j, \alpha, \beta} \, (\bar{d'}_R)^{\alpha i} \, \gamma^\mu \, (u'{}^c_L)^{\beta j}
+ \eta_1 \,  (V_{\rm CKM})^k_j \, (\tilde{V}^2_\mu)_{i,k, \alpha, \beta} \, (\bar{d'}_R)^{\alpha i} \, \gamma^\mu \, (d'{}^{c}_L)^{\beta j} + h.c.. \label{X6:EqBreak}
\eea
with $(\tilde V_\mu^1)$ and $(\tilde V_\mu^2)$ the charge and neutral vector fields, respectively. 

\subsection{Fields $\rm XI_{\bar{3},6}$}
The analysis is very similar to case X. The couplings to quarks, vector fields mass spectra and splittings can be obtained from the discussion  on case X with the  replacements  $d_R \leftrightarrow u_R$, $Y_D \leftrightarrow Y_U$. For instance, the quark coupling Lagrangian terms are (suppressing the $\rm SU(2)_L$ index)
\bea
\mathcal{L}_{\rm XI_{\bar{3}}} &=&  \eta_1 \, \epsilon_{\alpha \beta \gamma} \, (V_\mu)_{i,j}^{\gamma} \, (\bar{u}_R)^{\alpha i} \, \gamma^\mu \, (Q^c_L)^{\beta j} + h.c., \\
\mathcal{L}_{\rm XI_{6}} &=&  \eta_1 \, (V_\mu)_{i,j, \alpha, \beta} \, (\bar{u}_R)^{\alpha i} \, \gamma^\mu \, (Q^c_L)^{\beta j}+ h.c..
\eea

\subsection{Scalar Fields I, V, VI, IX, X}
For convenience, the interaction Lagrangians for the scalar models under discussion, in the $\rm G_F$ symmetric limit, are provided below:
\beq
{\cal L}_{\rm I} = \eta \,( \,S^0_{ij}  \bar{u}_{i\,L} u_{j \,R}   + S_{i j}^-  \bar{d}_{i\,L} u_{j\,R}  ) + h.c.  \,,   \label{eq:SILag:app} \eeq
\beq
 {\cal L}_{\rm V}=\eta  \,\epsilon^{\alpha \beta\gamma} \epsilon_{ijk} \,S_{\alpha}^i \, u_{R\beta}^j \,u_{R\gamma}^k  ~+~h.c.\,,~~~~~
{\cal L}_{\rm VI}=\eta \,  S^{a,b} (\hat T^a)_{ij} (\hat {\cal T}^b)^{\alpha \beta}\,u_{R\alpha}^i \,u_{R\beta}^j  ~+~h.c., 
\label{eq:SVandSVILag} 
 \eeq
\beq
{\cal L}_{\rm IX} = \eta \,  \epsilon^{\alpha \beta\gamma} S_{ij}^{\alpha  }  d_{R\beta}^i \,u_{R\gamma}^j ~+~h.c.\,,~~~~~
{\cal L}_{\rm X} = \eta \,   S_{ij}^a (\hat T^a)^{\alpha \beta }  \, d_{R\alpha}^i \,u_{R\beta}^j  ~+~h.c., 
\label{eq:SIXandSXLag} 
\eeq
where the $\eta$ are dimensionless flavor universal couplings, Latin (Greek) indices label flavor (color), and Lorentz spinor indices are suppressed. The generators $\hat T^a$ and $\hat {\cal T}^b$ of the symmetric sextet flavor and color representations are the symmetric $3\times 3$ matrices, normalized such that $\Tr(\hat T^a \hat T^b)=2\delta^{ab}$ and $a,b=1,\cdots, 6$.
The scalar fields appearing in  $ {\cal L}_{\rm I, V, VI, IX,X}$ are canonically normalized.  
 In ${\cal L}_{\rm I}$ we show explicitly the 
Yukawa couplings of the charged and neutral components of the scalar doublets.
The relevant NP cross section formulae for top quark pair production can be found in Appendix B.
The decay widths of the scalars are $\kappa \,\eta^2 \,m_S /16 \pi $,
where  $\kappa = 1,\,4,\, 2,\,1,\,1$ in models $\rm S_V$, $\rm S_{VI}$, $\rm S_{IX}$, $\rm S_X$, $\rm S_I$, 
respectively (assuming all quark decay channels are open and ignoring phase space effects).  
\section{Details on $2 \rightarrow 2,3$ scattering calculations and phenomenology}
\label{Apptevatron}
In this Appendix we give some details of the calculations for $t\bar t$ and light quark $q \bar q$ production in the scalar and vector MFV models. The comparisons with experimental data at the Tevatron and LHC were done in Sections \ref{Tevatron} and \ref{tevatron-constraints}.

\subsubsection{$2 \rightarrow 2$ scattering in the scalar models}

The relevant interaction Lagrangians for the scalar models are given in 
Eqs. \eqref{eq:SVandSVILag} -- \eqref{eq:SILag:app}.
The general formula for the spin averaged matrix element squared for $q\bar q \to t\bar t$, including
interference with the SM, for the exchange of a color triplet or sextet in the $u$-channel is \cite{Shu:2009xf}
\bea\label{scalarNPcross}
\sum_{ij} \, |\mathcal{M}|^2 = 64 \sum_{ij}  \left[\frac{g_s^2 \, \eta^2 \, C_0}{s} \, \left(\frac{(s \, m_t^2 + u_t^2)(u - m_s^2)}{(u - m_s^2)^2 + \Gamma^2 \, m_s^2}\right) +  \frac{4 \, \eta^4 \, C_2 \, u_t^2}{(u - m_s^2)^2 + \Gamma^2 \, m_s^2}\right]\,,
\eea
where $v_t \equiv v - m_t^2$ for $v = \{s,t,u\}$.
For the exchange of a color singlet or octet in the $t$-channel one has to make the replacement $u \to  t$ above.
The color factors for an exchange of a $\bf (1, \bar{3}, 6, 8)$ of $\rm SU(3)_c$ are given by
$C_0 = (4,1,-1,-2/3)$ and $C_2 = (9,3/4,3/2,2)$.

The differential partonic  $2 \rightarrow 2$ cross section averaged over initial spins and colors and summed over final states is given by
\bea\label{basic}
 \frac{d \, \hat{\sigma}}{d \, z} = \frac{\beta}{32 \, \pi \, s} \, \,  \overline{\sum_{ij}} |\mathcal{M}_{ij}|^2.
\eea
where $  \overline{\sum_{ij}} \, |\mathcal{M}_{ij}|^2=\frac{1}{4} \, \frac{1}{9} \sum_{ij} \, |\mathcal{M}_{ij}|^2$. 
Here $\beta = \sqrt{1 - 4 m^2/s}$ is the velocity of the final state quark (with mass $m$) in the initial state partonic rest frame while $z = \cos\theta$ and $\theta$ is the scattering angle in the partonic CM frame. The sum in Eq. (\ref{basic}) is over all contributing matrix elements. The initial states are weighted
with the appropriate PDFs to obtain the hadronic cross section.  When evaluating the dijet constraints
we use MSTW2008 PDF sets \cite{Martin:2009iq}.
The renormalization scale is set to the $p_T$ of the jets, $\mu = M_{jj} \, \sqrt{1 - z^2}/2$,
where $z = \cos \theta$ is the partonic scattering angle from the beam line and $M_{jj}$ is the invariant mass of the dijets.

In the case of light quark pair production, in models $\rm S_{V}$ and $\rm S_{VI}$ there are $u$-channel exchange contributions to $u \, \bar{u} \rightarrow c \, \bar{c}, c \, \bar{c}  \rightarrow u \, \bar{u}$,
and in $\rm S_{VI}$ there are additional $u$-channel exchange contributions to $u \, \bar{u} \rightarrow u \, \bar{u},~ c \, \bar{c}  \rightarrow c \, \bar{c}$. The corresponding matrix elements squared can be obtained from Eq. \eqref{scalarNPcross}, taking care to include factors of 2 and 4, respectively, in the first and second terms for the latter subprocesses. 
$s$-channel exchange in models ${\rm S}_{V}$ and $\rm S_{VI}$ contribute to 
 $u \, c \rightarrow u \, c + {\rm c.c.}$, and in $\rm S_{VI}$ there are additional contributions to 
  $u \, u \rightarrow u \, u + {\rm c.c.}, ~~c c\rightarrow c c +{\rm  c.c.}$. 
The corresponding cross sections can be obtained via crossing
symmetry from the above, and including $t$ channel SM contributions, also see \cite{Ligeti:2011vt}.

Following \cite{Lane:1991qh}, experimental cuts on the rapidity $y$ and the rapidity separation $\Delta y$ of the two leading jets in dijet production are implemented at the partonic level in the $2 \rightarrow 2$ subprocesses via
\bea
\frac{d \sigma}{d M_{jj}} = \sigma_N \frac{M_{jj}}{s} \int_{-y_B^0}^{y_B^0} d \, y_B  \int^{z_0}_{-z_0} d \, z \, \sum_{ij} f_i(\tau \,  e^{y_B}) \, f_j(\tau \,  e^{-y_B})  \frac{d \, \hat{\sigma}}{d \, z}
\eea
where $\sigma_N=0.3894 \cdot10^9 {\rm pb}$,  $s$ is the hadronic center of mass energy, $\tau = \sqrt{M_{jj}^2/s}$, $f_{i,j}$ are the MSTW parton densities, and 
$Y_B $ is the boost rapidity of the subprocess frame. The rapidity cuts
$|y_{1,2}| < Y_{\rm max}$ correspond to the integration limits
\beq
y_B^0 = {\rm Min}[Y_{\rm max}, -\log[M_{jj}^2/s]/2]
\eeq
and a cut $\Delta y < (\Delta y)^{\rm max}$ enters the angular integration limits as
\beq
z^0 = {\rm Min}[z^{\rm max}, \tanh(Y_{\rm max} - |y_B|)].
\eeq
where $z^{\rm max} =({\rm Exp}[(\Delta y)^{\rm max}]-1 )/({\rm Exp}[(\Delta y)^{\rm max}]+1 )$.

\subsubsection{Vector models $2 \rightarrow 2$ production}

The vector models can be divided  into two sets: 
the cases I-IX that couple only right-handed or left-handed quarks, and the models X and XI, where the right-handed quarks are coupled to left-handed (charge conjugated) quarks. 
To obtain a general equation for the $q \bar q \to t \bar t$ cross section for the first subset of models I-IX, we consider a generic Lagrangian (with trivial changes in notation if vectors couple to right-handed quarks)
\beq\label{generic-V-Lagr}
{\cal L}= f_q \bar q_L \slashed V q_L + f_t \bar t_L \slashed V t_L + [f_{qt} \bar q_L\slashed V t_L+ h.c.],
\eeq
where we have suppressed flavor, weak and color indices. This gives for the weighted average of the amplitude squared for the $q\bar q\to t\bar t$ scattering 
 \beq
 \begin{split}
\label{MaverageV}
\overline\sum|{\cal M}|^2 &=  {\cal C}_1 { |f_q f_t|^2  \over 36}  {s^2 (1+\beta_\theta )^2 \over s_V^2 + \Gamma_V^2 m_V^2 } +  {\cal C}_3  {|f_{qt}|^4 \over  36 } {s^2 \over  t_V^2 + \Gamma_V^2 m_V^2  }  \left( (1+\beta_\theta)^2   + { m_t^4 \over 4 m_V^4 }\left[(1-\beta_\theta )^2 + 16 {m_V^2 \over s}\right] \right) \\
&+ {\cal C}_2    {|f_{qt}|^2  \over 18}  {  f_q f_t  (s_V t_V  + \Gamma_V^2 m_V^2  ) \over  (s_V^2 + \Gamma_V^2 m_V^2 )(t_V^2 + \Gamma_V^2 m_V^2 )  }  s^2 \left(   (1+\beta_\theta )^2  + 2  {m_t^4 \over s m_V^2 }\right) \end{split}
\eeq
while the interference with the LO SM one gluon exchange gives
\beq
\begin{split}
\label{MinterfV}
\overline\sum 2 {\rm Re}({\cal M}{\cal M}_{SM}^\ast) &={8 \pi  \alpha_S \over 9 } {\cal C}_4  |f_{qt}|^2  {t_V \over t_V^2 + \Gamma_V^2  m_V^2 }  s
\left ({ m_t^2 \over s}  + {1\over 4} (1+\beta_\theta )^2 +  { m_t^2 \over 2 m_V^2 }   \left[ {m_t^2 \over s } + {1\over 4 }  (1-\beta_\theta)^2 \right]\right) \\
&+{8 \pi  \alpha_S \over 9 } {\cal C}_5 {f_q f_t s_V  \over s_V^2 + \Gamma_V^2 m_V^2 } 
s \left[ {m_t^2 \over s} + {1\over 4} (1+ \beta_\theta )^2 \right].
\end{split}
\eeq
Here $s_V=s-m_V^2$, and similarly $t_V=t-m_V^2, u_V= u-m_V^2$. Explicitly one has 
\beq
t = m_t^2 -\frac{1}{2} s (1 - \beta_\theta),\qquad  u= m_t^2 -\frac{1}{2} s (1 + \beta_\theta),
\eeq
with $\beta_\theta=\beta \cos\theta$, while $\beta=\sqrt{1-4m_t^2/s}$ is the top velocity in the $t \bar t$ rest frame. We have used the same notational conventions as in the scalar case above. 
This formula should be summed over all possible initial quark states $q$, which depends on the particular MFV vector field considered. 

\begin{table}[t]
\begin{center}
\begin{tabular}[t]{c|c|ccc|ccccc|ccccc}
  \hline
  \hline
   Case &Couples to $q$& $f_q$ & $f_t$ &$f_{qt}$  & ${\cal C}_1^s$ &  ${\cal C}_2^s$ &${\cal C}_3^s$ &${\cal C}_4^s$ &${\cal C}_5^s$ & ${\cal C}_1^o$ &  ${\cal C}_2^o$ &${\cal C}_3^o$ &${\cal C}_4^o$ &${\cal C}_5^o$\\ 
  \hline
$\rm I_{s,o}$ & $d,s,b$& $\eta_1$ & $0$ & $0$ & $9$  & $0$ & $0$ & $0$ & $0$ & $2$  & $0$ & $0$ & $0$ & $2$\\
$\rm II_{s,o}$ & $u,c$& $\eta_1$ & $\eta_1+ y_t^2 \eta_2 $ & $0$ &  $9$  & $0$ & $0$ & $0$ & $0$ & $2$  & $0$ & $0$ & $0$ & $2$\\
$\rm III_{s,o}$ & $u,d,s,c,b$& $\eta_1$ & $\eta_1+  y_t^2 \eta_2 $ & $0$ &  $9$  & $0$ & $0$ & $0$ & $0$ & $2$  & $0$ & $0$ & $0$ & $2$\\
$\rm IV_{s,o}$ & $u,c$ & $\eta_1$ & $\eta_1+  y_t^2 \eta_2 $ & 0 &$ 9 $ & $-6$ & $36$ & $8$ & $0$ &
$2$ & ${4\over 3}$ & $8$ & $-{4\over 3} $ &  $2$\\ 
& $d,s,b$ & $\eta_1$ & $\eta_1+  y_t^2 \eta_2 $ & $\eta_1 \delta_{q,b}$ &$ 9 $ & $-6$ & $36$ & $8$ & $0$ &
$2$ & ${4\over 3}$ & $8$ & $-{4\over 3} $ & $-2$\\ 
\hline
$\rm V_{s,o}$ & $d,s,b$& $\eta_1$ & $0$ &$0$ &$\frac{1}{4}$  & $-\frac{1}{4}$ & $\frac{9}{4}$ & $2$ & $0$ & $\frac{1}{18}$  & $\frac{1}{18}$ & $\frac{1}{2}$ & $-\frac{1}{3}$ & $-\frac{1}{3}$\\
$\rm VI_{s,o}$ & $u,c$& $\eta_1$ & $\eta_1+2 y_t^2 Re(\eta_2 )$ &$\eta_1+y_t^2 \eta_2$ &$\frac{1}{4}$  & $-\frac{1}{4}$ & $\frac{9}{4}$ & $2$ & $0$ & $\frac{1}{18}$  & $\frac{1}{18}$ & $\frac{1}{2}$ & $-\frac{1}{3}$ & $-\frac{1}{3}$\\
$\rm VII_{s,o}$ & $d,s,b$& $0$ & $0$ &$\eta_1+y_t^2\eta_2$ &$0$ & $0$ & $\frac{9}{4}$  & $2$ & $0$ & $0$ & $0$ & $\frac{1}{2}$  & $-\frac{1}{3}$ & $0$\\
$\rm VIII_{s,o}$ & $u,d,s,c,b$& $\eta_1+y_t^2 2 Re(\eta_2 )\delta_{q,b} $ & $\eta_1+y_t^2 2 Re(\eta_2 )$ &$\eta_1+y_t^2\eta_2$& $\frac{1}{4}$ & $\frac{-1}{4}$ & $\frac{9}{4}$  & $2$ & $0$ & $\frac{1}{18}$ & $\frac{1}{18}$ & $\frac{1}{2}$  & $-\frac{1}{3}$ & $-\frac{1}{3}$\\
$\rm IX_{s,o}$ & $u,c$& $\eta_1+y_t^2 2 Re(\eta_2 )\delta_{q,b} $ & $\eta_1+y_t^2 2 Re(\eta_2 )$ &$\eta_1+y_t^2\eta_2$& $\frac{1}{4}$ & $-\frac{1}{4}$& $\frac{9}{4}$  & $2$ & $0$ & $\frac{1}{18}$ & $\frac{1}{18}$ & $\frac{1}{2}$  & $-\frac{1}{3}$ & $-\frac{1}{3}$\\
 & $d,s$& $\eta_1+y_t^2 2 Re(\eta_2 )\delta_{q,b} $ & $\eta_1+y_t^2 2 Re(\eta_2 )$ &$\eta_1+y_t^2\eta_2$& $\frac{1}{4}$ & $\frac{1}{2}$& $9$  & $4$ & $0$ & $\frac{1}{18}$ & $-\frac{1}{9}$ & $2 $  & $-\frac{2}{3}$ & $\frac{1}{3}$\\
  & $b$& $\eta_1+y_t^2 2 Re(\eta_2 )\delta_{q,b} $ & $\eta_1+y_t^2 2 Re(\eta_2 )$ &$\eta_1+y_t^2\eta_2$& $1$ & -$\frac{2}{3}$& $4$  & $\frac{8}{3}$ & $0$ & $\frac{2}{9}$ & $\frac{4}{27}$ & $\frac{8}{9} $  & $-\frac{4}{9}$ & $-\frac{2}{3}$\\

\hline
  \hline
\end{tabular}
\end{center}
\caption{ The contributions to $t\bar t$ cross section from MFV vector resonances. The coefficients ${\cal C}_i$ are to be used in Eqs. \eqref{MaverageV}, \eqref{MinterfV}, where we have kept for simplicity only leading flavor breaking in the couplings of quarks to the vector, while neglecting the flavor breaking in the vector propagators (and in models IV, IX, X, XI also neglecting the electroweak 
symmetry breaking contributions to vector masses).}
\label{table-int-vector}
\end{table}

The expressions for the contributions to the $t\bar t$ cross section for the second subset, cases X and XI, are very similar to the above. The important difference is that both left-handed and right-handed quarks couple to the vectors at the same time. The general form of the interaction Lagrangian in this case is
\beq
{\cal L}=\bar t\gamma^\mu(f_L P_L+f_RP_R) V_\mu \, q^c +h.c.,
\eeq
which gives for the weighted average of the amplitude squared contribution to $q\bar q\to t\bar t$ scattering
\beq
\begin{split}\label{MaverageVsecond}
\overline\sum|{\cal M}|^2 &={\cal C}_1' (|f_{L}|^4+|f_{R}|^4) \frac{12 s^2}{36  u_V^2} \left[\frac{m_t^4}{4 m_V^4}\left((1+\beta_\theta)^2+16 \frac{m_V^2}{s}\right)+ (1-\beta_\theta)^2\right]+\\
&+{\cal C}_2'  2|f_{L}|^2|f_{R}|^2 \frac{12 s^2}{36  u_V^2} \left[\frac{m_t^4}{4 m_V^4}\left((1+\beta_\theta)^2+16 \frac{m_V^2}{s}\right)+ 4\left(1- 2\frac{m_t^2}{s}\right)^2\right].
\end{split}
\eeq
The interference with the SM one gluon exchange is
\beq
\begin{split}\label{MinterfVsecond}
\overline\sum 2 Re({\cal M}{\cal M}_{SM}^\ast) &=\frac{-1}{9}\pi \alpha_S {\cal C}_3' (|f_L|^2+|f_R|^2) \frac{s}{u_V} \left[(1-\beta_\theta)^2+8 \frac{m_t^2}{s}+\frac{m_t^2}{m_V^2}(1+\beta_\theta)^2+4 \frac{m_t^4}{m_V^2 s}\right].
\end{split}
\eeq

For the coefficients in these equations we have ${\cal C}_1'={\cal C}_2'=1(1/2)$, ${\cal C}_3'=2(1)$ for $\bf \bar{3} (6)$ cases, while  $f_L=\eta_1+y_t^2\eta_2$, $f_R=0$ for model X and $f_L=\eta_1+y_t^2\eta_2$, $f_R=\eta_1$ for model XI.

For light quark pair production 
we use the generic form of the interaction Lagrangian \eqref{generic-V-Lagr}, but replace the top quarks with light quarks. 
We denote by $f_q$ the flavor diagonal coupling of vectors to quarks
(these will contribute in the $s$-channel).  The flavor off-diagonal couplings for terms contributing in the $t$-channel are denoted as $f_{q \, q'}$.  
In the numerical analysis we set  $f_q = f_{q q'} $, which follows from $\rm H_F$ symmetry.

We examine dijet constraints on three sample models, $\rm II_{o}$, ${\rm VI}_s$ and $\rm VI_{o}$. 
The spin and color averaged amplitudes for $u \bar u  \to c \bar c $ or $c \bar c \to u \bar{u}$ processes are
 \beq
 \begin{split}
\label{MaverageVuubar}
\overline\sum|{\cal M}|^2 =  { |f_u f_c|^2  \over 9}  {{\cal C}_1 u^2 \over s_V^2 + \Gamma_V^2 m_V^2 } 
+   {2 |f_{uc}|^2  \over 9}  { {\cal C}_2  f_u f_c  (s_V t_V  + \Gamma_V^2 m_V^2  )\over  (s_V^2 + \Gamma_V^2 m_V^2 )(t_V^2 + \Gamma_V^2 m_V^2 )  } \, u^2
+  {|f_{uc}|^4 \over  9 } { {\cal C}_3  u^2 \over  t_V^2 + \Gamma_V^2 m_V^2  }  \end{split}
\eeq
and 
\beq
\label{MinterfVuubar}
\overline\sum \, 2 {\rm Re}({\cal M}{\cal M}_{SM}^\ast) ={8 \pi  \alpha_S \over 9 }\left( {\cal C}_4  |f_{uc}|^2  {t_V \over t_V^2 + \Gamma_V^2  m_V^2 }  +   {\cal C}_5 { f_u f_c  s_V  \over s_V^2 + \Gamma_V^2 m_V^2 } \right) {u^2  \over s},
\eeq
where $ t =   -s (1 -\cos\theta)/2, \,\,   u= -s (1 + \cos\theta)/2$.
The $C_i$'s are 
\beq 
{\cal C}_1 =  \frac{1}{18} ,~~~{\cal C}_2 = \frac{1}{18},~~~{\cal C}_3 = \frac{1}{2},~~~{\cal C}_4 = -{1\over 3},~~~{\cal C}_5=-\frac{1}{3}\,
\eeq
for model $\rm VI_o$,
\beq 
{\cal C}_1 =  \frac{1}{4} ,~~~{\cal C}_2 = -\frac{1}{4},~~~{\cal C}_3 = \frac{9}{4},~~~{\cal C}_4 =2,~~~{\cal C}_5=0\,,
\eeq
for model $\rm VI_s$, and
\beq 
{\cal C}_1 =  2 ,~~~{\cal C}_2 = 0,~~~{\cal C}_3 =0,~~~{\cal C}_4 = 0,~~~{\cal C}_5=2
\eeq
for model $\rm II_o $.

For the processes $u \bar u  \to u \bar{u} $ or $c \bar c \to c \bar{c}$ the spin and color averaged amplitudes squared are
\beq
\label{MaverageVuubaruubar}
\overline\sum|{\cal M}|^2 ={f_q^4 \over 9} u^2 \left( { {\cal C}_1  \over s_V^2 + \Gamma_V^2 m_V^2 }+ {2 \,  {\cal C}_2   \, {   s_V t_V  + \Gamma_V^2 m_V^2  \over  (s_V^2 + \Gamma_V^2 m_V^2 )(t_V^2 + \Gamma_V^2 m_V^2 ) } + { {\cal C}_3  \over  t_V^2 + \Gamma_V^2 m_V^2  } }\right)
\eeq
We have to add new terms to the interference between the NP and SM,
because now there is also a $t$-channel contribution in the SM.   The modified equation is
\beq
\label{MinterfVuubaruubar}
\overline\sum 2 {\rm Re}({\cal M}{\cal M}_{SM}^\ast) ={8 \pi  \alpha_S \over 9 }\, u^2 \, f_q^2 \left[{t_V \over t_V^2 + \Gamma_V^2  m_V^2 }\left({  C_4^s \over s} + {C_4^t \over t} \right) + { s_V  \over s_V^2 + \Gamma_V^2 m_V^2} \left({  C_5^s \over s} + {C_5^t \over t} \right)\right]
\eeq
The ${\cal C}_i$'s are
\beq 
{\cal C}_1 =  \frac{2}{9} ,~~~{\cal C}_2 = -\frac{2}{27},~~~{\cal C}_3 = \frac{2}{9},~~~{\cal C}^s_4 = -{2\over 9},
~~~{\cal C}^t_4 = {2\over 3},~~~{\cal C}^s_5=\frac{2}{3},~~~{\cal C}^t_5 = -{2\over 9}
\eeq
in model ${\rm VI_o}$,
\beq 
{\cal C}_1 = 1 ,~~~{\cal C}_2 = \frac{1}{3},~~~{\cal C}_3 =1,~~~{\cal C}^s_4 = {4\over 3},
~~~{\cal C}^t_4 = 0,~~~{\cal C}^s_5=0,~~~{\cal C}^t_5 = {4\over 3}
\eeq
in model ${\rm VI_s}$, and 
\beq 
{\cal C}_1 =  2 ,~~~{\cal C}_2 = -\frac{2}{3},~~~{\cal C}_3 =2,~~~{\cal C}^s_4 = -{2\over 3},
~~~{\cal C}^t_4 = {2},~~~{\cal C}^s_5=2,~~~{\cal C}^t_5 = -{2\over 3}
\eeq
for model ${\rm II_o}$

Using crossing symmetry, the processes $u \bar{c}\to u \bar{c} + {\rm c.c.}$
are obtained from Eqs. \eqref{MaverageVuubar}, \eqref{MinterfVuubar}
via the substitutions $s\to t, ~t\to s,~u\to u$; the processes
$u c \to u c + {\rm c.c.}$ are obtained from Eqs. \eqref{MaverageVuubar}, \eqref{MinterfVuubar}
via the substitutions $s\to t, ~t\to u,~u\to s$; and the processes 
$u u \to u u + {\rm c.c.}$, $cc\to cc+ {\rm c.c.}$
are obtained from Eqs. \eqref{MaverageVuubaruubar}, \eqref{MinterfVuubaruubar} via the substitutions $s\to t, ~t\to u,~u\to s$.

\subsubsection{LEP $e^+ \, e^- \rightarrow f \, \bar{f} V$ production}
Consider a vector Lagrangian of the
form $\mathcal{L} =
\eta \, V^\mu \, \bar{\psi}_R \, \gamma_\mu \, \psi_R $. Then the amplitude
squared for the production of a color singlet vector $V$ through $e^+ \, e^- \rightarrow \gamma^\star \rightarrow V \bar{f}
\, f$ is given by
\begin{multline}
  \overline{|\mathcal{M}|^2} =-\frac{8e^4Q^2N_c|\eta|^2}{s_1^2s_2^2}\bigg[
  \frac{1}{s^2}\left(2 m_v^2 ({s_1}{s_2}-s_1{t_1}-{s_2} {t_2})^2\right)\\
  -\frac{1}{s}\bigg(2 {s_1} {s_2} m_v^4- 2 ({s_1}+{s_2}) ({s_1} {t_1}+{s_2} {t_2}) m_v^2
  +{s_1} {s_2}  \big({s_1}^2-2 {t_2} {s_1}+{s_2}^2-2 {s_2}
    {t_1}+2 ({t_1}^2+{t_2}^2)\big)\bigg)\\
+m_v^2 ({s_1}-{s_2})^2+2 {s_1} {s_2}
({s_1}+{s_2}-{t_1}-{t_2}) -2 s {s_1} {s_2}\bigg].
\end{multline}
Here we have used the three body final state Mandelstam variables
using the notation of \cite{particalkin}, with the momenta assignments
$e^-(p_a),e^+(p_b),\bar{f}(p_1),V(p_2),f(p_3)$ and $e$ is the
electromagnetic coupling constant while $Q$ is the charge of a
particular quark. The corresponding production cross section at LEP is
given by
\bea 
  \sigma(e^+ \, e^- \rightarrow \gamma^\star \rightarrow V \bar{f}\, f) 
  = \frac{1}{2^{10}\pi^4s^2}
   \int \frac{ds_1 \, ds_2 \, dt_1 \, dt_2 \, \overline{|\mathcal{M}|}^2 \, \theta \left[-
    \mathcal{G} \right]}{ \sqrt{- \mathcal{G}}}, 
\eea
where the physical region that determines the integration range
satisfies the condition that the Gram determinant $\mathcal{G} \leq
0$. The Gram determinant for the $2 \rightarrow 3$ production is
explicitly given by
\bea
\mathcal{G} &=& \frac{1}{16} \left[m_v^4 s^2-2 m_v^2 s (s \, t_1+s \, t_2+s_1 (s_2-t_1)-s_2 t_2+2 t_1 t_2) \right.  \\
&\,& \left.+2 s_1 (s (s_2 (t_1+t_2)+t_1 (t_2-t_1))+s_2 t_2 (t_1-s_2))+(s (t_1-t_2)+s_2 t_2)^2+s_1^2 
   (s_2-t_1)^2\right]. \nn
\eea

\subsubsection{LHC $gg \rightarrow SS$ production}

For a field in a flavor representation $F$, and color representation $R$, the cross section for $g \, g \rightarrow S\, S$ is
\bea
\frac{d \sigma}{d t} = \frac{{\rm dim}(F) \, \pi \, \alpha_s^2}{8^2 \, s^2} \left[ C_1 \, F_1  + C_2 \, F_2 +  C_3 \, F_3+  C_4 \, F_4 \right].
\eea
Here the functions $F_i$ are
\beq
\begin{split}
F_1 &= 2 + \frac{s - 2 \, m_s^2}{t - m_s^2} + \frac{s - 2 \, m_s^2}{u - m_s^2}, \\
F_2 &= \frac{(u-t)^2}{2 \, s^2} + \left[\frac{(m_s^2 - u)(4 m_s^2 - s) + (t-u)(s - 2 m_s^2)/2}{s (t-m_s^2)} + u\leftrightarrow t \right],\\
F_3 &= 4 \, m_s^4 \left(\frac{1}{(t-m_s^2)^2} + \frac{1}{(u-m_s^2)^2} \right), \\
F_4 &= \frac{2 (s - 2 m_s^2)^2}{(u - m_s^2)(t - m_s^2)}.
\end{split}
\eeq
The color factors, for the $R = ({\bf 3,6,8})$ representations ($G = {\bf{8}}$) are
\beq
\begin{split}
C_1 &= C(R) {\rm dim}(G) (2 C_2(R) - C_2(G)/2)=  (14/3, 310/3, 108), \\
C_2 &= C(R) C_2(G) {\rm dim}(G) = (12,60,72), \\
C_3 &= [C_2(R)]^2 {\rm dim}(R) = (16/3, 200/3, 72), \\
C_4 &=  C(R) {\rm dim}(G)  (C_2(R) - C_2(G)/2) =  (-2/3, 110/3, 36).
\end{split}
\eeq

\section{Low Energy Constraints}
\label{AppFCNCs}
The main motivation for studying flavor symmetric extensions of the SM  is to have
an effective way of suppressing FCNCs allowing new states of fairly
low mass. In this appendix we give  a brief account of the constraints
on our models from low energy data. We focus on processes that can be
very restrictive,  $K-\bar K$, $B_{s,d}-\bar B_{s,d}$ and $D-\bar D$ mixing.

The effective weak Hamiltonian describing the $B_s-\bar B_s$ mixing in the SM is
\beq
{\cal H}_{\rm eff}^{\rm SM}=\eta_W  \eta_{RG}  \frac{\big( V_{ts}^* V_{tb}\big)^2}{\Lambda_{SM}^2} \big(\bar s_L \gamma^\mu b_L)^2, \label{SMeffH}
\eeq
where 
\beq
\Lambda_{SM}= \frac{2 \pi}{G_F m_W  \sqrt{S_0(x_t)}}=4.39~{\rm TeV}.
\eeq
This is larger than $m_W$ by an order of magnitude because SM contribution arises at one loop level from box diagrams with $t$ and $W$ exchanges. The factor $\eta_W$  encodes the QCD corrections to the matching at the hard scale $\mu_W\sim m_W$, while $\eta_{RG}$ is the correction due to the running from $\mu_W$ to $\mu_b$. In NDR at NLO they are $\eta_W=0.970$, $\eta_{RG}=0.860$ (using  $\alpha_S(m_z)=0.1184$, $\mu_W=m_W$, $\mu_b=m_b=4.2$ GeV and $\bar m_t=164$ GeV for which $S_0(x_t)=2.33$). 
The SM weak hamiltonians for the $B_d-\bar B_D$, $K-\bar K$ and $D-\bar D$ mixing follow from \eqref{SMeffH}  with the obvious replacements in quark flavors. In addition $K$ and $D$ mixing amplitudes receive large nonlocal contributions in the SM.

NP contributions can modify the effective weak Hamiltonian to ${\cal H}_{\rm eff}^{\rm SM}+\Delta {\cal H}_{\rm eff}^{\rm NP}$. To make the comparison with the experiment easier we normalize the NP contributions to the SM predictions. For $B_{s,d}$ mixing we thus define
\beq
h_{s,d} e^{i 2\sigma_{s,d}}\equiv \frac{\langle B_{s,d}|\Delta {\cal H}_{\rm eff}^{\rm NP}|\bar B_{s,d}\rangle}{\langle B_{s,d}|{\cal H}_{\rm eff}^{SM}|\bar B_{s,d}\rangle}. \label{hsd-definitions}
\eeq
Note that $\sigma_{s,d}$ measures the deviation of NP weak phase from the SM one (for $B_s$ mixing the SM weak phase is very small, $O(1^\circ)$, while for $B_d$ mixing it is $2\beta\sim 45^\circ$). Since the (real part) of $\langle K^0|{\cal H}_{\rm eff}^{SM}|\bar K^0\rangle$ receives potentially large long distance contributions from $\Delta S=1$ operators, we normalize the NP matrix elements to the measured mass difference
\beq
h_K e^{i 2\sigma_K}\equiv  \frac{M_{12}^{NP}}{\Delta m_K}= \frac{1}{\Delta m_K}\cdot \frac{\langle K^0|\Delta {\cal H}_{\rm eff}^{\rm NP}|\bar K^0\rangle}{2 m_K}.
\eeq
Similarly there are potentially large long distance effects in $D-\bar D$ mixing amplitude, so that we define in the same way
\beq
h_D e^{i 2\sigma_D}\equiv  \frac{M_{12}^{NP}}{\Delta m_D}=\frac{1}{\Delta m_D}\cdot \frac{\langle D^0|\Delta {\cal H}_{\rm eff}^{\rm NP}|\bar D^0\rangle}{2 m_D}.
\eeq
Note that $\sigma_K$ and $\sigma_D$ measure the {\it total} weak phase of the NP contributions to the mixing, not just the deviation from the SM one. 
No NP in the mixing corresponds to $h_s=h_d=k_K=h_D=0$.  

 From the agreement of measured and predicted values of $\epsilon_K$ we obtain the bound at 95\% C.L. 
 \beq
 |h_K\sin(2\sigma_K)|\lesssim 1.3 \times 10^{-3}.
 \eeq
Conservatively we use the the prediction $|\epsilon_K|=(2.01^{+0.59}_{-0.66})\cdot 10^{-3}$ from \cite{Lenz:2010gu}, which is in good agreement with the measured value $|\epsilon_K|=(2.229\pm0.010)\cdot 10^{-3}$ (for a different treatment of lattice inputs see \cite{Lunghi:2010gv}). The bound on $h_K \cos 2\sigma_K$ is more uncertain because of potentially large long distance contributions. Assuming conservatively that $\Delta m_K$ is saturated by NP gives $
  |h_K \cos 2\sigma_K|\lesssim 0.5$. The CP violation in $D-\bar D$ mixing is well constrained and so 
$|h_D\sin(2 \sigma_D)|\lesssim 0.3$,
at 95 \% C.L. \cite{Kagan:2009gb} (the NP contribution to $\phi_{12}^D$ is $2\sigma_D$, while the SM CP violating contribution is negligible). Conservatively assuming that NP saturates the mass difference $\Delta m_D$ we obtain 
$h_D\lesssim 0.5$.
 
The NP contributions to the effective weak Hamiltonian in \eqref{SMeffH}, $\Delta {\cal H}_{\rm eff}^{\rm NP}$, are described by local operators
\beq
\Delta {\cal H}_{\rm eff}^{\rm NP}=\sum_iC_i Q_i+\sum_i \tilde C_i \tilde Q_i
\eeq 
where the sum runs over the full set of dimension 6 local operators. For $B_s$ mixing these are \cite{Bona:2007vi}
\beq
\begin{split}
Q_1=(\bar s_L\gamma_\mu b_L)&(\bar s_L\gamma_\mu b_L), \qquad Q_{2,4}=(\bar s_R b_L)(\bar s_{R,L} b_{L,R}), \qquad Q_{3,5}=(\bar s_R^\alpha b_L^\beta)(\bar s_{R,L}^\beta b_{L,R}^\alpha),\label{Qi}
\end{split}
\eeq
Above $P_{L,R}=(1\pm \gamma_5)/2$, while color indices are not displayed, if the summation is within brackets. The operators $\tilde Q_i$   are obtained from Eq. \eqref{Qi} by making a $P_R\leftrightarrow P_L$ replacement. Note that $Q_1$ is the operator in the SM effective weak Hamiltonian Eq. \eqref{SMeffH}.

The SM and NP weak hamiltonians for the $B_d$, $K$ and $D$ mixing follow from \eqref{SMeffH} and \eqref{Qi} with the obvious replacements in quark flavors. In addition $K$ and $D$ mixing amplitudes receive large nonlocal contributions in the SM.
We evaluate the matrix elements for the operators $Q_i$  in the same way as described in Eqs. (8)-(10) of \cite{Bona:2007vi}, including the numerical values, except for the updated values of $f_{B_d}=205\pm 12$ MeV, $f_{B_s}=250\pm 12$ MeV \cite{Laiho:2009eu} and the quark masses, for which we use the PDG 2010 values  $\beta=(21.8\pm0.9)^\circ$ and $\gamma=(67\pm4)^\circ$ \cite{Charles:2004jd}. The matrix elements for $\tilde Q_i$ are the same as for $Q_i$ since QCD interactions are parity conserving. 

\subsection{Models I-IX}
The models I-IX give contributions to the effective Hamiltonian for $B_s$ mixing that at scale $\mu= m_b$ take the form
\beq\label{eq:deltaHB2}
\Delta {\cal H}_{\rm eff}^{{\rm NP},B_s}=\frac{\kappa_s}{M_V^2}(y_t^2 V_{tb} V_{ts}^*)^2 \eta' \eta_{RG} \big[\bsbs\big],
\eeq 
or of its parity image: replacing $\bsbsR$ for $\bsbs$.  Parity symmetry
of the strong interactions implies that these operators have identical
matrix elements between one particle states. The value of model dependent constant  $\kappa_s$ is listed in Table \ref{kappa-table1}, while $M_V$ stands for the average mass of the MFV-vector
exchanged. The factor $\eta'$ is a QCD correction that arises from
running from the scale of the new physics, $M_V$, to the electroweak
scale, $M_W$. In calculating bounds we set $\eta'=1$ incurring an
error of order $[\alpha_S(M_W)/4\pi]\ln(M_V/M_W)\approx
0.01\ln(M_V/M_W) $.  

The contributions to $B_d$, $D$ and $K$ mixing we define analogously (for simplicity of notation setting $\eta' \eta_{RG}=1$),
\begin{align}
\Delta {\cal H}_{\rm eff}^{{\rm NP},B_d}=&\frac{\kappa_d}{M_V^2}(y_t^2 V_{tb} V_{td}^*)^2  (\bar d_L \gamma^\mu b_L)^2,
\qquad  \Delta {\cal H}_{\rm eff}^{{\rm NP},K}=\frac{\kappa_K}{M_V^2}(y_t^2 V_{ts} V_{td}^*)^2  (\bar d_L \gamma^\mu s_L)^2,\label{HeffBd}\\
\Delta {\cal H}_{\rm eff}^{{\rm NP},D}=&\frac{\kappa_D}{M_V^2}(y_b^2 V_{ub}^* V_{cb})^2  (\bar c_L \gamma^\mu u_L)^2.\label{HeffD}
\end{align} 
The coefficients $\kappa_{s,d,K,D}$ are collected in Table  \ref{kappa-table1}.
 In models V, VIII and IX several vector fields mediate mixing and we indicate their separate contributions. This requires we
distinguish among their masses , with  $M_V$ then denoting their average mass in the defining formulae \eqref{eq:deltaHB2}-\eqref{HeffD} for $\kappa_i$.

\renewcommand\arraystretch{1.1}
\begin{table}[t]
\begin{center}
\begin{tabular}[t]{c|ccc}
\hline\hline
  \multirow{2}{*}{Case}   & \multicolumn{3}{c}{$\Delta\mathcal{L}$}\\[-2mm]
  & $\hspace{1.7cm}\kappa_{s,d}/C_{s,o}\hspace{1.7cm}$ &  $\hspace{1.7cm}\kappa_{K}/C_{s,o}\hspace{1.7cm}$ & $\hspace{1.7cm}\kappa_{D}/C_{s,o}\hspace{1.7cm}$\\ 
  \hline\hline
 \multirow{2}{*}{$\rm I_{s,o}$}  &         \multicolumn{3}{c}{$\eta_1\, \bar d_R (\slashed V Y_D  \Delta_U Y_D^\dagger )d_R+\big[\eta_2\, \bar d_R( \slashed V Y_D\Delta_U \Delta_D Y_D^\dagger) d_R +h.c.\big]$}     \\ 
& $ y_{s,d}^2 y_b^2(\eta_1+\eta_2 y_b^2)^2 $  & $ y_{d}^2 y_s^2 (\eta_1+\eta_2 y_b^2)^2 $ & $0$ \\[1mm]
\hline
\multirow{2}{*}{$\rm II_{s,o}$}  &         \multicolumn{3}{c}{$\eta_1\, \bar u_R (\slashed V Y_U  \Delta_D Y_U^\dagger )u_R+\big[\eta_2\, \bar u_R( \slashed V Y_U\Delta_U \Delta_D Y_U^\dagger) u_R +h.c.\big]$}     \\   
& $0$  & $0$ & $y_u^2 y_{c}^2 (\eta_1+\eta_2 y_c^2)^2 $ \\[1mm]
\hline
\multirow{2}{*}{$\rm III_{s,o}$}  &         \multicolumn{3}{c}{$\eta_1 \bar Q_L (\slashed V \Delta_U) Q_L +\eta_1'  \bar Q_L (\slashed V\Delta_D) Q_L +\big[\eta_2 \bar Q_L (\slashed V \Delta_U \Delta_D) Q_L +h.c.\big]$}     \\   
& $\big(\eta_1+\eta_2 y_b^2)^2$  & $\eta_1^2$ & $\eta_1'{}^2$ \\[1mm]
\hline
\multirow{2}{*}{$\rm IV_{s,o}$}  &         \multicolumn{3}{c}{$\eta_0 \bar Q_L \slashed V  Q_L +\eta_1 \bar Q_L (\slashed V \Delta_U) Q_L +\eta_1'  \bar Q_L (\slashed V\Delta_D) Q_L +\big[\eta_2 \bar Q_L (\slashed V \Delta_U \Delta_D) Q_L +h.c.\big]$}     \\   
& $\big(\eta_1+\eta_2 y_b^2)^2+\Delta_{CC}$  & $\eta_1^2+\Delta_{CC}$ & $\eta_1'^2+\Delta_{CC}$ \\[1mm]
\hline
\multirow{2}{*}{$\rm V_{s,o}$}  &         \multicolumn{3}{c}{$\eta_1 \bar d_R \slashed V d_R+\big[\eta_2 \bar d_R \,\, \slashed V \big(Y_D \Delta_U Y_D^\dagger\big) d_R +h.c.\big]+\eta_3 \bar d_R   \big(Y_D \Delta_U Y_D^\dagger\big)  \slashed V \big(Y_D \Delta_U Y_D^\dagger\big) d_R$}     \\   
& $ y_{s,d}^2 y_b^2\left[  \frac{\eta_2^2}{4 r_3^2}+
\frac{(\eta_2-2 \eta_2^*)^2}{12 r_8^2} +\frac{ \eta_1\eta_3}{r_{6,4}^2}\right]$  & $ y_{d}^2 y_s^2 
\left[  \frac{(\eta_2-\eta_2^*)^2}{4 r_3^2}+
\frac{(\eta_2+ \eta_2^*)^2}{12 r_8^2} +\frac{ \eta_1\eta_3}{r_{1}^2}\right]$ & $0$ \\[1mm]
\hline
\multirow{2}{*}{$\rm VI_{s,o}$}  &         \multicolumn{3}{c}{$\eta_1 \bar u_R \slashed V u_R+\big[\eta_2 \bar u_R \,\, \slashed V \big(Y_U \Delta_D Y_U^\dagger\big) u_R +h.c.\big]+\eta_3 \bar u_R   \big(Y_U \Delta_D Y_U^\dagger\big)  \slashed V \big(Y_U \Delta_D Y_U^\dagger\big) u_R$}     \\   
& $ 0$  & $ 0$ & $y_{u}^2 y_c^2\left[  \frac{(\eta_2^*-\eta_2)^2}{4r_3^2}+
\frac{(\eta_2+\eta_2^*)^2}{12 r_8^2} +\frac{ \eta_1\eta_3}{r_{1}^2}\right]$ \\[1mm]
\hline
\multirow{2}{*}{$\rm VII_{s,o}$}  &         \multicolumn{3}{c}{$\eta_1 \; \bar d_R \slashed{V}  u_R
+ \eta_2 \; \bar d_R (Y_D\Delta_U Y_D^\dagger) \slashed{V}  u_R + \text{h.c.}$}     \\   
& $ \Delta_{CC}$  & $ \Delta_{CC}$ & $\Delta_{CC}$ \\[1mm]
\hline
\multirow{2}{*}{$\rm VIII_{s,o}$}  &         \multicolumn{3}{c}{$\eta_1 \bar Q_L \slashed V Q_L +[\eta_2 \bar Q_L \Delta_U \slashed V Q_L +\eta_2' \bar Q_L \Delta_D \slashed V Q_L +h.c.]+\eta_3 \bar Q_L \Delta_U \slashed V \Delta_U Q_L +\eta_3' \bar Q_L \Delta_D \slashed V \Delta_D Q_L $}     \\   
& $ \frac{\eta_2^{*2}}{4r_3^2}+
\frac{(\eta_2^*-2 \eta_2)^2}{12 r_8^2}+\frac{ \eta_1\eta_3}{r_{6,4}^2}$  & 
$ \frac{(\eta_2^*-\eta_2)^2}{4r_3^2}+
\frac{(\eta_2+ \eta_2^*)^2}{12 r_8^2}+\frac{\eta_1\eta_3}{r_1^2}$ & 
$ \frac{(\eta_2'-\eta_2'{}^*)^2}{4r_3^2}+
\frac{(\eta_2'+ \eta_2'{}^*)^2}{12 r_8^2}+\frac{\eta_1\eta_3'}{r_1^2}$ \\[1mm]
\hline
\multirow{2}{*}{$\rm IX_{s,o}$}  &         \multicolumn{3}{c}{$\eta_1 \bar Q_L \slashed V Q_L+[\eta_2 \bar Q_L \Delta_U \slashed V Q_L +\eta_2' \bar Q_L \Delta_D \slashed V Q_L +h.c.]+\cdots$}     \\   
& $ \frac{\eta_2^{*2}}{4r_3^2}+
\frac{(\eta_2^*-2 \eta_2)^2}{12 r_8^2}+\Delta_{CC}$  & 
$ \frac{(\eta_2^{*}-\eta_2)^2}{4r_3^2}+
\frac{(\eta_2+ \eta_2^*)^2}{12 r_8^2}+\Delta_{CC}$ & 
$ \frac{(\eta_2'-\eta_2'{}^*)^2}{4r_3^2}+
\frac{(\eta_2'+ \eta_2'{}^*)^2}{12 r_8^2}+\Delta_{CC}$ \\[1mm]
\hline
\hline
\end{tabular}
\end{center}
\caption{The part of interaction Lagrangian $\Delta\mathcal{L}$ relevant for meson mixing amplitudes and coefficients $\kappa_i$ in \eqref{eq:deltaHB2}-\eqref{HeffD} for each of the models I-IX. We use the abbreviations $\Delta_U=Y_U^\dagger Y_U$, $\Delta_D=Y_D^\dagger Y_D$ and $r_i=m_{V_i}/m_V$ for the ratios of vector meson masses in the flavor multiplet to the average vector meson mass $m_V$, while $\Delta_{\text{CC}}$ stands for a 1-loop contribution, a box diagram, from exchange of charged MFV-vectors; see text. The factors of
  $C_s=1 (C_o=\frac13) $ must be included for color singlet (octet) vectors. Due to lack of space we display for model IX results only for leading complex parameters (the undisplayed leading real terms are the same as for model VIII). }
\label{kappa-table1}
\end{table}

Numerically, for $B_{d,s}$ mixing contributions we have
\begin{equation}
h_{d,s}e^{2i\sigma_{d,s}}= \kappa_{d,s} y_t^4 \frac{\eta'}{\eta_W}\frac{\Lambda_{SM}^2}{ M_V^2}
\approx 20\times \kappa_{d,s} y_t^4\left(\frac{1~\text{TeV}}{M_V}\right)^2\,.
\end{equation}
The contribution to the $K$ mixing is
\beq
h_K e^{i 2\sigma_K}=\kappa_K y_t^4 (V_{ts} V_{td}^*)^2\frac{m_K}{3\Delta m_K}\frac{f_K^2 \hat B_K}{m_V^2}\approx 0.1 \times \kappa_Ke^{i2\beta}  y_t^4 \left(\frac{1~\text{TeV}}{M_V}\right)^2,
\eeq
and for the $D$ mixing
\beq
h_D e^{i 2\sigma_D}=\kappa_D y_b^4 (V_{ub}^* V_{cb})^2\frac{m_D}{3\Delta m_D}\frac{f_D^2 B_1^D}{m_V^2}\approx (0.3\cdot 10^{-2}) \times \kappa_D e^{i2\gamma}  y_b^4 \left(\frac{1~\text{TeV}}{M_V}\right)^2.
\eeq
In the numerics above we use $f_K=156$ MeV, $\hat B_K=0.79$ \cite{Bona:2007vi}, $B_1^D=0.865$ \cite{Bona:2007vi}, the averages of CKM elements from CKMfitter ICHEP2010 update \cite{Charles:2004jd}. The standard CKM unitarity triangle angles have the values $\beta=(21.8\pm0.9)^\circ$ and $\gamma=(67.2\pm3.9)^\circ$. We have also included the effect of running from electroweak scale to $m_D$ for $D-\bar D$ mixing, however, this is numerically unimportant and results in a percent level shift in $h_D$. 

The flavor constraints thus require that (for $M_V=1$ TeV) the $\kappa_i$ coefficients for the particular model are below $\kappa_d \lesssim 0.02$, $\kappa_K \lesssim 10^{-2}$ (unless the phase of $\kappa_K$ is finely tuned not to give a contribution to $\epsilon_K$), and $\kappa_D\lesssim 20$. In order to explain the hint for nonzero phase of $B_s$ mixing, on the other hand, requires $h_s\sim 0.02$ or $h_s\sim 0.1$ for the two solutions, with appropriate weak phases. 

We make the following observations regarding models I-IX:
\begin{itemize}
\item
We can group the models into two categories. The first category form ``universal" models, where the contributions to the mixing amplitudes are due to class-1 operators in the notation of \cite{Kagan:2009bn}. For universal models we have 
\beq
\kappa_s=\kappa_d\simeq\kappa_K\sim \kappa_D,
\eeq
 where the last approximate equalities are valid, if all the couplings $\eta_i$ are of the same size. In addition in $y_b\to 0$ limit $\kappa_{s,d}=\kappa_K$. 
 \item
 The second category form the ``Yukawa suppressed" models, where the contributions to the mixing come from operators with additional Yukawa suppression (the so-called class-2 operators in the notation of  \cite{Kagan:2009bn}). These are  the models ${\rm II_{s,o}}$, ${\rm VI_{s,o}}$, that contribute only to $D-\bar D$ mixing and the models ${\rm I_{s,o}}$,  ${\rm V_{s,o}}$,  for which 
 \beq
\kappa_d:\kappa_s:\kappa_K=y_d^2:y_s^2:\frac{y_s^2 y_d^2}{y_b^2},
\eeq
 (the relation to $\kappa_K$ is only approximate for model V). 
 \item
Universal models can require small values for $\eta_i\sim 0.1 \cdot (M_V/1{\rm ~TeV})$ in order to avoid $K-\bar K$ and $B_d-\bar B_d$ mixing constraints. This is an order of magnitude smaller then the flavor-conserving couplings typically required to obtain large $A_{FB}^{t\bar t}$. This by itself is not a problem as the flavor violating couplings could be loop suppressed. 
\item
Yukawa suppressed models have no problems satisfying FCNC bounds. The typical sizes of contributions to $h_i$ are
\begin{align}
h_s e^{i 2\sigma_s}&\simeq 0.09\times y_t^4 \frac{\kappa_s}{y_s^2}  \left(\frac{y_s}{0.02}\right)^2\left(\frac{0.3{\rm ~TeV}}{m_V}\right)^2,\\
h_d e^{i 2\sigma_d}&\simeq (2\cdot 10^{-4})\times y_t^4 \frac{\kappa_d}{y_d^2}  \left(\frac{y_d}{10^{-3}}\right)^2\left(\frac{0.3{\rm ~TeV}}{m_V}\right)^2,\\
h_K e^{i 2\sigma_K}&\simeq (5 \cdot 10^{-10})\times y_t^4 \frac{\kappa_K}{y_d^2 y_s^2}  \left(\frac{y_d}{10^{-3}}\right)^2  \left(\frac{y_s}{0.02}\right)^2 \left(\frac{0.3{\rm ~TeV}}{m_V}\right)^2,\\
h_D e^{i 2\sigma_D}&\simeq (8 \cdot 10^{-16})\times y_b^4 \frac{\kappa_D}{y_u^2 y_c^2}   \left(\frac{y_u}{10^{-5}}\right)^2 \left(\frac{y_c}{5\cdot 10^{-3}}\right)^2\left(\frac{0.3{\rm ~TeV}}{m_V}\right)^2,
\end{align}
where we have normalized $y_{s,d}$ to values typical for large $\tan\beta$ $(y_b\sim 1)$. The contributions to $B_d$, $K$ and $D$ mixing are completely negligible for these models.
\item
Yukawa suppressed models ${\rm I_{s,o}}$,  ${\rm V_{s,o}}$ can potentially explain $B_s$ mixing anomaly, if $y_b\sim {\mathcal O}(1)$ and $\eta_i\sim {\mathcal O}(1)$, while vectors need to have electroweak scale masses of no more than a few 100 GeV. For this a flavor diagonal weak phase is needed \cite{Ligeti:2010ia}. 
A flavor diagonal weak phase arises if the coefficients $\eta_i$ are complex. For models I-VI the coefficients can be real only for terms that are higher order in Yukawas and therefore non-hermitian: $\eta_{0,1,3}$ and $\eta_1'$ are all real, while $\eta_2$ can be complex. In models ${\rm VII_{s,o}}$ all $\eta_i$ can be complex. In models V, VIII and IX the contributions to the mixing carry a weak phase only if the vector mesons are not degenerate.
 \end{itemize}
 
 \begin{figure}
\includegraphics[width=3.0in]{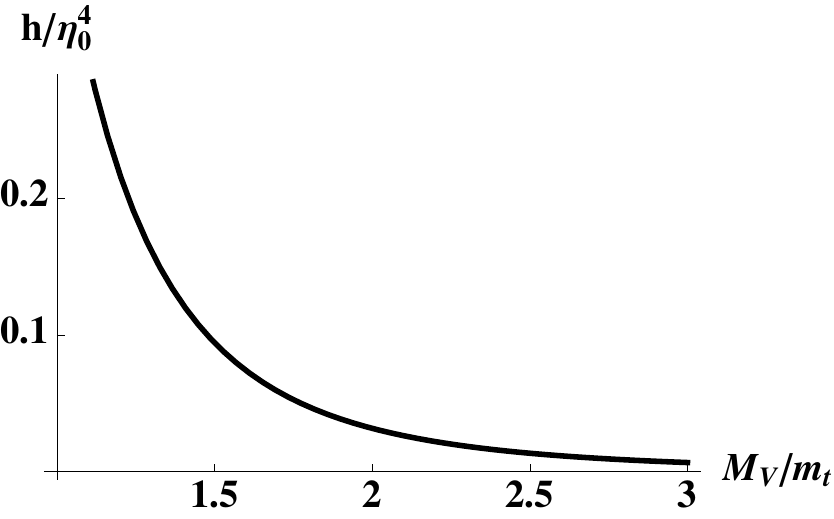}
\vspace{-0.1 cm}
\caption{\small Correction $h$, in units of $\eta_0^4$, to BB mixing
  in Model ${\rm IV_s}$ from double exchange of charged currents as a
  function of vector mass in units of top mass. }
\label{ratioBoxDiagrams}
\end{figure}
 
Finally, in models IV, VII and IX we include
a contribution $\Delta_{\text{CC}}$ from charged vectors that produce
mixing through a 1-loop box diagram, much like in the standard model.
For example, for model ${\rm IV_s}$ we find for the contribution to $B_s$, $B_d$ and $K$ mixing
\begin{equation}
\Delta_{\text{CC}}=\frac1{128\pi^2}\frac{\eta_0^4}{y_t^4}S_0(x_{tV})\,,
\end{equation}
where $x_{tV}=(m_t/M_V)^2$ and \cite{Inami:1980fz}
\begin{equation}
\label{BB:InamiLim}
S_0(x)=x\left[\frac14+\frac94\frac{1}{1-x}-\frac32\frac1{(1-x)^2}\right]-\frac32\left(\frac{x}{1-x}\right)^3\ln(x)\,.
\end{equation}
For model ${\rm VII_s}$ the $B_s$ mixing contribution is 
\begin{equation}
  \Delta_{\text{CC}}=\frac1{64\pi^2}(y_by_{s})^2S_0(x_{tV})
  \left[|\eta_1\eta_2|^2+\text{Re}(\eta_1^*\eta_2)^2\right]\,,
\end{equation}
while the $B_d$ and $K$ mixing contributions are obtained through the replacements $y_s\to y_d$ and $y_b\to y_d$, respectively. 
Above, we have assumed degenerate MFV-vectors for simplicity. For ${\rm IV_o}$ and
${\rm VII_o}$ insert an additional factor of $11/6$ in these
formulae. Fig.~\ref{ratioBoxDiagrams} shows the correction $h=h_s=h_d$ in
model ${\rm IV_s}$ resulting only from $\Delta_{\rm CC}$, as a
function of the top to charged vector mass ratio $m_t/M_V$. The figure
is for $\eta_0=1$, but can easily be rescaled as indicated. The bound
$h<0.2$ gives $M_V>200$~GeV (using $m_t=164$~GeV) for $\eta_0=1$,
increasing to $M_V>420$~GeV for $\eta_0=2$. In estimating this effect
we have assumed the box diagram for these vectors is precisely as for
the case of $W$'s in the SM, because the GIM mechanism renders the
diagram finite even in unitary gauge.

 In the $D-\bar D$ case, in
contrast to the $B-\bar B$ case, the long distance contribution is
believed to dominate the double charge current exchange
amplitude. Since this is not computable, we settle for estimating
$\Delta_{\rm CC}$ by comparing the perturbative box diagram to the
corresponding one in the SM. Since all internal quarks are light the
computation simplifies: for example, in model ${\rm IV_s}$ the ratio
is
\begin{equation}
\frac{{\rm NP_{Box}}}{{\rm SM_{Box}}}=
\left(\frac{\eta_0}{g_2}\right)^4 \left(\frac{M_W}{M_V}\right)^4
\approx\left(\frac{190 \eta_0}{M_V({\rm GeV})}\right)^4\,,
\end{equation}
so that, for example,  the new physics is a factor of 10 suppressed relative the SM
for $M_V/\eta_0\gtrsim 340$~GeV. 

\subsection{Models X and XI}

The vector models give a contribution to the effective Hamiltonian for $B_s$ mixing that at the scale $\mu\sim m_V$ takes the form
\beq
\Delta {\cal H}_{\rm eff}^{{\rm NP}, B_s}=\frac{2 \kappa_s}{M_V^2} (y_t^2 V_{tb} V_{ts}^*)^2 \left(Q_4-Q_5\right),
\eeq
where the operators $Q_{4,5}$ are given in \eqref{Qi}. The contributions to $B_d$, $K$ and $D$ mixing are defined in the same way, with appropriate replacement of the CKM factors and quark fields, as in \eqref{HeffBd}, \eqref{HeffD}. The above expression is for color triplets, ${\rm X_{\bar 3}, XI_{\bar 3}}$. The result for color sextets,  ${\rm X_{6}, XI_{6}}$, is obtained with a replacement $(Q_4-Q_5)\to -(Q_4+Q_5)/2$. The factors $\kappa_i$ are collected in Table \ref{kappa-table2}. 

To obtain the low energy predictions, we include the RG running  using the equations and numerical values for the matrix elements given in \cite{Bona:2007vi}. As before, we use the updated values of CKM elements from CKMFitter, ICHEP2010 update \cite{Charles:2004jd}. We obtain
\begin{align}
h_s e^{i 2\sigma_s}&\simeq 1.7(-2.1)\times y_t^4 \frac{\kappa_s}{y_s}  \left(\frac{y_s}{0.02}\right)\left(\frac{1{\rm ~TeV}}{m_V}\right)^2,\\
h_d e^{i 2\sigma_d}&\simeq0.09(-0.11)\times y_t^4 \frac{\kappa_d}{y_d}  \left(\frac{y_d}{10^{-3}}\right)\left(\frac{1{\rm ~TeV}}{m_V}\right)^2,\\
h_K e^{i 2\sigma_K}&\simeq 5 (-4) \cdot 10^{-3}\times  e^{i2\beta}  y_t^4\frac{\kappa_K}{y_d y_s}   \left(\frac{y_s}{0.02}\right)\left(\frac{y_d}{10^{-3}}\right)\left(\frac{1{\rm ~TeV}}{m_V}\right)^2,\\
h_D e^{i 2\sigma_D}&\simeq 1.4 (-1.5) \cdot 10^{-8}\times e^{i 2\gamma} y_b^4 \frac{\kappa_D}{y_u y_c}   \left(\frac{y_u}{10^{-5}}\right) \left(\frac{y_c}{5\cdot 10^{-3}}\right)\left(\frac{1{\rm ~TeV}}{m_V}\right)^2,
\end{align}
where the first values are for color triplet models and in the brackets for the color sextet models. 

\renewcommand\arraystretch{1.5}
\begin{table}[t]
\begin{center}
\begin{tabular}[t]{c|ccc}
\hline\hline
  \multirow{2}{*}{Case}   & \multicolumn{3}{c}{$\Delta\mathcal{L}$}\\[-4mm]
  & $\hspace{2.1cm}\kappa_{s,d}\hspace{2.3cm}$ & $\hspace{2.3cm}\kappa_{K}\hspace{2.1cm}$&$\hspace{2.1cm}\kappa_{D}\hspace{2.1cm}$\\ 
  \hline\hline
 \multirow{2}{*}{$\rm X_{\bar 3,6}$}  &         \multicolumn{3}{c}{$\eta_1 \bar d_R  \slashed V Q_L^c + \eta_2 \bar d_R  \slashed V \Delta_U^* Q_L^c+\eta_3 \bar d_R (Y_D \Delta_U Y_D^\dagger) \slashed V Q_L^c+\eta_4 \bar d_R (Y_D \Delta_U Y_D^\dagger) \slashed V \Delta_U^* Q_L^c +h.c.$}     \\ 
& $y_{s,d}\,y_b\left[\frac{\eta_1\eta_4^*}{r_{ii}^2}+\frac{\tilde \eta_1^*\eta_4}{r_{33}^2}+\frac{\eta_2\tilde \eta_3^*}{r_{i3}}+\frac{\tilde \eta_2^*\eta_3}{r_{3i}}\right]$   & 
$y_{s}\,y_d\left[\frac{\eta_1\eta_4^*+2 \Re \eta_2\eta_3^*}{r_{ii}^2}+\frac{\tilde \eta_1^*\eta_4}{r_{33}^2}\right]$   & $0$ \\
\hline
 \multirow{2}{*}{$\rm XI_{\bar 3,6}$}  &         \multicolumn{3}{c}{$\eta_1 \bar u_R  \slashed V Q_L^c + \eta_2 \bar u_R  \slashed V \Delta_D^* Q_L^c+\eta_3 \bar u_R (Y_U \Delta_D Y_U^\dagger) \slashed V Q_L^c+\eta_4 \bar u_R (Y_U \Delta_D Y_U^\dagger) \slashed V \Delta_D^* Q_L^c +h.c.$}     \\ 
& $0$  & $0$  & $y_{u}\,y_c\left[\frac{\eta_1^*\eta_4+2\Re \eta_2\eta_3^*}{r_{ii}^2}+\frac{\tilde \eta_1'\eta_4^*}{r_{33}^2}\right]$\\
\hline
\hline
\end{tabular}
\end{center}
\caption{The same as Table \ref{kappa-table1}, but for models X and XI. We further defined $\tilde \eta_1=\eta_1+\eta_2 y_t^2+\eta_3 y_t^2 y_b^2+\eta_4 y_t^4 y_b^2$, $\tilde \eta_2=\eta_2+\eta_4 y_t^2$, $\tilde \eta_3=\eta_3+\eta_4 y_t^2$,  and $\tilde \eta_1'=\eta_1+\eta_2 y_b^2+\eta_3 y_b^2 y_t^2+\eta_4 y_b^4 y_t^2$. The results are written in the limit of degenerate vectors mediating transitions between the first two generations. Above the index $i$ stands for $i=1,2$.}
\label{kappa-table2}
\end{table}

We make the following observations:
\begin{itemize}
\item
Models X and XI are of the "Yukawa suppressed" category, but the class-2 operators that they give rise to are linear in $y_s$, $y_d$ suppression, which is quite atypical. Commonly the suppression in class-2 operators is quadratic in light Yukawas. The ratios of $\kappa_i$ for models X are
\beq
\kappa_d:\kappa_s:\kappa_K=y_d:y_s:\frac{y_s  y_d}{y_b},
\eeq
where the ratio with $\kappa_K$ is only approximate. 
\item
Models ${\rm X_{\bar 3, 6}}$ offer the possibility to explain $A_{FB}^{t\bar t}$ and $B_s$ mixing anomaly at the same time, if flavor violating couplings $\eta_i$ are a factor few smaller than the corresponding couplings in the flavor conserving part of the Lagrangian.
\end{itemize}

\subsection{Models ${\rm S_{I-IV}}$}
The models ${\rm S_{I-IV}}$ give contributions to $B_s$ mixing of the form
\beq
\Delta {\cal H}_{\rm eff}^{{\rm NP}, B_s}=-\frac{\kappa_s}{M_S^2} (y_t^2 V_{tb} V_{ts}^*)^2 Q_4,
\eeq
where the operators $Q_{4,5}$ are given in \eqref{Qi}. The contributions to $B_d$, $K$ and $D$ mixing are defined in the same way, with appropriate replacement of the CKM factors and quark fields, as in \eqref{HeffBd}, \eqref{HeffD}. The above expression is for models $\rm S_I$ and $\rm S_{III}$, while for models $\rm S_{II}$ and $\rm S_{IV}$ the result is obtained with a replacement $Q_4\to -Q_4/6+Q_5/2$. The factors $\kappa_i$ are collected in Table \ref{kappa-table3}. Numerically these models are very similar to the MFV models $\rm X$ and $\rm XI$,
\begin{align}
-h_s e^{i 2\sigma_s}&\simeq 1.5(0.05)\times y_t^4 \frac{\kappa_s}{y_s}  \left(\frac{y_s}{0.02}\right)\left(\frac{1{\rm ~TeV}}{m_S}\right)^2,\\
-h_d e^{i 2\sigma_d}&\simeq7(0.3)\cdot 10^{-2}\times y_t^4 \frac{\kappa_d}{y_d}  \left(\frac{y_d}{10^{-3}}\right)\left(\frac{1{\rm ~TeV}}{m_S}\right)^2,\\
-h_K e^{i 2\sigma_K}&\simeq 3.6(-0.1) \cdot 10^{-3}\times e^{i 2\beta} y_t^4 \frac{\kappa_K}{y_d y_s}   \left(\frac{y_s}{0.02}\right)\left(\frac{y_d}{10^{-3}}\right)\left(\frac{1{\rm ~TeV}}{m_S}\right)^2,\\
-h_D e^{i 2\sigma_D}&\simeq 1.1( 0.01)\cdot 10^{-8}\times e^{i 2\gamma} y_b^4 \frac{\kappa_D}{y_u y_c}   \left(\frac{y_u}{10^{-5}}\right) \left(\frac{y_c}{5\cdot 10^{-3}}\right)\left(\frac{1{\rm ~TeV}}{m_S}\right)^2,
\end{align}
where the values in the brackets are for the models $\rm S_{II}$ and $\rm S_{IV}$, for which numerical cancellation between the matrix elements of the two operators occurs. 

\renewcommand\arraystretch{1.5}
\begin{table}[t]
\begin{center}
\begin{tabular}[t]{c|ccc}
\hline\hline
  \multirow{2}{*}{Case}   & \multicolumn{3}{c}{$\Delta\mathcal{L}$}\\[-4mm]
  & $\hspace{2.1cm}\kappa_{s,d}\hspace{2.3cm}$ & $\hspace{2.3cm}\kappa_{K}\hspace{2.1cm}$&$\hspace{2.1cm}\kappa_{D}\hspace{2.1cm}$\\ 
  \hline\hline
 \multirow{2}{*}{$\rm I, II$}  &         \multicolumn{3}{c}{$\eta_0 \bar u_R  S Q_L + \eta_1 \bar u_R S \Delta_D Q_L+\eta_2 \bar u_R (Y_U \Delta_D Y_U^\dagger) S Q_L+\eta_3 \bar u_R (Y_U \Delta_D Y_U^\dagger) S \Delta_D Q_L +h.c.$}     \\ 
& $0$   & $0$   & $y_u y_c \big[2 \frac{ \Re(\eta_0 \eta_3^*+\eta_1 \eta_2^*)}{r_{ii}^2}+\frac{|y_t\eta_3|^2}{r_{33}^2}\big]$ \\
\hline
  \multirow{2}{*}{$\rm III, IV$}  &         \multicolumn{3}{c}{$\eta_0 \bar d_R  S Q_L + \eta_1 \bar d_R S \Delta_U Q_L+\eta_2 \bar d_R (Y_D \Delta_U Y_D^\dagger) S Q_L+\eta_3 \bar d_R (Y_D \Delta_U Y_D^\dagger) S \Delta_U Q_L +h.c.$}     \\ 
& $y_{d,s} y_b \big[\frac{\eta_1 \eta_2^*}{r_{ii}^2}+\frac{\tilde \eta_0\eta_3^*}{r_{i3}^2}+\frac{\eta_3\eta_0^*}{r_{3i}^2}+\frac{\tilde \eta_2 \tilde \eta_1^*}{r_{33}^2}\big]$   & $y_{d} y_s \big[2 \frac{ \Re(\eta_0 \eta_3^*+\eta_1 \eta_2^*)}{r_{ii}^2}+\frac{|y_b\eta_3|^2}{r_{33}^2}\big]$   & $0$ \\
\hline
\hline
\end{tabular}
\end{center}
\caption{The same as Table \ref{kappa-table1}, but for models  ${\rm S_{I-IV}}$. We also use the abbreviations $\tilde \eta_0=\eta_0+\eta_1 y_t^2$, $\tilde \eta_1=\eta_1+\eta_3 y_b^2 y_t^2$, $\tilde \eta_2=\eta_2+\eta_3 y_t^2$. The results are written in the limit of degenerate scalars mediating transitions between the first two generations. Above the index $i$ stands for $i=1,2$. In all the models there is also a 1-loop contribution $\Delta_{CC}$ (not displayed).}
\label{kappa-table3}
\end{table}

\subsection{Models  ${\rm S_{V-XIV}}$}
Of the remaining models only models ${\rm S_{VI,VIII,XIV}}$ lead to tree level FCNCs. The models $\rm S_{V,VII, IX, XI}$ and $\rm S_{XIII}$ do not lead to tree level FCNCs because of its color triplet structure (and thus antisymmetric contraction of indices). The models $\rm S_X$ and $\rm S_{XII}$ lead to charge currents and thus give FCNCs at loop level only. 

We next focus on the models that lead to FCNCs at tree level. They give contributions to $B_s$ mixing of the form
\beq
\Delta {\cal H}_{\rm eff}^{{\rm NP}, B_s}=-\frac{4 \kappa_s}{M_S^2} (y_t^2 V_{tb} V_{ts}^*)^2 \tilde Q_1,
\eeq
where the operator $\tilde Q_{1}$ is replaced by $Q_1$ for model $\rm S_{XIV}$. The contributions to $B_d$, $K$ and $D$ mixing are defined in the same way, with appropriate replacement of the CKM factors and quark fields, as in \eqref{HeffBd}, \eqref{HeffD}. The factors $\kappa_i$ are collected in Table \ref{kappa-table4}. Numerically the analysis is exactly the same as for MFV vector models $\rm S_{I-X}$. The Yukawa suppressed (class-2 operator) models are $\rm S_{VI}$ and $\rm S_{VIII}$, while $\rm S_{XIV}$ is an example of the universal model (class-1 operator model). 

\renewcommand\arraystretch{1.5}
\begin{table}[t]
\begin{center}
\begin{tabular}[t]{c|ccc}
\hline\hline
  \multirow{2}{*}{Case}   & \multicolumn{3}{c}{$\Delta\mathcal{L}$}\\[-4mm]
  & $\hspace{2.1cm}\kappa_{s,d}\hspace{2.3cm}$ & $\hspace{2.3cm}\kappa_{K}\hspace{2.1cm}$&$\hspace{2.1cm}\kappa_{D}\hspace{2.1cm}$\\ 
  \hline\hline
 \multirow{2}{*}{$\rm VI$}  &         \multicolumn{3}{c}{$\eta_0 \bar u_R^c  S u_R + \eta_1 \bar u_R^c S (Y_U \Delta_D Y_U^\dagger) u_R+\eta_1' \bar u_R^c (Y_U \Delta_D Y_U^\dagger)^T S u_R+\eta_2 \bar u_R^c (Y_U \Delta_D Y_U^\dagger)^T S (Y_U \Delta_D Y_U^\dagger) u_R +h.c.$}     \\ 
& $0$   & $0$   & $2 \Re (\eta_1\eta_1'{}^*+\eta_0\eta_2'{}^*) y_u^2 y_c^2$ \\
\hline
  \multirow{2}{*}{$\rm VIII$}  &         \multicolumn{3}{c}{$\eta_0 \bar d_R^c  S d_R + \eta_1 \bar d_R^c S (Y_D \Delta_U Y_D^\dagger) d_R+\eta_1' \bar d_R^c (Y_D \Delta_U Y_D^\dagger)^T S d_R+\eta_2 \bar d_R^c (Y_D \Delta_U Y_D^\dagger)^T S (Y_D \Delta_U Y_D^\dagger) d_R +h.c.$}     \\ 
& $y_{d,s}^2 y_b^2 \big[\frac{2\Re \eta_1 \eta_1'{}^*}{r_{i3}^2}+\frac{\eta_0\eta_2^*}{r_{ii}^2}+\frac{\eta_2\eta_0^*}{r_{33}^2}\big]$   & $y_{d}^2 y_s^2 2\Re( \eta_1 \eta_1'{}^*+ \eta_0\eta_2^*)$   & $0$ \\
\hline
  \multirow{2}{*}{$\rm XIV$}  &         \multicolumn{3}{c}{$\eta_0 \bar Q_L^c  S Q_L + \eta_1 \bar Q_L^c S \Delta_U Q_L+\eta_1' \bar Q_L^c  \Delta_U^T S Q_L+\eta_2 \bar Q_L^c  \Delta_U^T S \Delta_U Q_L +h.c.$}     \\ 
& $ \big[\frac{2\Re \eta_1 \eta_1'{}^*}{r_{i3}^2}+\frac{\eta_0\eta_2^*}{r_{ii}^2}+\frac{\eta_2\eta_0^*}{r_{33}^2}\big]$   & $2\Re( \eta_1 \eta_1'{}^*+ \eta_0\eta_2^*)$   & $2\Re( \tilde \eta_1 \tilde \eta_1'{}^*+ \eta_0\tilde \eta_2^*)$ \\
\hline
\hline
\end{tabular}
\end{center}
\caption{The same as Table \ref{kappa-table1}, but for models $\rm S_{V-XIV}$. Only models that lead to tree level FCNCs are shown. The results are written in the limit of degenerate scalars mediating transitions between the first two generations. Above the index $i$ stands for $i=1,2$. For model $\rm S_{XIV}$ the terms that contribute to $D-\bar D$ mixing have not been written out explicitly in $\Delta {\cal L}$. They can be obtained with $\Delta_U\to \Delta_D$ and have coupling $\tilde\eta_i$.}
\label{kappa-table4}
\end{table}

\subsection{Models ${\rm S_{H,8}}$}
The models ${\rm S_{H,8}}$ give contributions to $B_s$ mixing of the form
\beq
\Delta {\cal H}_{\rm eff}^{{\rm NP}, B_s}=\kappa_s \left(\frac{1}{m_{S_1}^2} -\frac{1}{m_{S_2}^2}\right) (y_t^2 V_{tb} V_{ts}^*)^2 (\tilde C_2 \tilde Q_2 + \tilde C_3 \tilde Q_3),
\eeq
where the operators $\tilde Q_{2,3}$ are given in \eqref{Qi}, $m_{S_{1,2}}$ are the masses of CP even and odd neutral scalars, and $\tilde C_2=1/2(1/4)$, $\tilde C_3=0(-1/12)$ for models ${\rm S_{H}}$ (${\rm S_{8}}$). The contributions to $B_d$, $K$ and $D$ mixing are defined in the same way, with appropriate replacement of the CKM factors and quark fields, as in \eqref{HeffBd}, \eqref{HeffD}. Numerically (taking $m_{S_2}\gg m_{S_1}$ limit for simplicity),
\begin{align}
h_{s,d} e^{i 2\sigma_{s,d}}&\simeq -15 (5)\times y_t^4 \kappa_{s,d}  \left(\frac{1{\rm ~TeV}}{m_{S_1}}\right)^2,\\
h_K e^{i 2\sigma_K}&\simeq 1.1(-0.3) \cdot 10^{-2}\times e^{i 2\beta} y_t^4 \frac{\kappa_K}{y_s^2}   \left(\frac{y_s}{0.02}\right)^2 \left(\frac{1{\rm ~TeV}}{m_{S_1}}\right)^2,\\
h_D e^{i 2\sigma_D}&\simeq -1.1( 0.3)\cdot 10^{-6}\times e^{i 2\gamma} y_b^4 \frac{\kappa_D}{y_c^2}   \left(\frac{y_c}{5\cdot 10^{-3}}\right)^2 \left(\frac{1{\rm ~TeV}}{m_{S_1}}\right)^2,
\end{align}
where the values in the brackets are for model $\rm S_{8}$. 

\renewcommand\arraystretch{1.5}
\begin{table}[t]
\begin{center}
\begin{tabular}[t]{c|ccc}
\hline\hline
  \multirow{2}{*}{Case}   & \multicolumn{3}{c}{$\Delta\mathcal{L}$}\\[-4mm]
  & $\hspace{2.1cm}\kappa_{s,d}\hspace{2.3cm}$ & $\hspace{2.3cm}\kappa_{K}\hspace{2.1cm}$&$\hspace{2.1cm}\kappa_{D}\hspace{2.1cm}$\\
  \hline\hline
 \multirow{2}{*}{$\rm H, 8$}  &         \multicolumn{3}{c}{$\eta_0 \bar d_R  Y_D S Q_L+ \eta_0' \bar u_R  Y_U S Q_L   + \eta_1 \bar d_R (Y_D \Delta_U )S Q_L+ \eta_1' \bar u_R (Y_U \Delta_D )S Q_L+h.c.$}     \\ 
& $y_b^2 \eta_1^2+\Delta_{CC}$   & $y_s^2 \eta_1^2+\Delta_{CC}$   & $y_c^2 \eta_1'{}^2+\Delta_{CC}$ \\
\hline
\hline
\end{tabular}
\end{center}
\caption{The same as Table \ref{kappa-table1}, but for models  ${\rm S_{H,8}}$.  The 1-loop contributions $\Delta_{CC}$ arise from charge currents.}
\label{kappa-table5}
\end{table}




\begin{thebibliography}{99}




\bibitem{Glashow:1976nt}
  S.~L.~Glashow and S.~Weinberg,
  Phys.\ Rev.\  D {\bf 15}, 1958 (1977).
  
  
\bibitem{Agashe:2005hk}
  K.~Agashe, M.~Papucci, G.~Perez and D.~Pirjol,
  arXiv:hep-ph/0509117.

  
\bibitem{Arnold:2009ay}
  J.~M.~Arnold, M.~Pospelov, M.~Trott and M.~B.~Wise,
  JHEP {\bf 1001}, 073 (2010)
  [arXiv:0911.2225 [hep-ph]].
  
\bibitem{Grossman:2007bd}
  Y.~Grossman, Y.~Nir, J.~Thaler, T.~Volansky and J.~Zupan,
  Phys.\ Rev.\  D {\bf 76}, 096006 (2007)
  [arXiv:0706.1845 [hep-ph]].

\bibitem{Arnold:2010vs}
  J.~M.~Arnold, B.~Fornal and M.~Trott,
  JHEP {\bf 1008}, 059 (2010)
  [arXiv:1005.2185 [hep-ph]].
   
\bibitem{Aaltonen:2011kc}
  T.~Aaltonen {\it et al.}  [The CDF Collaboration],
  arXiv:1101.0034 [hep-ex].
  
 
\bibitem{Abazov:2011rq}
  V.~M.~Abazov {\it et al.} [ D0 Collaboration ],
  [arXiv:1107.4995 [hep-ex]].  
  
\bibitem{CDF-dilepton} 
CDF Collaboration, CDF note 10398, {\rm http://www-cdf.fnal.gov/physics/new/top/2011/DilAfb/}


  
\bibitem{Abazov:2010hv}
  V.~M.~Abazov {\it et al.}  [D0 Collaboration],
  arXiv:1005.2757.

\bibitem{Williams:2011nc}
  M.~R.~J.~Williams {\it et al.} [on behalf of D0 Collaboration],
  [arXiv: 1106.6308 [hep-ex]].


\bibitem{Chivukula:1987py}
  R.~S.~Chivukula and H.~Georgi,
  Phys.\ Lett.\  B {\bf 188} (1987) 99.

  \bibitem{D'Ambrosio:2002ex}
  G.~D'Ambrosio et al.,
  Nucl.\ Phys.\  B {\bf 645} (2002) 155
  [arXiv:hep-ph/0207036].


  \bibitem{Kagan:2009bn}
  A.~L.~Kagan, G.~Perez, T.~Volansky and J.~Zupan,
  arXiv:0903.1794 [hep-ph].


\bibitem{Manohar:2006ga}
  A.~V.~Manohar and M.~B.~Wise,
  Phys.\ Rev.\  D {\bf 74}, 035009 (2006)
  [arXiv:hep-ph/0606172].
  
  
\bibitem{Cao:2010zb}
  Q.~H.~Cao, D.~McKeen, J.~L.~Rosner, G.~Shaughnessy and C.~E.~M.~Wagner,
  Phys.\ Rev.\  D {\bf 81}, 114004 (2010)
  [arXiv:1003.3461 [hep-ph]].
  
\bibitem{Blum:2011up}
  K.~Blum, C.~Delaunay, O.~Gedalia, Y.~Hochberg, S.~J.~Lee, Y.~Nir, G.~Perez, Y.~Soreq,
  [arXiv:1102.3133 [hep-ph]].
  
\bibitem{Tavares:2011zg}
  G.~M.~Tavares, M.~Schmaltz,
  [arXiv:1107.0978 [hep-ph]].



\bibitem{Shu:2009xf}
  J.~Shu, T.~M.~P.~Tait and K.~Wang,
  Phys.\ Rev.\  D {\bf 81}, 034012 (2010)
  [arXiv:0911.3237].


  \bibitem{Feldmann:2008ja}
  T.~Feldmann and T.~Mannel,
  Phys.\ Rev.\ Lett.\  {\bf 100} (2008) 171601
  [arXiv:0801.1802 [hep-ph]].

  \bibitem{Feldmann:2009dc}
  T.~Feldmann, M.~Jung and T.~Mannel,
  arXiv:0906.1523 [hep-ph].
  
\bibitem{Barbieri:2011ci}
R.~Barbieri, G.~Isidori, J.~Jones-Perez, P.~Lodone and D.~M.~Straub,
arXiv:1105.2296 [hep-ph].


\bibitem{Jung:2009jz}
  S.~Jung, H.~Murayama, A.~Pierce and J.~D.~Wells,
  Phys.\ Rev.\  D {\bf 81}, 015004 (2010)
  [arXiv:0907.4112 ].
  
  \bibitem{Cheung:2009ch}
  K.~Cheung, W.~Y.~Keung and T.~C.~Yuan,
  Phys.\ Lett.\  B {\bf 682}, 287 (2009)
  [arXiv:0908.2589 ].
  
\bibitem{Jung:2011id}
S.~Jung, A.~Pierce and J.~D.~Wells,
arXiv:1108.1802 [hep-ph].
  
  \bibitem{Ferrario:2009bz}
  P.~Ferrario, G.~Rodrigo,
  Phys.\ Rev.\  {\bf D80}, 051701 (2009).
  [arXiv:0906.5541 [hep-ph]].

  
\bibitem{Frampton:2009rk}
  P.~H.~Frampton, J.~Shu and K.~Wang,
  Phys.\ Lett.\  B {\bf 683}, 294 (2010)
  [arXiv:0911.2955 [hep-ph]].

  
\bibitem{Arhrib:2009hu}
  A.~Arhrib, R.~Benbrik and C.~H.~Chen,
  Phys.\ Rev.\  D {\bf 82}, 034034 (2010)
  [arXiv:0911.4875 ].

\bibitem{Dorsner:2009mq}
  I.~Dorsner, S.~Fajfer, J.~F.~Kamenik and N.~Kosnik,
  Phys.\ Rev.\  D {\bf 81}, 055009 (2010)
  [arXiv:0912.0972 [hep-ph]].


\bibitem{Kamenik:2011wt}
J.~F.~Kamenik, J.~Shu and J.~Zupan,
arXiv:1107.5257 [hep-ph].

\bibitem{Jung:2011zv}
S.~Jung, A.~Pierce and J.~D.~Wells,
arXiv:1103.4835 [hep-ph].

\bibitem{Shelton:2011hq}
J.~Shelton and K.~M.~Zurek,
Phys.\ Rev.\  D {\bf 83} (2011) 091701
[arXiv:1101.5392 [hep-ph]].

\bibitem{Nelson:2011us}
  A.~E.~Nelson, T.~Okui and T.~S.~Roy,
  arXiv:1104.2030 [hep-ph].


\bibitem{Babu:2011yw}
  K.~S.~Babu, M.~Frank and S.~K.~Rai,
  Phys.\ Rev.\ Lett.\  {\bf 107}, 061802 (2011)
  [arXiv:1104.4782 [hep-ph]].


\bibitem{Shu:2011au}
J.~Shu, K.~Wang and G.~Zhu,
arXiv:1104.0083 [hep-ph].

\bibitem{Grinstein:2011yv}
B.~Grinstein, A.~L.~Kagan, M.~Trott and J.~Zupan,
Phys.\ Rev.\ Lett.\  {\bf 107} (2011) 012002
[arXiv:1102.3374 [hep-ph]].

\bibitem{Ligeti:2011vt}
  Z.~Ligeti, M.~Schmaltz and G.~M.~Tavares,
  arXiv:1103.2757 [hep-ph].

\bibitem{Blum:2011fa}
  K.~Blum, Y.~Hochberg and Y.~Nir,
  arXiv:1107.4350 [hep-ph].

\bibitem{Gresham:2011dg}
M.~I.~Gresham, I.~W.~Kim and K.~M.~Zurek,
arXiv:1102.0018 [hep-ph].

\bibitem{Cheung:2011qa}
  K.~Cheung and T.~C.~Yuan,
  Phys.\ Rev.\  D {\bf 83}, 074006 (2011)
  [arXiv:1101.1445 [hep-ph]].
  

\bibitem{Barger:2010mw}
  V.~Barger, W.~Y.~Keung and C.~T.~Yu,
  Phys.\ Rev.\  D {\bf 81}, 113009 (2010)
  [arXiv:1002.1048 [hep-ph]].
  
 
\bibitem{Barger:2011ih}
  V.~Barger, W.~Y.~Keung and C.~T.~Yu,
  Phys.\ Lett.\  B {\bf 698}, 243 (2011)
  [arXiv:1102.0279 [hep-ph]].
  

\bibitem{Berger:2011ua}
  E.~L.~Berger, Q.~H.~Cao, C.~R.~Chen, C.~S.~Li and H.~Zhang,
  Phys.\ Rev.\ Lett.\  {\bf 106}, 201801 (2011)
  [arXiv:1101.5625  ].
\bibitem{Bhattacherjee:2011nr}
  B.~Bhattacherjee, S.~S.~Biswal and D.~Ghosh,
  Phys.\ Rev.\  D {\bf 83}, 091501 (2011)
  [arXiv:1102.0545 [hep-ph]].
  
\bibitem{Gresham:2011fx}
M.~I.~Gresham, I.~W.~Kim and K.~M.~Zurek,
arXiv:1107.4364 [hep-ph].

\bibitem{Degrande:2010kt}
  C.~Degrande, J.~-M.~Gerard, C.~Grojean, F.~Maltoni, G.~Servant,
  JHEP {\bf 1103}, 125 (2011).
  [arXiv:1010.6304 [hep-ph]].
  
\bibitem{AguilarSaavedra:2011vw}
  J.~A.~Aguilar-Saavedra, M.~Perez-Victoria,
  JHEP {\bf 1105}, 034 (2011).
  [arXiv:1103.2765 [hep-ph]].

\bibitem{Cheung:1995nt}
  K.~-m.~Cheung,
  Phys.\ Rev.\  {\bf D53}, 3604-3615 (1996).
  [hep-ph/9511260].

\bibitem{Antipin:2008zx}
  O.~Antipin, G.~Valencia,
  Phys.\ Rev.\  {\bf D79}, 013013 (2009).
  [arXiv:0807.1295 [hep-ph]].

\bibitem{Gupta:2009wu}
  S.~K.~Gupta, A.~S.~Mete, G.~Valencia,
  Phys.\ Rev.\  {\bf D80}, 034013 (2009).
  [arXiv:0905.1074 [hep-ph]].

\bibitem{Hioki:2009hm}
  Z.~Hioki, K.~Ohkuma,
  Eur.\ Phys.\ J.\  {\bf C65}, 127-135 (2010).
  [arXiv:0910.3049 [hep-ph]].

\bibitem{Choudhury:2009wd}
  D.~Choudhury, P.~Saha,
  [arXiv:0911.5016 [hep-ph]].

\bibitem{Hioki:2010zu}
  Z.~Hioki, K.~Ohkuma,
  Eur.\ Phys.\ J.\  {\bf C71}, 1535 (2011).
  [arXiv:1011.2655 [hep-ph]].

\bibitem{HIOKI:2011xx}
  Z.~HIOKI, K.~OHKUMA,
  Phys.\ Rev.\  {\bf D83}, 114045 (2011).
  [arXiv:1104.1221 [hep-ph]].

\bibitem{Jung:2009pi}
  D.~-W.~Jung, P.~Ko, J.~S.~Lee, S.~-h.~Nam,
  Phys.\ Lett.\  {\bf B691}, 238-242 (2010).
  [arXiv:0912.1105 [hep-ph]].

\bibitem{Zhang:2010dr}
  C.~Zhang, S.~Willenbrock,
  Phys.\ Rev.\  {\bf D83}, 034006 (2011).
  [arXiv:1008.3869 [hep-ph]].

\bibitem{Delaunay:2011gv}
  C.~Delaunay, O.~Gedalia, Y.~Hochberg, G.~Perez, Y.~Soreq,
  JHEP {\bf 1108}, 031 (2011).
  [arXiv:1103.2297 [hep-ph]].

\bibitem{Cao:2009uz}
  J.~Cao, Z.~Heng, L.~Wu, J.~M.~Yang,
  Phys.\ Rev.\  {\bf D81}, 014016 (2010).
  [arXiv:0912.1447 [hep-ph]].

\bibitem{Xiao:2010hm}
  B.~Xiao, Y.~-k.~Wang, S.~-h.~Zhu,
  Phys.\ Rev.\  {\bf D82}, 034026 (2010).
  [arXiv:1006.2510 [hep-ph]].

\bibitem{Cao:2011ew}
  J.~Cao, L.~Wang, L.~Wu, J.~M.~Yang,
  [arXiv:1101.4456 [hep-ph]].

\bibitem{Patel:2011eh}
  K.~M.~Patel, P.~Sharma,
  JHEP {\bf 1104}, 085 (2011).
  [arXiv:1102.4736 [hep-ph]].

\bibitem{Barreto:2011au}
  E.~R.~Barreto, Y.~A.~Coutinho, J.~Sa Borges,
  Phys.\ Rev.\  {\bf D83}, 054006 (2011).
  [arXiv:1103.1266 [hep-ph]].

\bibitem{Craig:2011an}
  N.~Craig, C.~Kilic, M.~J.~Strassler,
  [arXiv:1103.2127 [hep-ph]].

\bibitem{Buckley:2011vc}
  M.~R.~Buckley, D.~Hooper, J.~Kopp, E.~Neil,
  Phys.\ Rev.\  {\bf D83}, 115013 (2011).
  [arXiv:1103.6035 [hep-ph]].

\bibitem{Jung:2011ua}
  S.~Jung, A.~Pierce, J.~D.~Wells,
  [arXiv:1104.3139 [hep-ph]].
  
  
\bibitem{Fox:2011qd}
  P.~J.~Fox, J.~Liu, D.~Tucker-Smith, N.~Weiner,
  [arXiv:1104.4127 [hep-ph]].

\bibitem{Cui:2011xy}
  Y.~Cui, Z.~Han, M.~D.~Schwartz,
  JHEP {\bf 1107}, 127 (2011).
  [arXiv:1106.3086 [hep-ph]].

\bibitem{Duraisamy:2011pt}
  M.~Duraisamy, A.~Rashed, A.~Datta,
  [arXiv:1106.5982 [hep-ph]].

\bibitem{AguilarSaavedra:2011ug}
  J.~A.~Aguilar-Saavedra, M.~Perez-Victoria,
  [arXiv:1107.0841 [hep-ph]].

\bibitem{Dorsner:2010cu}
  I.~Dorsner, S.~Fajfer, J.~F.~Kamenik, N.~Kosnik,
  Phys.\ Rev.\  {\bf D82}, 094015 (2010).
  [arXiv:1007.2604 [hep-ph]].

\bibitem{Delaunay:2011vv}
  C.~Delaunay, O.~Gedalia, S.~J.~Lee, G.~Perez, E.~Ponton,
  [arXiv:1101.2902 [hep-ph]].

\bibitem{Isidori:2011dp}
  G.~Isidori, J.~F.~Kamenik,
  Phys.\ Lett.\  {\bf B700}, 145-149 (2011).
  [arXiv:1103.0016 [hep-ph]].

\bibitem{Chivukula:2010fk}
  R.~S.~Chivukula, E.~H.~Simmons, C.~-P.~Yuan,
  Phys.\ Rev.\  {\bf D82}, 094009 (2010).
  [arXiv:1007.0260 [hep-ph]].

\bibitem{Bai:2011ed}
  Y.~Bai, J.~L.~Hewett, J.~Kaplan, T.~G.~Rizzo,
  JHEP {\bf 1103}, 003 (2011).
  [arXiv:1101.5203 [hep-ph]].

\bibitem{Xiao:2010ph}
  B.~Xiao, Y.~-k.~Wang, S.~-h.~Zhu,
  [arXiv:1011.0152 [hep-ph]].

\bibitem{Ferrario:2009ee}
  P.~Ferrario, G.~Rodrigo,
  JHEP {\bf 1002}, 051 (2010).
  [arXiv:0912.0687 [hep-ph]].

\bibitem{Martynov:2010ed}
  M.~V.~Martynov, A.~D.~Smirnov,
  Mod.\ Phys.\ Lett.\  {\bf A25}, 2637-2643 (2010).
  [arXiv:1006.4246 [hep-ph]].

\bibitem{Bauer:2010iq}
  M.~Bauer, F.~Goertz, U.~Haisch, T.~Pfoh, S.~Westhoff,
  JHEP {\bf 1011}, 039 (2010).
  [arXiv:1008.0742 [hep-ph]].

\bibitem{Chen:2010hm}
  C.~-H.~Chen, G.~Cvetic, C.~S.~Kim,
  Phys.\ Lett.\  {\bf B694}, 393-397 (2011).
  [arXiv:1009.4165 [hep-ph]].

\bibitem{Burdman:2010gr}
  G.~Burdman, L.~de Lima, R.~D.~Matheus,
  Phys.\ Rev.\  {\bf D83}, 035012 (2011).
  [arXiv:1011.6380 [hep-ph]].

\bibitem{Choudhury:2010cd}
  D.~Choudhury, R.~M.~Godbole, S.~D.~Rindani, P.~Saha,
  Phys.\ Rev.\  {\bf D84}, 014023 (2011).
  [arXiv:1012.4750 [hep-ph]].

\bibitem{Cao:2010nw}
  J.~Cao, L.~Wu, J.~M.~Yang,
  Phys.\ Rev.\  {\bf D83}, 034024 (2011).
  [arXiv:1011.5564 [hep-ph]].

\bibitem{Foot:2011xu}
  R.~Foot,
  Phys.\ Rev.\  {\bf D83}, 114013 (2011).
  [arXiv:1103.1940 [hep-ph]].
  
\bibitem{Haisch:2011up}
U.~Haisch and S.~Westhoff,
arXiv:1106.0529 [hep-ph].


\bibitem{Dobrescu:2010rh}
  B.~A.~Dobrescu, P.~J.~Fox and A.~Martin,
  Phys.\ Rev.\ Lett.\  {\bf 105}, 041801 (2010)
  [arXiv:1005.4238 [hep-ph]].

\bibitem{Buras:2010mh}
  A.~J.~Buras, M.~V.~Carlucci, S.~Gori and G.~Isidori,
  JHEP {\bf 1010}, 009 (2010)
  [arXiv:1005.5310 [hep-ph]].
  
\bibitem{Jung:2010ik}
  M.~Jung, A.~Pich and P.~Tuzon,
  JHEP {\bf 1011}, 003 (2010)
  [arXiv:1006.0470 [hep-ph]].
  
\bibitem{Chen:2010aq}
  C.~H.~Chen, C.~Q.~Geng and W.~Wang,
  JHEP {\bf 1011}, 089 (2010)
  [arXiv:1006.5216 [hep-ph]].
  
\bibitem{Blum:2010mj}
  K.~Blum, Y.~Hochberg and Y.~Nir,
  JHEP {\bf 1009}, 035 (2010)
  [arXiv:1007.1872 [hep-ph]].
  
\bibitem{Buras:2010pz}
  A.~J.~Buras, K.~Gemmler and G.~Isidori,
  Nucl.\ Phys.\  B {\bf 843}, 107 (2011)
  [arXiv:1007.1993 [hep-ph]].
  
\bibitem{Buras:2010zm}
  A.~J.~Buras, G.~Isidori and P.~Paradisi,
  Phys.\ Lett.\  B {\bf 694}, 402 (2011)
  [arXiv:1007.5291 [hep-ph]].

\bibitem{Trott:2010iz}
  M.~Trott and M.~B.~Wise,
  JHEP {\bf 1011}, 157 (2010)
  [arXiv:1009.2813 [hep-ph]].
  
  

\bibitem{Ligeti:2010ia}
  Z.~Ligeti, M.~Papucci, G.~Perez and J.~Zupan,
  Phys.\ Rev.\ Lett.\  {\bf 105}, 131601 (2010)
  [arXiv:1006.0432 []].



  
\bibitem{Ahrens:2011uf}
V.~Ahrens, A.~Ferroglia, M.~Neubert, B.~D.~Pecjak and L.~L.~Yang,
arXiv:1106.6051 [hep-ph].



  \bibitem{Martin:2009iq}
  A.~D.~Martin, W.~J.~Stirling, R.~S.~Thorne and G.~Watt,
  Eur.\ Phys.\ J.\  C {\bf 63}, 189 (2009)
  [arXiv:0901.0002].

\bibitem{Hollik:2011ps}
  W.~Hollik and D.~Pagani,
  arXiv:1107.2606 [hep-ph].

\bibitem{Kidonakis:2011zn}
  N.~Kidonakis,
  [arXiv:1105.5167 [hep-ph]].

  
\bibitem{Antunano:2007da}
  O.~Antunano, J.~H.~Kuhn and G.~Rodrigo,
  Phys.\ Rev.\  D {\bf 77}, 014003 (2008)
  [arXiv:0709.1652 [hep-ph]].

\bibitem{Bowen:2005ap}
  M.~T.~Bowen, S.~D.~Ellis and D.~Rainwater,
  Phys.\ Rev.\  D {\bf 73}, 014008 (2006)
  [arXiv:hep-ph/0509267].

\bibitem{Kuhn:1998kw}
  J.~H.~Kuhn and G.~Rodrigo,
  Phys.\ Rev.\  D {\bf 59}, 054017 (1999)
  [arXiv:hep-ph/9807420].
  
  
  
\bibitem{Moch:2008ai}
  S.~Moch and P.~Uwer,
  Nucl.\ Phys.\ Proc.\ Suppl.\  {\bf 183}, 75 (2008)
  [arXiv:0807.2794 [hep-ph]].
  
  
\bibitem{Czakon:2009zw}
  M.~Czakon, A.~Mitov, G.~F.~Sterman,
  Phys.\ Rev.\  {\bf D80}, 074017 (2009).
  [arXiv:0907.1790 [hep-ph]].

\bibitem{Beneke:2009ye}
  M.~Beneke, M.~Czakon, P.~Falgari, A.~Mitov, C.~Schwinn,
  Phys.\ Lett.\  {\bf B690}, 483-490 (2010).
  [arXiv:0911.5166 [hep-ph]].

\bibitem{Kidonakis:2010dk}
  N.~Kidonakis,
  Phys.\ Rev.\  {\bf D82}, 114030 (2010).
  [arXiv:1009.4935 [hep-ph]].

  
\bibitem{Kidonakis:2008mu}
N.~Kidonakis and R.~Vogt,
Phys.\ Rev.\  D {\bf 78} (2008) 074005
[arXiv:0805.3844 [hep-ph]].

\bibitem{Cacciari:2008zb}
M.~Cacciari, S.~Frixione, M.~L.~Mangano, P.~Nason and G.~Ridolfi,
JHEP {\bf 0809} (2008) 127
[arXiv:0804.2800 [hep-ph]].


\bibitem{Ahrens:2011mw}
  V.~Ahrens, A.~Ferroglia, M.~Neubert, B.~D.~Pecjak, L.~L.~Yang,
  [arXiv:1103.0550 [hep-ph]].
    


\bibitem{Kidonakis:2011jg}
  N.~Kidonakis,
  [arXiv:1105.3481 [hep-ph]].
  
       
\bibitem{Aaltonen:2009iz}
T.~Aaltonen {\it et al.}  [CDF Collaboration],
Phys.\ Rev.\ Lett.\  {\bf 102} (2009) 222003
[arXiv:0903.2850 [hep-ex]].

  



\bibitem{Ahrens:2010zv}
V.~Ahrens, A.~Ferroglia, M.~Neubert, B.~D.~Pecjak and L.~L.~Yang,
JHEP {\bf 1009} (2010) 097
[arXiv:1003.5827].
 

\bibitem{Gresham:2011pa}
M.~I.~Gresham, I.~W.~Kim and K.~M.~Zurek,
arXiv:1103.3501 [hep-ph].


     
\bibitem{Alwall:2007st}
  J.~Alwall, P.~Demin, S.~de Visscher, R.~Frederix, M.~Herquet, F.~Maltoni, T.~Plehn, D.~L.~Rainwater {\it et al.},
  JHEP {\bf 0709}, 028 (2007).
  [arXiv:0706.2334 [hep-ph]].


\bibitem{Sjostrand:2006za}
  T.~Sjostrand, S.~Mrenna, P.~Z.~Skands,
  JHEP {\bf 0605}, 026 (2006).
  [hep-ph/0603175].


\bibitem{Aaltonen:2008hc}
T.~Aaltonen {\it et al.}  [CDF Collaboration],
Phys.\ Rev.\ Lett.\  {\bf 101} (2008) 202001
[arXiv:0806.2472 [hep-ex]].



  
  \bibitem{AguilarSaavedra:2011hz}
  J.~A.~Aguilar-Saavedra and M.~Perez-Victoria,
  arXiv:1105.4606 [hep-ph].

  
  \bibitem{Zhu:2011ww}
  G.~Zhu,
  arXiv:1104.3227 [hep-ph].

  
 \bibitem{Abazov:2011af}
  V.~M.~Abazov  [D0 Collaboration],
  Phys.\ Rev.\ Lett.\  {\bf 107}, 011804 (2011)
  [arXiv:1106.1921 [hep-ex]].

\bibitem{Aaltonen:2011mk}
  T.~Aaltonen {\it et al.} [ CDF Collaboration ],
  Phys.\ Rev.\ Lett.\  {\bf 106}, 171801 (2011).
  [arXiv:1104.0699 [hep-ex]]. 
  
  \bibitem{UA2} J. Alitti et al. [UA2 Collaboration], Nucl. Phys. B 400,
3 (1993).

\bibitem{atlasnote}
ATLAS Collaboration, ATLAS-CONF-2011-087,
http://cdsweb.cern.ch/record/1356196/files/ATLAS-CONF-2011-087.pdf

\bibitem{cmsnote}
CMS Collaboration, CMS-PAS-TOP-10-007, 
http://cdsweb.cern.ch/record/1335720?ln=en

\bibitem{CMS-Mulders-talk}
M. Mulders, talk on behalf of the CMS Collaboration presented  at EPS High Energy Phsyics Conference 2011, July 21-27 Grenoble, France. 

\bibitem{Lenz:2010gu}
A.~Lenz {\it et al.},
Phys.\ Rev.\  D {\bf 83} (2011) 036004
[arXiv:1008.1593 [hep-ph]].



\bibitem{CDF-10206}
 CDF Collaboration, CDF Note 10206
 
 

\bibitem{D0-6098-CONF}
 D0 Collaboration, D0 Note 6098-CONF
 
 
\bibitem{Burgess:2009wm}
  C.~P.~Burgess, M.~Trott and S.~Zuberi,
  JHEP {\bf 0909}, 082 (2009)
  [arXiv:0907.2696 [hep-ph]].


     
\bibitem{Carena:2004xs}
  M.~S.~Carena et al.
  Phys.\ Rev.\  D {\bf 70}, 093009 (2004)
  [arXiv:hep-ph/0408098].

\bibitem{Barbieri:2000gf}
  R.~Barbieri, A.~Strumia,
  [hep-ph/0007265].
  

\bibitem{Giudice:2011ak}
  G.~F.~Giudice, B.~Gripaios and R.~Sundrum,
  arXiv:1105.3161 [hep-ph].
  


\bibitem{Aaltonen:2008dn}
  T.~Aaltonen {\it et al.}  [CDF Collaboration],
  Phys.\ Rev.\  D {\bf 79}, 112002 (2009)
  [arXiv:0812.4036 [hep-ex]].
  
  
  
\bibitem{Khachatryan:2010jd}
  V.~Khachatryan {\it et al.}  [CMS Collaboration],
  Phys.\ Rev.\ Lett.\  {\bf 105}, 211801 (2010)
  [arXiv:1010.0203 [hep-ex]].


\bibitem{Aad:2011aj}
  G.~Aad {\it et al.}  [ATLAS Collaboration],
  New J.\ Phys.\  {\bf 13}, 053044 (2011)
  [arXiv:1103.3864 [hep-ex]].

\bibitem{ATLASnote095}  ATLAS~Collaboration,  ATLAS conference note ATLAS-CONF-2011-095.

\bibitem{CMSCollaboration:2011ns} CMS~Collaboration, CMS-EXO-11-015,
  arXiv:1107.4771 [hep-ex].
  
  \bibitem{:2009mh}
  V.~M.~Abazov {\it et al.}  [D0 Collaboration],
  Phys.\ Rev.\ Lett.\  {\bf 103}, 191803 (2009)
  [arXiv:0906.4819 [hep-ex]] .
  "EAPS Document with data: E-PRLTAO-103-025946"
   
\bibitem{Khachatryan:2011as}
  V.~Khachatryan {\it et al.}  [CMS Collaboration],
  arXiv:1102.2020 [hep-ex].
  
\bibitem{Chatrchyan:2011dk}
  S.~Chatrchyan {\it et al.}  [CMS Collaboration],
  arXiv:1106.2142 [hep-ex].
  
  
\bibitem{Kagan:2011yx}
A.~L.~Kagan, J.~F.~Kamenik, G.~Perez and S.~Stone,
arXiv:1103.3747 [hep-ph].

\bibitem{Arguin:2011xm}
  J.~F.~Arguin, M.~Freytsis and Z.~Ligeti,
  arXiv:1107.4090 [hep-ph].

  
  \bibitem{Lane:1991qh}
  K.~D.~Lane and M.~V.~Ramana,
  Phys.\ Rev.\  D {\bf 44}, 2678 (1991).
  
  
\bibitem{Lunghi:2010gv}
E.~Lunghi and A.~Soni,
Phys.\ Lett.\  B {\bf 697} (2011) 323
[arXiv:1010.6069 [hep-ph]].


    
\bibitem{particalkin}
E. Byckling, K. Kajantie, Particle Kinematics, John Wiley \& Sons Ltd (January 1, 1973)


\bibitem{Kagan:2009gb}
A.~L.~Kagan and M.~D.~Sokoloff,
Phys.\ Rev.\  D {\bf 80} (2009) 076008
[arXiv:0907.3917 [hep-ph]].


  
\bibitem{Bona:2007vi}
M.~Bona {\it et al.}  [UTfit Collaboration],
JHEP {\bf 0803} (2008) 049
[arXiv:0707.0636 [hep-ph]].
      
\bibitem{Laiho:2009eu}
J.~Laiho, E.~Lunghi and R.~S.~Van de Water,
Phys.\ Rev.\  D {\bf 81} (2010) 034503
[arXiv:0910.2928 [hep-ph]]; and 2011 updates at {\tt http://krone.physik.unizh.ch/$\sim$lunghi/webpage/LatAves/index.html}



\bibitem{Charles:2004jd}
J.~Charles {\it et al.}  [CKMfitter Group],
Eur.\ Phys.\ J.\  C {\bf 41} (2005) 1
[arXiv:hep-ph/0406184], we are using the ICHEP2010 update from {\tt http://ckmfitter.in2p3.fr}.

\bibitem{Inami:1980fz}
  T.~Inami, C.~S.~Lim,
  Prog.\ Theor.\ Phys.\  {\bf 65}, 297 (1981).
  



\bibitem{Zhang:2010kr}
  H.~Zhang, E.~L.~Berger, Q.~H.~Cao, C.~R.~Chen and G.~Shaughnessy,
  Phys.\ Lett.\  B {\bf 696}, 68 (2011)
  [arXiv:1009.5379 [hep-ph]].
  
 

  
\end{thebibliography}
\end{document}